\documentclass[aphy]{apjrnl}
\usepackage{amssymb}
\usepackage{amsfonts}
\usepackage{amsmath}
\usepackage{axodraw}
\usepackage{epsfig}
\usepackage{mcite}

\allowdisplaybreaks
\equationsection

{\theoremstyle{plain}%

{\theoremstyle{remark}

}
{\theoremstyle{definition}

}
\def\Mathematica{{\tt Mathematica}}
\def\Tracer{{\tt Tracer}}
\newcommand{\epsp}[3][ ]{\ensuremath{\epsilon^{#2}_{#3}(#1)}}
\newcommand{\epspm}{\ensuremath{\epsp[q]{\mu}{m}}}
\newcommand{\epspn}{\ensuremath{\epsp[p]{\nu}{n}}}
\newcommand{\epspmp}{\ensuremath{{\epsp[q]{\mu^\prime}{m^\prime}}}}
\newcommand{\epspnp}{\ensuremath{{\epsp[p]{\nu^\prime}{n^\prime}}}}
\newcommand{\Norm}{\ensuremath{\tilde{N}_q}\,}
\newcommand{\Normh}{\ensuremath{\tilde{N}_h}\,}

\newcommand{\Normc}{\ensuremath{\frac{Q^2}{4 \pi^2 \alpha}}}
\newcommand{\vgvg}{\ensuremath{\gamma^\star(Q^2)\gamma^\star(P^2)
\rightarrow q \bar{q}}}
\newcommand{\vgvgmu}{\ensuremath{\gamma^\star(Q^2)\gamma^\star(P^2)
\rightarrow \mu^+ \mu^-}}
\newcommand{\vgg}{\ensuremath{\gamma^\star(Q^2)\gamma \rightarrow q \bar{q}}}
\newcommand{\mev}{\ensuremath{\mathrm{MeV}}}

\newcommand{\gevsq}{\ensuremath{\mathrm{GeV^2}}}

\newcommand{\me}{\ensuremath{m_{\mathrm{e}}}}
\newcommand{\lamQCD}{\Lambda_{\rm QCD}}
\newcommand{\vone}{\ensuremath{\beta}}
\newcommand{\vtwo}{\ensuremath{\bar{\beta}}}
\newcommand{\xp}{\ensuremath{\delta}}

\newcommand{\vep}{\ensuremath{{\varepsilon}}}
\newcommand{\Ord}{\ensuremath{{\cal O}}}

\newcommand{\DISeg}{\ensuremath{\mathrm{DIS}_{e\gamma}}}
\newcommand{\X}{\ensuremath{\mathrm{X}}}

\newcommand{\PWTT}{\ensuremath{\mathrm{P^{\mu^\prime\nu^\prime,\mu\nu}_{\WTT}}}}
\newcommand{\PWTS}{\ensuremath{\mathrm{P^{\mu^\prime\nu^\prime,\mu\nu}_{\WTS}}}}
\newcommand{\PWST}{\ensuremath{\mathrm{P^{\mu^\prime\nu^\prime,\mu\nu}_{\WST}}}}
\newcommand{\PWSS}{\ensuremath{\mathrm{P^{\mu^\prime\nu^\prime,\mu\nu}_{\WSS}}}}
\newcommand{\PtTT}{\ensuremath{\mathrm{P^{\mu^\prime\nu^\prime,\mu\nu}_{\tTT}}}}
\newcommand{\PtTS}{\ensuremath{\mathrm{P^{\mu^\prime\nu^\prime,\mu\nu}_{\tTS}}}}
\newcommand{\PaTT}{\ensuremath{\mathrm{P^{\mu^\prime\nu^\prime,\mu\nu}_{\aTT}}}}
\newcommand{\PaTS}{\ensuremath{\mathrm{P^{\mu^\prime\nu^\prime,\mu\nu}_{\aTS}}}}
\newcommand{\ronepp}{\ensuremath{\rho_1^{++}}}
\newcommand{\ronemm}{\ensuremath{\rho_1^{--}}}
\newcommand{\ronepm}{\ensuremath{\rho_1^{+-}}}
\newcommand{\ronezz}{\ensuremath{\rho_1^{00}}}
\newcommand{\ronepz}{\ensuremath{\rho_1^{+0}}}
\newcommand{\rtwopp}{\ensuremath{\rho_2^{++}}}
\newcommand{\rtwomm}{\ensuremath{\rho_2^{--}}}
\newcommand{\rtwopm}{\ensuremath{\rho_2^{+-}}}
\newcommand{\rtwozz}{\ensuremath{\rho_2^{00}}}
\newcommand{\rtwopz}{\ensuremath{\rho_2^{+0}}}

\newcommand{\ri}{\ensuremath{\rho_{\mathrm{i}}}}
\newcommand{\ripp}{\ensuremath{\rho_{\mathrm{i}}^{++}}}
\newcommand{\ripm}{\ensuremath{\rho_{\mathrm{i}}^{+-}}}
\newcommand{\ripz}{\ensuremath{\rho_{\mathrm{i}}^{+0}}}
\newcommand{\rizz}{\ensuremath{\rho_{\mathrm{i}}^{00}}}

\newcommand{\ricontra}{\ensuremath{\rho_{\mathrm{i}}^{\alpha \beta}}}

\newcommand{\sigiil}{\ensuremath{\sigma_{\mathrm{2L}}}}
\newcommand{\sigiit}{\ensuremath{\sigma_{\mathrm{2T}}}}

\newcommand{\sigtt}{\ensuremath{\sigma_{\mathrm{TT}}}}
\newcommand{\sigtl}{\ensuremath{\sigma_{\mathrm{TL}}}}
\newcommand{\siglt}{\ensuremath{\sigma_{\mathrm{LT}}}}
\newcommand{\sigll}{\ensuremath{\sigma_{\mathrm{LL}}}}
\newcommand{\tautt}{\ensuremath{\tau_{\mathrm{TT}}}}
\newcommand{\tautl}{\ensuremath{\tau_{\mathrm{TL}}}}
\newcommand{\tauatt}{\ensuremath{\tau^{\mathrm{a}}_{\mathrm{TT}}}}
\newcommand{\tauatl}{\ensuremath{\tau^{\mathrm{a}}_{\mathrm{TL}}}}

\newcommand{\WIIT}{\ensuremath{W_{\mathrm{2T}}}}
\newcommand{\WIIS}{\ensuremath{W_{\mathrm{2L}}}}
\newcommand{\WTT}{\ensuremath{W_{\mathrm{TT}}}}
\newcommand{\WTS}{\ensuremath{W_{\mathrm{TL}}}}
\newcommand{\WST}{\ensuremath{W_{\mathrm{LT}}}}
\newcommand{\WSS}{\ensuremath{W_{\mathrm{LL}}}}
\newcommand{\tTT}{\ensuremath{W^\tau_{\mathrm{TT}}}}
\newcommand{\tTS}{\ensuremath{W^\tau_{\mathrm{TL}}}}
\newcommand{\aTT}{\ensuremath{W^{\mathrm{a}}_{\mathrm{TT}}}}
\newcommand{\aTS}{\ensuremath{W^{\mathrm{a}}_{\mathrm{TL}}}}

\newcommand{\Wcontra}{\ensuremath{W^{\mu^\prime\nu^\prime,\mu\nu}}}
\newcommand{\Wco}{\ensuremath{W_{\mu^\prime\nu^\prime,\mu\nu}}}
\newcommand{\Tcontra}{\ensuremath{T^{\mu^\prime\nu^\prime,\mu\nu}}}
\newcommand{\pisl}{\ensuremath{{p\!\!\!\!\;/}_{i}}}
\newcommand{\pipsl}{\ensuremath{{p\!\!\!\!\;/}_{i}^\prime}}
\newcommand{\e}{\ensuremath{e}}
\newcommand{\ee}{\ensuremath{\e^+ \e^-}}
\newcommand{\eetoeeX}{\ensuremath{\e^+\e^- \to \e^+ \e^- X}}
\newcommand{\twogam}{\ensuremath{\e^+\e^- \to \e^+\e^- \gam^\star \gam^\star \to \e^+ \e^- X}}
\newcommand{\Fee}{\ensuremath{F_{\e\e}}}
\newcommand{\ft}{\ensuremath{f_{\gamma_T/ \e}}}
\newcommand{\fl}{\ensuremath{f_{\gamma_L/ \e}}}
\newcommand{\qp}{\ensuremath{\xi_q^+}}
\newcommand{\qm}{\ensuremath{\xi_q^-}}
\newcommand{\ppl}{\ensuremath{\xi_p^+}}
\newcommand{\pmi}{\ensuremath{\xi_p^-}}

\newcommand{\qt}{\ensuremath{q_\perp}}
\newcommand{\pt}{\ensuremath{p_\perp}}
\newcommand{\piit}{\ensuremath{{p_2}_\perp}}
\newcommand{\pit}{\ensuremath{{p_1}_\perp}}

\newcommand{\AWTT}{\ensuremath{A_{W_{\mathrm{TT}}}}}
\newcommand{\CWTT}{\ensuremath{C_{W_{\mathrm{TT}}}}}
\newcommand{\EWTT}{\ensuremath{E_{W_{\mathrm{TT}}}}}

\newcommand{\AWTS}{\ensuremath{A_{W_{\mathrm{TL}}}}}
\newcommand{\CWTS}{\ensuremath{C_{W_{\mathrm{TL}}}}}
\newcommand{\EWTS}{\ensuremath{E_{W_{\mathrm{TL}}}}}

\newcommand{\AWST}{\ensuremath{A_{W_{\mathrm{LT}}}}}
\newcommand{\CWST}{\ensuremath{C_{W_{\mathrm{LT}}}}}
\newcommand{\EWST}{\ensuremath{E_{W_{\mathrm{LT}}}}}

\newcommand{\AWSS}{\ensuremath{A_{W_{\mathrm{LL}}}}}
\newcommand{\CWSS}{\ensuremath{C_{W_{\mathrm{LL}}}}}
\newcommand{\EWSS}{\ensuremath{E_{W_{\mathrm{LL}}}}}

\newcommand{\AtTT}{\ensuremath{A_{W^\tau_{\mathrm{TT}}}}}
\newcommand{\CtTT}{\ensuremath{C_{W^\tau_{\mathrm{TT}}}}}
\newcommand{\EtTT}{\ensuremath{E_{W^\tau_{\mathrm{TT}}}}}
\newcommand{\AtTS}{\ensuremath{A_{W^\tau_{\mathrm{TL}}}}}
\newcommand{\CtTS}{\ensuremath{C_{W^\tau_{\mathrm{TL}}}}}
\newcommand{\EtTS}{\ensuremath{E_{W^\tau_{\mathrm{TL}}}}}

\newcommand{\AaTT}{\ensuremath{A_{W^{\mathrm{a}}_{\mathrm{TT}}}}}
\newcommand{\CaTT}{\ensuremath{C_{W^{\mathrm{a}}_{\mathrm{TT}}}}}
\newcommand{\EaTT}{\ensuremath{E_{W^{\mathrm{a}}_{\mathrm{TT}}}}}
\newcommand{\AaTS}{\ensuremath{A_{W^{\mathrm{a}}_{\mathrm{TL}}}}}
\newcommand{\CaTS}{\ensuremath{C_{W^{\mathrm{a}}_{\mathrm{TL}}}}}
\newcommand{\EaTS}{\ensuremath{E_{W^{\mathrm{a}}_{\mathrm{TL}}}}}

\newcommand{\sfs}[3][F]{\ensuremath{#1_{\text{#2}}^{#3}}}
\newcommand{\gam}[1][]{\ensuremath{\gamma_{\mathrm{#1}}}}

\begin{document}
\begin{titlepage} 
\setlength{\parskip}{0.25cm} 
\setlength{\baselineskip}{0.25cm} 
\begin{flushright} 
DO-TH 2002/08\\ 
\vspace{0.2cm} 
hep-ph/0205301\\ 
\vspace{0.2cm} 
May 2002 
\end{flushright} 
\vspace{1.0cm} 
\begin{center} 
\LARGE 
{\bf Two-Photon Processes and Photon Structure}
\vspace{1.5cm} 
 
\large 
I.\ Schienbein\\ 
\vspace{1.0cm} 
 
\normalsize 
{\it Universit\"{a}t Dortmund, Institut f\"{u}r Physik,}\\ 
{\it D-44221 Dortmund, Germany} \\ 
\vspace{0.5cm}

\vspace{1.5cm} 
\end{center} 
 
\begin{abstract} 
\noindent 
In this article aspects of photon-photon physics related to
the structure of real and virtual photons are reviewed. 
A re-calculation of the virtual photon-photon box
is performed and some discrepancies in the literature are clarified.
A useful compilation of various relevant limits derived
from the most general expressions is provided.
Furthermore, structure functions of spin-averaged, transverse and
longitudinal virtual target photons are defined and discussed.
Finally, the factorization of two-photon processes
in the Bjorken limit is demonstrated to hold
also for the case of virtual target photons. 
\end{abstract} 
\end{titlepage} 

%%%%%%%%%%%%%%%%%%%%%%%%%%%%%%%%%%%%%%%%%%%%%%%%%
%\title{Two-Photon Processes and Photon Structure}

%\author{I.\ Schienbein
%\\Department of Physics
%\\University of Dortmund       %As many lines as needed
%\\D-44221 Dortmund, Germany
%\\E-mail: schien@zylon.physik.uni-dortmund.de
%}

%\date{22.05.2002} 

%\maketitle

%\begin{abstract}  
%In this article aspects of photon-photon physics related to
%the structure of real and virtual photons are reviewed. 
%A re-calculation of the virtual photon-photon box
%is performed and some discrepancies in the literature are clarified.
%A useful compilation of various relevant limits derived
%from the most general expressions is provided.
%Furthermore, structure functions of spin-averaged, transverse and
%longitudinal virtual target photons are defined and discussed.
%Finally, the factorization of two-photon processes
%in the Bjorken limit is demonstrated to hold
%also for the case of virtual target photons. 
%\end{abstract}
%\begin{keywords}
%two-photon processes, photon structure, 
%photon-photon box, doubly virtual box, quark parton model,
%virtual photons, 
%photon structure functions,
%factorization
%\end{keywords}

\section{Introduction}\label{chap:twogam}
At present, two-photon processes 
$\twogam$ in $\ee$ collider experiments 
are the main source of information
on the (hadronic and QED) structure of the photon. 
As has been first realized in \cite{Brodsky:1971vm,*Walsh:1971xy}
such processes {\em factorize} in the Bjorken limit 
into a flux of (quasi-)real target photons 
which can then be probed in deep inelastic
electron-photon scattering ($\DISeg$).
Analogously to the familiar
deep inelastic $ep$ scattering,
the unpolarized cross section for $\DISeg$, $e \gamma \to e X$, can be 
described in terms of two independent photon structure functions, say
$F_2^\gamma(x,Q^2)$ and $F_L^\gamma(x,Q^2)$. 
(There are two more structure functions
in the case of a (quasi-)real target photon:
$g_1^\gamma(x,Q^2)$ which can be measured in 
{\em polarized} $\DISeg$ again completely analogous to
the case of deep inelastic scattering off spin-$1/2$ targets 
like protons and a further structure function involving 
helicity-flips of the target photon's helicity
which is
accessible by the azimuthal angular dependence of 
the $e^+e^-$ cross section.)
Due to experimental constraints the $\DISeg$ cross sections are measured
at small values of the variable $y$ (defined as usual) where 
$F_L^\gamma(x,Q^2)$ is kinematically suppressed due to a pre-factor 
$\propto y^2$ in the cross section and it is the structure function
$F_2^\gamma(x,Q^2)$ which is measured.
The $\DISeg$ data on the photon structure function $F_2^\gamma(x,Q^2)$
are mainly sensitive to the up-quark density $u^\gamma(x,Q^2)$ 
in the photon
as can
be seen from the parton model expression for $F_2^\gamma$ (in LO),
$F_2^\gamma \propto 4 u^\gamma + d^\gamma + s^\gamma$,
whereas the gluon distribution $g^\gamma(x,Q^2)$ is only indirectly 
constrained at small values of $x$ due to the evolution.
Complementary information has become available in the last years due to 
'resolved photon processes' 
(e.g.\ production of (di-)jets or hadrons with large $p_T$ ($E_T$),
heavy quark production, isolated prompt photon production)
in $\gamma p$ and $\gamma \gamma$ collisions at the $ep$ collider HERA 
and the $e^+e^-$ collider LEP, respectively,
which are mainly sensitive to the gluon distribution 
in the photon. 
Triggered by 
the wealth of experimental results from the $e^+ e^-$ collider LEP  
and the $ep$ collider HERA \cite{Nisius:1999cv,Krawczyk:2000mf} 
which became available in the last decade several
QCD analyses of photonic parton distributions 
\cite{cit:GRV-9201,*cit:GRV-9202,cit:AFG-9401,cit:SaS-9501,*Schuler:1996fc,
Gordon:1997pm,cit:GRSc99} 
have been performed.
Recently some effort has also been devoted to the polarized structure
function $g_1^\gamma$ 
\cite{Gluck:1992fy,*Gluck:1993zq,*Gluck:1994ee,Stratmann:1996an,
Gluck:2001az,*Gluck:2001rn}.
The polarized parton distributions $\Delta f^\gamma(x,Q^2)$ will be 
measurable at future polarized $\ee$ and
$\e p$ colliders, see  
\cite{Stratmann:1999bv,*Stratmann:1997xy,*Stratmann:2000yd} and
references therein.

So far, the discussion has referred to the most important case
of (quasi-)real photon targets.
However, there are good reasons for considering
the general case of virtual photon targets $\gam(P^2)$ where
$P^2$ is the virtuality of the target photon:
% (i)
First of all, 
at present $e^+ e^-$ and $ep$ collider experiments the photon beams
consist of bremsstrahlung radiated off the incident lepton beam
resulting in a {\em continuous spectrum} of target photons $\gamma(P^2)$. 
The bremsstrahlung spectrum is proportional to $1/P^2$ such that the
bulk of target photons is produced at $P^2 \simeq P^2_{\rm min} \simeq 0$.
The parton content of such (quasi-)real photons is 
well established both experimentally and theoretically
and 
in general one expects
\cite{Borzumati:1993za,Drees:1994eu,cit:GRS95,
cit:SaS-9501,*Schuler:1996fc,cit:GRSc99}
%\cite{Borzumati:1993za,Drees:1994eu,cit:GRS95,
%cit:SaS-9501,*Schuler:1996fc,cit:GRSc99,
%Chyla:2000hp,*Chyla:2000cu,*Chyla:2000ue}
also {\em virtual} photons to possess a parton content
smoothly depending on $P^2$.
In this sense the real photon $\gamma \equiv \gamma(P^2 \simeq 0)$ is
just a
'primus inter pares' and
unified approaches to the parton content
of virtual photons $\gamma(P^2)$ which comprise the real photon
case in the limit $P^2 \to 0$ 
\cite{cit:GRSc99,cit:GRS95,cit:SaS-9501,*Schuler:1996fc,
Gluck:2001az,*Gluck:2001rn}
are highly desirable.
This is also reflected by the fact that measurements of
the real photon structure function $F_2^\gamma$ in single-tag events
integrate over the bremsstrahlung spectrum from $P^2_{\rm min}$ up to
a $P^2_{\rm max}$ which depends on the experimental details.
For instance, at LEP1(LEP2) $P^2_{\rm max}$ is as large as
$P^2_{\rm max} \simeq 1.5\ \gevsq$($4.5\ \gevsq$),
cf.\ Section 2.2 in \cite{Nisius:1999cv}.
Although the bulk of photons is produced at $P^2 \simeq P^2_{\rm min}$
the amount of ignorance of the $P^2$-dependence, mainly in the
range $P^2 \lesssim \Lambda^2$, 
where $\Lambda$ is a typical hadronic scale,
feeds back on the determination
of the structure function $F_2^\gamma$ (parton distributions) 
of (quasi-)real photons.
% (ii)
Furthermore, the case of deeply virtual target photons
$Q^2 \gg P^2 \gg \Lambda^2$
has attracted a lot of theoretical interest 
in the unpolarized 
\cite{Uematsu:1981qy,*Uematsu:1982je,
Rossi:1984xz,cit:Rossi-PhD,Ibes:1990pj}
as well as in the polarized case
\cite{Sasaki:1998vb,*Sasaki:1999py,*Sasaki:2000zc,Gluck:2001az,*Gluck:2001rn}
because it is purely perturbative allowing for absolute QCD
predictions.

% aim of the work
The aim of this work is to review
aspects of two-photon physics related to
the structure of real and virtual photons. 
The description of virtual target photons
necessitates a more involved theoretical framework
due to the longitudinal photon polarization and the more
complicated kinematics. 
The paper is organized as follows:
In Section \ref{sec:twogam} we give a comprehensible introduction 
into two-photon processes.
After defining the required kinematical variables, we discuss
the 4-photon amplitude and 
derive the general cross section for two-photon processes.
Furthermore, we perform a re-calculation of the
doubly virtual box $\vgvg$ in lowest order perturbation theory.
Thereby we clarify two discrepancies in the literature 
\cite{cit:Bud-7501,cit:Rossi-PhD}.
The general results of this calculation 
are casted in a form which easily allows to read off various important
limits, e.g., the quark-parton model (QPM) results
for the structure functions of real and virtual photons and the
heavy quark contributions to the photon structure functions.
A useful compilation of these limits can be found in 
Appendix \ref{app:limits}.
The notation of this section follows mainly the 
report of Budnev et al.~\cite{cit:Bud-7501}.
Next, in Sec.~\ref{sec:photonsfs} we define structure functions
for virtual photons. Beside the spin-averaged target usually considered
in the literature we discuss the possibility of defining structure
functions for transverse and longitudinal target photons in view
of the recent approaches in the literature to the parton content
of virtual longitudinal photons 
\cite{Chyla:2000hp,*Chyla:2000cu,*Chyla:2000ue,Friberg:2000nx}.
This will also be of importance in 
Sec.~\ref{sec:fact1} where we investigate the factorization 
of two-photon processes 
in the 'generalized' Bjorken limit $P^2 \ne 0$, 
$Q^2 \to \infty, x = \text{fixed}$.
As we will see, the cross section factorizes also in this case
of non-zero virtualities $P^2$ of the 'target photon'
into fluxes of transversely and longitudinally polarized photons times
the corresponding cross sections for deep inelastic scattering off 
these target photons up to terms of the order $\Ord(\sqrt{P^2/Q^2})$.
It should be noted that the factorization is essential for 
a theoretical description of two-photon processes in terms of 
(virtual) photon structure functions which can be measured in
deep inelastic electron-(virtual) photon scattering.
Finally, in Sec.~\ref{sec:conclusions} we summarize the main 
results and draw some conclusions.
\section{Two-Photon Processes}\label{sec:twogam}
\subsection{Kinematics}\label{sec:kinematics}
\begin{figure}[ht]
\begin{center}
\begin{picture}(200,200)(0,0)
% upper electron
\ArrowLine(20,180)(100,160)
\ArrowLine(100,160)(180,180)
\Vertex(100,160){1.5}
% upper photon
\Photon(100,160)(100,120){4}{5}
% lower electron
%\ArrowLine(20,20)(100,40)
%\ArrowLine(100,40)(180,20)
\ArrowLine(180,20)(100,40)
\ArrowLine(100,40)(20,20)
\Vertex(100,40){1.5}
% lower photon
\Photon(100,40)(100,80){4}{5}
% blub
\Line(120,100)(150,100)
\Line(117,110)(150,115)
\Line(117,90)(150,85)
\GCirc(100,100){20}{0.5}
%
%\Text(80,60)[l]{\large $p \uparrow$}
%\Text(80,140)[l]{\large $q \downarrow$}
\Text(95,60)[r]{\large $p \uparrow$}
\Text(95,140)[r]{\large $q \downarrow$}
\Text(160,100)[l]{$\Bigg\}\ $\large \X}
\Text(0,180)[l]{\large $p_1$}
\Text(0,20)[l]{\large $p_2$}
\Text(190,180)[l]{\large $p_1^\prime$}
\Text(190,20)[l]{\large $p_2^\prime$}
\Text(120,140)[l]{$-q^2 = Q^2 \equiv Q_1^2$}
\Text(-10,100)[l]{$W^2 = (q+p)^2$}
\Text(120,60)[l]{$-p^2 = P^2 \equiv Q_2^2$}
\end{picture}
\end{center}
\caption{\sf Two-photon particle production. The solid lines are the incoming and
outgoing leptons and the wavy lines are virtual photons which produce
a final state $\X$ consisting of hadrons (or leptons).}
\label{fig:kin}
\end{figure}
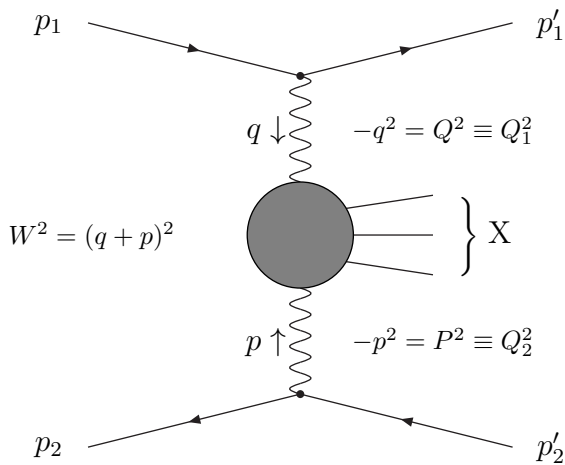

The kinematics of particle production via photon-photon scattering in
$\ee$ collisions 
\begin{equation}
\begin{aligned}
\e^{-}(p_1)\e^{+}(p_2)\rightarrow 
\e^{-}(p_1^{\prime})\e^{+}(p_2^{\prime}) \gamma^\star(q)\gamma^\star(p) 
\rightarrow 
\e^{-}(p_1^{\prime})\e^{+}(p_2^{\prime}) \X(p_{\X})
\end{aligned}
\label{eq:twogam}
\end{equation}
is depicted in Fig.~\ref{fig:kin}.
% Kinematics of e^- + e^+ -> e^- + e^+ + \gamma\gamma
The momenta of the incoming and outgoing leptons are denoted by
$p_i \equiv (E_i,\vec{p_i})$ and 
$p_i^\prime \equiv (E_i^\prime,\vec{p_i}^\prime)$ ($i = 1,2$) respectively and
the momenta of the photons are given by
\begin{equation}\label{eq:kinematics1}
\begin{aligned}
q & \equiv p_1 - p_1^\prime\ ,\, \qquad \, Q^2 &= -q^2 \equiv Q_1^2&  \ ,
\\
p & \equiv p_2 - p_2^\prime\ ,\, \qquad \, P^2 &= -p^2  \equiv Q_2^2 &\ .
\end{aligned}
\end{equation} 
%In general both photons have space-like momenta 
%and the particles are ordered such that $P^2$ refers to the photon 
%with smaller virtuality ($0 < P^2 \le Q^2$).
In general both photons have space-like momenta 
and $P^2$ refers to the photon with smaller virtuality ($P^2 \le Q^2$).
%($0 < P^2 \le Q^2$).
\X\ denotes the final state produced in the 
$\gamma^\star(q)+\gamma^\star(p) \rightarrow \X$ subprocess.
For later use, we define the following variables:
\begin{gather}\label{eq:variables}
\nu = p \cdot q\ ,\, \quad
x = \frac{Q^2}{2 \nu}\ ,\, \quad \, \xp = \frac{P^2}{2 \nu}\ ,\, \quad
y_1 =  \frac{p \cdot q}{p \cdot p_1}\ ,\, \quad \, 
y_2 = \frac{p \cdot q}{q \cdot p_2}\ ,
\nonumber\\
W^2 \equiv (p+q)^2 = 2 \nu (1-x-\xp) 
= Q^2 \frac{1-x}{x} - P^2 \ .
\end{gather}

% electron electron scattering
%\vspace*{1cm}
%\SetScale{1.0}
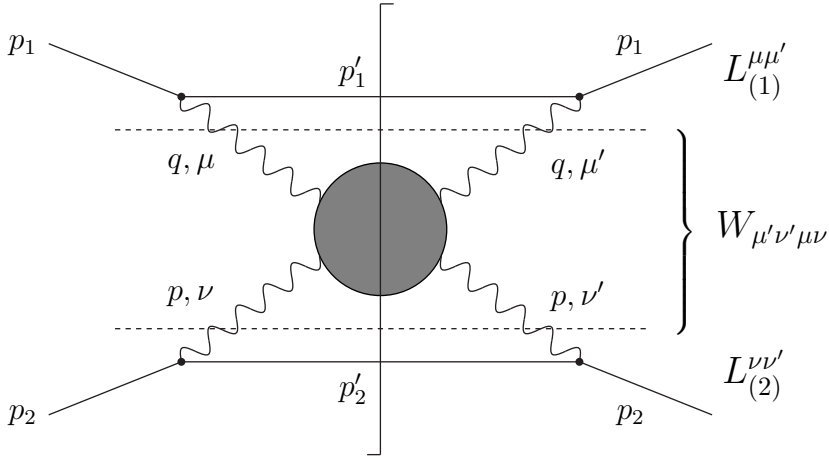
\begin{figure}[ht]
\begin{center}
\begin{picture}(300,200)(10,0)

\GCirc(150,100){25}{0.5}
%upper left photon
\Photon(75,150)(127,110){-4}{5}
%lower left photon
\Photon(75,50)(127,90){4}{5}
%upper right photon
\Photon(173,110)(225,150){4}{5}
%lower right photon
\Photon(173,90)(225,50){-4}{5}
%
% upper electron line
\Line(75,150)(225,150)
\Line(25,170)(75,150)
\Line(225,150)(275,170)
\Vertex(75,150){1.5}
\Vertex(225,150){1.5}
% lower electron line
\Line(75,50)(225,50)
\Line(25,30)(75,50)
\Line(225,50)(275,30)
\Vertex(75,50){1.5}
\Vertex(225,50){1.5}
%
% cut
\Line(150,185)(150,15)
\Line(150,185)(155,185)
\Line(150,15)(145,15)
\DashLine(50,137.5)(250,137.5){2}
\DashLine(50,62.5)(250,62.5){2}

\Text(280,160)[l]{\Large $L_{(1)}^{\mu\mu^\prime}$}
\Text(280,45)[l]{\Large $L_{(2)}^{\nu\nu^\prime}$}
%\Text(280,100)[l]{\Large $W_{\mu^{\prime}\nu^{\prime}\mu\nu}$}
\Text(260,100)[l]{\parbox{0.5cm}{$\left.\rule{0cm}{1.5cm} \right\}$}
{\Large $W_{\mu^{\prime}\nu^{\prime}\mu\nu}$}}

\Text(70,125)[l]{\large $q,\mu$}
\Text(70,75)[l]{\large $p,\nu$}
\Text(215,125)[l]{\large $q,\mu^\prime$}
\Text(215,75)[l]{\large $p,\nu^\prime$}
\Text(10,170)[l]{\large $p_1$}
\Text(10,30)[l]{\large $p_2$}
\Text(135,160)[l]{\large $p_1^{\prime}$}
\Text(135,40)[l]{\large $p_2^{\prime}$}
\Text(240,170)[l]{\large $p_1$}
\Text(240,30)[l]{\large $p_2$}

\end{picture}
\end{center}
\caption{\sf Squared matrix element of the process~(\ref{eq:twogam}).
Integration over the phase space of the system \X\ is implied
as indicated by the vertical cut.}
\label{fig:Xsec}
\end{figure}
It should be noted that two-photon processes comprise only a subset
of Feynman diagrams contributing to the physically observable
process $\eetoeeX$.
Other contributions arise from $t$- and $s$-channel bremsstrahlung
diagrams 
where the bremsstrahlung photon produces the hadronic final state
(see, e.g., Fig.~1 and the discussion in \cite{Bhattacharya:1977re})
and from diagrams where one (or both) of the photons is
replaced by a $Z$ boson. 
However, applying appropriate cuts these contributions are negligible.
At first, the $s$-channel diagram is suppressed by a factor $1/s$ and
is therefore very small.
Furthermore, $Z$-exchange can be neglected as long as the boson 
virtualities are not too large, say $Q^2 < 500\ \gevsq$.
On the other hand, the $t$-channel bremsstrahlung contribution 
deserves more care in certain kinematical regions.
However, as long as the scattering angle of at least one of the leptons is
small the direct two-photon process is strongly enhanced by 
a collinear logarithm $\ln E/m_e$ such that the cross section 
for $\eetoeeX$
is dominated by the two-photon mechanism.
In the following we do not distinguish between two-photon processes 
and the measurable $\eetoeeX$.

The cross section for the process~(\ref{eq:twogam})
is given by
\begin{equation}
d\sigma = \frac{1}{F_{ee}} |M|^2 dQ^{(n+2)}
\end{equation}
with the invariant matrix element $M$, the M{\o}ller flux factor of the two incoming leptons
\begin{equation}
\Fee = 4 \sqrt{(p_1 \cdot p_2)^2 - m_e^2 m_e^2}\ ,
\end{equation}
and the Lorentz-invariant $(n+2)$-particle phase space
\begin{equation}
\begin{aligned}
d Q^{(n+2)} & (p_1+p_2;p_1^{\prime},p_2^{\prime},k_1,\ldots,k_n)
\\
&=
\frac{d^3 p_1^{\prime}}{(2 \pi)^3 2 E_1^{\prime}}
\frac{d^3 p_2^{\prime}}{(2 \pi)^3 2 E_2^{\prime}}
d Q^{(n)}(q+p;k_1,\ldots,k_n)
\\
& =
\frac{d^3 p_1^{\prime}}{(2 \pi)^3 2 E_1^{\prime}}
\frac{d^3 p_2^{\prime}}{(2 \pi)^3 2 E_2^{\prime}}
%(2\pi)^4\delta^{(4)}\Big(q+p-\sum_{i=1}^n k_i\Big)
%\prod^n_{i=1}\frac{d^3 k_i}{(2\pi)^3 2 k_i^0}\ .
(2\pi)^4\delta^{(4)}\left(q+p-p_{\X} \right)d\Gamma\ .
\end{aligned}
\end{equation}
Here $p_{\X} = \sum_i k_i, i \in \X$ is the total momentum  
and $d\Gamma = \prod_i \frac{d^3 k_i}{2 k_i^0 (2 \pi)^3}, i \in \X$
the phase space volume of the produced system \X.

The cross section can be expressed in terms of the 
amplitudes $M^{\mu \nu}$ of the 
$\gamma^\star(q)+\gamma^\star(p) \rightarrow \X$ subprocess as follows
(see Fig.~\ref{fig:Xsec}):
\begin{equation}
d\sigma =    
\frac{d^3 p^{\prime}_1 d^3 p^{\prime}_2}{2 E^{\prime}_1 2 E^{\prime}_2 
(2 \pi)^6}
\frac{(4 \pi \alpha)^2}{Q^4 P^4} 
\frac{1} {F_{ee}} 
%\frac{1} {4 [(p_1 \cdot p_2)^2 - m_e^2 m_e^2]^{1/2}} 
L_{(1)}^{\mu \mu^\prime} L_{(2)}^{\nu \nu^\prime} 
\Wco
\label{eqn:eeXsec1}
\end{equation}
with %\cite{cit:Bud-7501,cit:Rossi-PhD} 
\begin{equation}
\Wcontra = \frac{1}{2} \int 
{M^\star}^{\mu^\prime \nu^\prime}M^{\mu\nu} (2 \pi)^4 \delta^{(4)}(q+p-p_{\X})
d\Gamma\ .
\label{eq:defw}
\end{equation}
For unpolarized leptons
the (leptonic) tensors $L_{(1)}$ and $L_{(2)}$ (see Fig.~\ref{fig:Xsec})
are given by
\begin{eqnarray}
L_{(i)}^{\alpha\beta} &=&
\frac{1}{2} Tr[(\pisl + \me) \gamma^\alpha (\pipsl + \me) \gamma^\beta]\ .
\label{eq:lalbe}
\end{eqnarray}
The factor $1/2$ is due to a spin average over the incoming leptons.
In addition, we introduce the dimensionless quantities \cite{cit:Bud-7501}
\begin{equation}
\ricontra = \frac{1}{Q_i^2} L_{(i)}^{\alpha\beta} 
= -\left(g^{\alpha \beta} - \frac{q_i^\alpha q_i^\beta}{q_i^2}\right) 
- \frac{(2 p_i - q_i)^\alpha (2 p_i - q_i)^\beta}{q_i^2}
\label{eq:rhoalbe}
\end{equation}
which have the interpretation of (unnormalized) density matrices for
the corresponding virtual photons.
\subsection{The Hadronic Tensor \Wcontra}
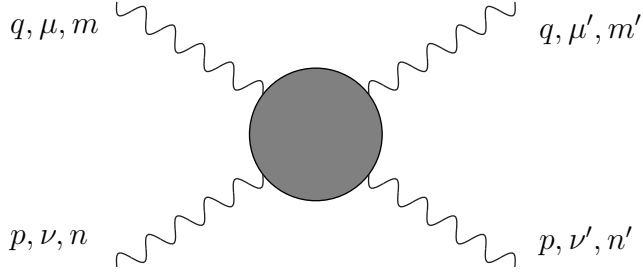
\begin{figure}[ht]
\begin{center}
\begin{picture}(300,100)(0,0)
\GCirc(150,50){25}{0.5}
%upper left photon
\Photon(75,100)(130,65){-4}{5}
%lower left photon
\Photon(75,0)(130,35){4}{5}
%upper right photon
\Photon(170,65)(225,100){4}{5}
%lower right photon
\Photon(170,35)(225,0){-4}{5}
\Text(35,90)[l]{\large $q,\mu,m$}
\Text(35,10)[l]{\large $p,\nu,n$}
\Text(235,90)[l]{\large $q,\mu^\prime,m^\prime$}
\Text(235,10)[l]{\large $p,\nu^\prime,n^\prime$}
\end{picture}
\end{center}
\caption{\sf The photon-photon forward scattering amplitude \Tcontra;
$m$, $n$, $m^\prime$, and $n^\prime$ are helicity indices. 
The tensor 
$\Wcontra$ defined in (\protect\ref{eq:defw}) is the absorptive part of
the $\gamma\gamma$ forward amplitude: $\Wcontra = \frac{1}{\pi} \mathrm{Im} \Tcontra$.}
\label{fig:walbemunu}
\end{figure}

According to the optical theorem \Wcontra\ is the absorptive part of the
virtual $\gamma\gamma$ forward amplitude shown in Fig.~\ref{fig:walbemunu}.
Taking into account P- and  T-invariance
(symmetry $\mu^\prime\nu^\prime \leftrightarrow \mu\nu$) and
gauge invariance, i.e.
\begin{displaymath}
q_\mu \Wcontra =  q_{\mu^\prime} \Wcontra = p_\nu \Wcontra =
p_{\nu^\prime} \Wcontra = 0\ ,
\end{displaymath}
the Lorentz-tensor $\Wcontra$ can be expanded in terms of a 
basis of 8 independent 
tensors constructed from the vectors $q$, $p$ and the metric tensor $g$.
%This expansion 
The choice of the tensor basis in which the expansion is carried out
is clearly arbitrary and different forms are 
discussed in the literature 
\cite{cit:Bud-7501,Brown:1971pk,Carlson:1971pk}.
In the following we will stick to the very physical expansion
given in \cite{cit:Bud-7501}:
\begin{equation}
\begin{aligned}
\Wcontra & = R^{{\mu^\prime}\mu}R^{{\nu^\prime}\nu}\WTT
+R^{{\mu^\prime}\mu}Q_2^\nu Q_2^{\nu^\prime} \WTS + Q_1^{\mu^\prime} Q_1^\mu
R^{{\nu^\prime}\nu} \WST 
\\
& + Q_1^{\mu^\prime} Q_1^\mu Q_2^{\nu^\prime} Q_2^\nu \WSS 
\\
& + \frac{1}{2}(R^{{\mu^\prime}{\nu^\prime}}R^{\mu\nu}
+R^{{\mu^\prime}\nu}R^{{\nu^\prime}\mu}
-R^{{\mu^\prime}\mu}R^{{\nu^\prime}\nu})\tTT\\
&-(R^{\mu\nu}Q_1^{\mu^\prime}
Q_2^{\nu^\prime}+R^{{\nu^\prime}\mu}Q_1^{\mu^\prime} Q_2^\nu
+R^{{\mu^\prime}{\nu^\prime}}Q_1^\mu
Q_2^\nu+R^{{\mu^\prime}\nu}Q_2^{\nu^\prime}
Q_1^\mu)\tTS \\ 
& +(R^{{\mu^\prime}{\nu^\prime}}R^{\mu\nu}-R^{{\nu^\prime}\mu}
R^{{\mu^\prime}\nu})\aTT
\\
& -(R^{\mu\nu}Q_1^{\mu^\prime}
Q_2^{\nu^\prime}-R^{{\nu^\prime}\mu}Q_1^{\mu^\prime} Q_2^\nu
+R^{{\mu^\prime}{\nu^\prime}}Q_1^\mu
Q_2^\nu-R^{{\mu^\prime}\nu}Q_2^{\nu^\prime} Q_1^\mu)\aTS\ 
\end{aligned}
\label{eq:budtensor}
\end{equation}
with $\vtwo^2 \equiv 1 - 4 x \xp$ and
\begin{gather}
R^{\alpha\beta}=  -g^{\alpha\beta}+
\frac{\nu (p^\alpha q^\beta+q^\alpha p^\beta)
-p^2 q^\alpha q^\beta-q^2 p^\alpha p^\beta}{\nu^2 \vtwo^2}\, ,
\nonumber\\
Q_1^\alpha=
\frac{\sqrt{-q^2}}{\nu \vtwo}
\left(p^\alpha-\frac{\nu}{q^2}q^\alpha\right)
\ , \ %\qquad
Q_2^\alpha=\frac{\sqrt{-p^2}}{\nu \vtwo}
\left(q^\alpha-\frac{\nu}{p^2}p^\alpha\right)\ .
\label{eq:rmunu}
\end{gather}

\subsubsection{Construction of the Tensor}\label{sec:construction}
Since it is quite insightful we now describe the main ideas of
the construction of the photon tensor from the 
photon photon helicity amplitudes 
\cite{cit:Bud-7501} which have simple physical 
interpretations. A nice discussion of the helicity amplitude formalism
can also be found in Ref.~\cite{Manohar:1992tz}.

In Fig.~\ref{fig:walbemunu}
$m, n$ and $m^\prime, n^\prime$ are the helicities of the
initial state and final state photons, respectively,
which can adopt the values $m, n, m^\prime, n^\prime = 0,\pm 1$
and are constrained by 
%total helicity conservation
angular momentum conservation: 
$m^\prime - n^\prime = m - n$.
The helicity amplitudes $W_{m^\prime n^\prime,mn}$ are related to the
photon tensor in the following way
\begin{equation}\label{eq:helicity}
W_{m^\prime n^\prime,mn} \equiv \epspmp^\star\ \epspnp^\star\ 
\Wco\ \epspm\ \epspn \, .
\end{equation}
Due to P- and  T-invariance and helicity conservation we have
\begin{equation*}
W_{m^\prime n^\prime,m n} 
\overset{P}{=} (-1)^{m^\prime-n^\prime+m-n} W_{-m^\prime -n^\prime,-m -n}
= W_{-m^\prime -n^\prime,-m -n} 
\overset{T}{=}W_{m n,m^\prime n^\prime}
\end{equation*}
such that there are 8 independent helicity amplitudes, say
$W_{++,++}$, $W_{+-,+-}$, $W_{+0,+0}$, $ W_{0+,0+}$, $W_{00,00}$, 
$W_{++,--}$, $W_{++,00}$, $W_{0+,-0}$
%$W_{++,++},\ W_{+-,+-},\ W_{+0,+0},\ W_{0+,0+},\ W_{00,00},\ 
%{\underbrace{W_{++,--},\ W_{++,00},
%\ W_{0+,-0}}_{\rm{helicity\ flips}}}$
where the latter three involve helicity flips. 
In addition, there are three independent 
positivity constraints on these
amplitudes due to the
Cauchy-Schwarz inequality 
$|W_{m^\prime n^\prime,mn}| \le 
\sqrt{W_{m^\prime n^\prime,m^\prime n^\prime}W_{mn,mn}}$
\cite{Sasaki:2001pc}.
For instance, we find for the above helicity amplitudes the relations
$|W_{++,--}| \le W_{++,++}$, 
$|W_{++,00}| \le \sqrt{W_{++,++}W_{00,00}}$, and 
$|W_{0+,-0}| \le \sqrt{W_{+0,+0}W_{0+,0+}}$.
Due to the completeness and orthonormality relations for
(space-like) polarization vectors
Eq.~\eqref{eq:helicity} can be 'inverted' resulting in a 
%most physical construction 
very nice representation
of $\Wco$ through the helicity amplitudes:
\begin{equation}\label{eq:wcontra}
\begin{aligned}
\Wcontra = &\sum_{{m^\prime,n^\prime,m,n}}
C(m^\prime,n^\prime,m,m)\ \epspmp\ \epspnp\ \epspm^\star\ \epspn^\star\
W_{m^\prime n^\prime,m n}
\end{aligned}
\end{equation}
where for a space-like target
$C(m^\prime,n^\prime,m,m) = (-1)^{m^\prime+n^\prime+m+n}=1$.
Note that gauge invariance is satisfied by construction due to
$q \cdot \epsilon(q) = p \cdot \epsilon(p)=0$.
Now the photon polarization vectors in the $\gam\gam$-CMS can be
written in a covariant way, 
see App.~B and C in \cite{cit:Bud-7501}:
\begin{gather}
\epsp[q]{\alpha}{0} = i Q_1^\alpha\ , \ %\qquad 
\epsp[p]{\alpha}{0} = -i Q_2^\alpha\ , \ %\qquad
\nonumber\\
\epsp[q]{\star \alpha}{\pm}\epsp[q]{\beta}{\pm} = 
\frac{1}{2} \left[R^{\alpha\beta} \pm i \frac{1}{\nu \vtwo} \varepsilon^{\alpha\beta\rho\sigma}
q_{\rho} p_{\sigma}\right]\ ,\ %\quad
\epsp[p]{\star \alpha}{\pm}\epsp[p]{\beta}{\pm} = 
\epsp[q]{\star \alpha}{\mp}\epsp[q]{\beta}{\mp}   
\label{eq:photonpol}
\end{gather}
with 
$R^{\alpha\beta}$, $Q_{1,2}^\alpha$ from Eq.~\eqref{eq:rmunu}.
Using these relations in Eq.~\eqref{eq:wcontra} we arrive at the
final tensor given in Eq.~\eqref{eq:budtensor}.
The dimensionless invariant functions $W_{ab}$ depend only on the
invariants $W^2$, $Q^2$ and $P^2$ and are related to the 
$\gamma\gamma$-helicity amplitudes
$W_{m^\prime n^\prime,mn}$ in the $\gamma\gamma$-CMS 
via \cite{cit:Bud-7501}
\begin{equation}
\begin{aligned}
\WTT &= \frac{1}{2}(W_{++,++}+W_{+-,+-})\ , %\ %\quad 
&\WTS &= W_{+0,+0}\ ,
\\
\WST &= W_{0+,0+}\ , &\WSS &= W_{00,00}\ ,
\\
\tTT &= W_{++,--}\ , &\tTS &= \frac{1}{2}(W_{++,00}+W_{0+,-0})\ ,
\\
\aTT &= \frac{1}{2}(W_{++,++}-W_{+-,+-})\ , 
&\aTS &= \frac{1}{2}(W_{++,00}-W_{0+,-0})\ .
\end{aligned}
\label{eq:amptoheli}
\end{equation}
The amplitudes $\tTT$, $\tTS$, and $\aTS$ correspond to transitions 
with spin flip for
each of the photons (with total helicity conservation).
As we will see in the next section only 6 of these amplitudes 
($\WTT$, $\WTS$, $\WST$, $\WSS$, $\tTT$, $\tTS$)
enter the cross section for unpolarized lepton beams because the tensors in
(\ref{eq:rhoalbe}) are symmetric whereas the tensor structures 
multiplying
$\aTT$ and $\aTS$ in (\ref{eq:budtensor}) are anti-symmetric 
such that these 
terms do not contribute when
the leptonic and the hadronic tensors are contracted.
Only if the initial leptons are polarized, can the amplitudes 
$\aTT$ and $\aTS$ 
be measured as well \cite{cit:Bud-7501}.

Finally, it is noteworthy that the helicity amplitudes obviously
satisfy the symmetry [just turn Fig.~\ref{fig:walbemunu} upside down] 
\begin{equation}\label{eq:upsidedown}
W_{m^\prime n^\prime,mn}(q,p) =
W_{n^\prime m^\prime,nm}(p,q) \ .
\end{equation}
As a consequence of this and the other above stated symmetries
we find that the invariant functions $W_{ab}$ are symmetric
under exchanging $q \leftrightarrow p$, i.e.\ $W_{ab}(q,p) = W_{ab}(p,q)$, 
with exception of
$\WST$ and $\WTS$ which satisfy $\WST(q,p) = \WTS(p,q)$.
\subsubsection{Projection Operators}
The photon momenta $q,p$, the unit vectors $Q_1,Q_2$, and the symmetric 
tensor $R^{\alpha \beta}$ satisfy the following (orthogonality) relations:
\begin{gather}
q \cdot Q_1 = p \cdot Q_2 = 0\ , \ %\qquad 
Q_{1,2}^2 = 1\ ,
\nonumber\\
q^{\alpha} R_{\alpha\beta} = p^{\alpha} R_{\alpha\beta} =
Q_{1,2}^{\alpha} R_{\alpha\beta} = 0
\ , \ %\qquad
R_{\alpha\beta} R^{\alpha\beta} = 2\ , \ %\qquad 
R^{\alpha}_{\beta} R^{\beta\gamma} = -R^{\alpha\gamma}\ .
\label{eq:orthogonality}
\end{gather}
With help of these relations it is easy to see that the various
tensors in front of the invariant functions $W_{ab}$ in 
Eq.~(\ref{eq:budtensor}) are mutually orthogonal and therefore 
can be used to project out the invariant functions.
This is also a direct consequence of the construction
in Eq.~\eqref{eq:wcontra} due to the orthonormality of the polarization
vectors.
%\clearpage
%\pagebreak
With obvious notation ($\PWTT \Wco = \WTT$ etc.) the projectors read:
\begin{align}
\PWTT &= \frac{1}{4} R^{{\mu^\prime}\mu}R^{{\nu^\prime}\nu}
\ , 
\nonumber\\
\PWTS &= \frac{1}{2} R^{{\mu^\prime}\mu}Q_2^\nu Q_2^{\nu^\prime} \ , 
\nonumber\\
\PWST &=  \frac{1}{2} Q_1^{\mu^\prime} Q_1^\mu R^{{\nu^\prime}\nu} 
\ , 
\nonumber\\
\PWSS &= Q_1^{\mu^\prime} Q_1^\mu Q_2^{\nu^\prime} Q_2^\nu \ , 
\nonumber\\
\PtTT &=   \frac{1}{4}(R^{{\mu^\prime}{\nu^\prime}}R^{\mu\nu}
+R^{{\mu^\prime}\nu}R^{{\nu^\prime}\mu}
-R^{{\mu^\prime}\mu}R^{{\nu^\prime}\nu})\ , 
\nonumber\\
\PtTS &= 
\frac{-1}{8}(R^{\mu\nu}Q_1^{\mu^\prime}
Q_2^{\nu^\prime}+R^{{\nu^\prime}\mu}Q_1^{\mu^\prime} Q_2^\nu
+R^{{\mu^\prime}{\nu^\prime}}Q_1^\mu
Q_2^\nu+R^{{\mu^\prime}\nu}Q_2^{\nu^\prime}
Q_1^\mu)\ , 
\nonumber\\ 
\PaTT &= 
\frac{1}{4}(R^{{\mu^\prime}{\nu^\prime}}R^{\mu\nu}-R^{{\nu^\prime}\mu}
R^{{\mu^\prime}\nu})\ , 
\nonumber\\
\PaTS &= 
\frac{-1}{8}(R^{\mu\nu}Q_1^{\mu^\prime}
Q_2^{\nu^\prime}-R^{{\nu^\prime}\mu}Q_1^{\mu^\prime} Q_2^\nu
+R^{{\mu^\prime}{\nu^\prime}}Q_1^\mu
Q_2^\nu-R^{{\mu^\prime}\nu}Q_2^{\nu^\prime} Q_1^\mu)\ .
\label{eq:budprojectors}
\end{align}
\subsection{Derivation of the Cross Section}
Using Eqs.~(\ref{eq:lalbe}), (\ref{eq:rhoalbe}) and (\ref{eq:budtensor}) 
one obtains by a straightforward (but tedious) calculation 
\begin{equation}
\begin{aligned}
L_{(1)}^{\mu \mu^\prime} L_{(2)}^{\nu \nu^\prime} \Wco & = Q^2 P^2
\Big[4 \ronepp \rtwopp \WTT + 2 |\ronepm \rtwopm| \tTT \cos 2 \bar{\phi}
\\
&\phantom{= Q^2 P^2\Big[} 
+ 2 \ronepp \rtwozz \WTS
+ 2\ronezz \rtwopp \WST 
\\
&\phantom{= Q^2 P^2\Big[}
+ \ronezz \rtwozz \WSS 
-8 |\ronepz \rtwopz| \tTS \cos \bar{\phi}\Big]
\label{eq:contraction}
\end{aligned}
\end{equation}
where 
$\bar{\phi}$ is the angle between the scattering planes of
the $\e^-$ and the $\e^+$ in the center-of-mass system (CMS)
of the colliding photons
and the \ri's are elements of the photon density matrix:
\begin{align}\label{eqn:rhos}
 2\ronepp  &= 2\ronemm = \rho_1^{\alpha\beta} R_{\alpha\beta} =
\frac{\left(2 p_1\cdot p-p\cdot q\right)^2}
                {(p\cdot q)^2 - Q^2 P^2} + 1 - 4\frac{\me^2}{Q^2}\, ,
\nonumber\\
 2\rtwopp  &= 2\rtwomm = \rho_2^{\alpha\beta} R_{\alpha\beta} =
\frac{\left(2 p_2\cdot q-p\cdot q\right)^2}
                     {(p\cdot q)^2 -Q^2P^2} + 1 - 4\frac{\me^2}{P^2}\, ,
\nonumber\\
 \ronezz   &= \rho_1^{\alpha\beta} {Q_1}_{\alpha}{Q_1}_{\beta} =
2\ronepp - 2 + 4\frac{\me^2}{Q^2}\, , 
\nonumber\\
 \rtwozz   &= \rho_2^{\alpha\beta} {Q_2}_{\alpha}{Q_2}_{\beta} =
2\rtwopp - 2 + 4\frac{\me^2}{P^2}\, , 
\nonumber\\
2 |\ronepm \rtwopm| \cos 2 \bar{\phi} & = 
\frac{C^2}{Q^2 P^2} - 2 (\ronepp-1)(\rtwopp-1)\, ,
\nonumber\\
8 |\ronepz \rtwopz| \cos \bar{\phi} &= \frac{4C}{\sqrt{Q^2 P^2}} 
\frac{\left(2 p_1\cdot p-p\cdot q\right)
\left(2 p_2\cdot q-p\cdot q\right)}{(p\cdot q)^2 - Q^2 P^2}\, 
\nonumber\\
\text{with} \quad C & = (2 p_1 - q)^\alpha (2 p_2 - p)^\beta R_{\alpha\beta} =
-(2 p_1 - q) \cdot (2 p_2 - p)
\nonumber\\*
&\quad + \frac{p \cdot q}{(p\cdot q)^2 - Q^2 P^2}
\left(2 p_1\cdot p-p\cdot q\right)\left(2 p_2\cdot q-p\cdot q\right)\, ,
\nonumber\\
 |\ripm|   &= \ripp - 1,
\nonumber\\
 |\ripz|   &= \sqrt{\left(\rizz+1\right)|\ripm|}.
\end{align}
Note that all these quantities are expressed in terms of the measurable 
momenta $p_1,p_2$ and $p_1^\prime,p_2^\prime$ (respectively $q,p$) only
and therefore are entirely known.

With help of Eqs.~(\ref{eqn:eeXsec1}) and (\ref{eq:contraction}) 
we easily find the 
fully general final result for the $ee \rightarrow ee X$ cross section 
\cite{cit:Bud-7501,cit:Berger-Rep,Nisius:1999cv}:
\begin{equation}
\begin{aligned}
%d^6\sigma &(ee \rightarrow ee X) =
d^6\sigma (ee \rightarrow ee X) &=
\frac{d^3 p^{\prime}_1 d^3 p^{\prime}_2}{E^{\prime}_1 E^{\prime}_2}
\frac{\alpha^2}{16 \pi^4 Q^2 P^2} 
%\left[\frac{(p \cdot q)^2 - Q^2 P^2} {(p_1 \cdot p_2)^2 - m_e^2 m_e^2} 
%\right]^{1/2}
\frac{F_{\gamma\gamma}}{F_{ee}} 
\Big[4 \ronepp \rtwopp \sigtt 
\\
&\phantom{= \Big[}
+ 2\ronezz \rtwopp \siglt 
+ 2 \ronepp \rtwozz \sigtl 
+ \ronezz \rtwozz \sigll 
\\
&\phantom{= \Big[}
+ 2 |\ronepm \rtwopm| \tautt \cos 2 \bar{\phi}
-8 |\ronepz \rtwopz| \tautl \cos \bar{\phi}\Big]
\\
& \text{with}
\\
\frac{F_{\gamma\gamma}}{F_{ee}} &=
\left[\frac{(p \cdot q)^2 - Q^2 P^2} {(p_1 \cdot p_2)^2 - m_e^2 m_e^2} 
\right]^{1/2}\, . 
\label{eq:eeXsec}
\end{aligned}
\end{equation}
Here the cross sections $\sigma_{ab}$ (used as a shorthand for
\sigtt, \sigtl, \siglt, \sigll, \tautt, \tautl, \tauatt, \tauatl)
are identical to the corresponding structure functions $W_{ab}$ 
up to a division by the appropriate flux factor of the two 
incoming photons\footnote{Note 
that a factor $\frac{1}{2}$ has already been absorbed into
the definition of \Wcontra\ in Eq.~(\ref{eq:defw}).}, i.e.
\begin{equation}\label{eq:sigma_ab}
\sigma_{ab} = \frac{1}{2 \sqrt{(p \cdot q)^2 - Q^2 P^2}} W_{ab}
= \frac{1}{2 \nu \vtwo} W_{ab} \ .
\end{equation}

The cross section in (\ref{eq:eeXsec}) considerably simplifies in certain
kinematical regions \cite{cit:Bud-7501,cit:Berger-Rep,Nisius:1999cv}.
For instance, if both photons are highly virtual
Eq.~(\ref{eq:eeXsec}) can be evaluated in the limit $Q^2, P^2 \gg \me^2$ 
\cite{Nisius:1999cv} in which some relations between the elements of the
photon density matrix exist.
Of special interest is the case where one of the 
lepton scattering angles becomes
small leading to a small virtuality $P^2\approx 0$ of the corresponding 
photon while the
other photon provides a hard scale $Q^2 \gtrsim 1\ \gevsq$.
In this limit the cross section factorizes into a product of a 
flux of quasi-real target photons
times the cross section for deep inelastic electron-photon scattering,
see for example \cite{Nisius:1999cv}.
This process is the classical way of measuring the structure of 
(quasi-real) photons.
The findings in the latter limit can be generalized to the case of 
photons with non-zero virtuality
$P^2 \ne 0$ as we will see in Section \ref{sec:fact1}.
This allows to study the structure of virtual photons in deep inelastic 
$\e \gamma(P^2)$ scattering processes in a continuous range of the scale $P^2$.
\subsection{The Doubly Virtual Box in LO}\label{sec:lobox}
\begin{figure}[h]
\begin{center}
\begin{picture}(140,100)(0,0)
%upper left photon
\Photon(10,90)(40,70){-4}{5}
%lower left photon
\Photon(10,10)(40,30){4}{5}
%upper right photon
\Photon(130,90)(100,70){4}{5}  
%lower right photon
\Photon(130,10)(100,30){-4}{5}
% linewidth, default = 0.5
\SetWidth{1.0}
\Line(40,70)(100,70)
\Line(40,30)(100,30)
\Line(40,70)(40,30)
\Line(100,70)(100,30)
\Vertex(40,70){1.5}
\Vertex(40,30){1.5}
\Vertex(100,30){1.5}
\Vertex(100,70){1.5}
\SetWidth{0.5}
% cut
\Line(70,5)(70,95)
\Line(70,95)(75,95)
\Line(70,5)(65,5)
%\Text(35,90)[l]{\large $q,\mu,m$}
%\Text(35,10)[l]{\large $p,\nu,n$}
%
%\Text(235,90)[l]{\large $q,\mu^\prime,m^\prime$}
%\Text(235,10)[l]{\large $p,\nu^\prime,n^\prime$}
\end{picture}
\end{center}
\caption{\sf Box-diagram for 
$\gamma^\star(q,\mu) \gamma^\star(p,\nu)\ \rightarrow
\gamma^\star(q,\mu^\prime) \gamma^\star(p,\nu^\prime)$ 
.There are 4 possibilities to attach the photons to the vertices.
(2 for the initial state times 2 for the final state.)}
\label{fig:Box}
\end{figure}
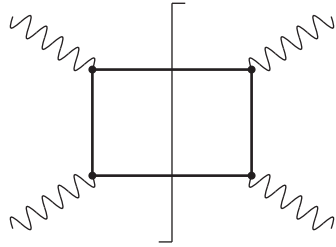

In this section we calculate the invariant amplitudes
$W_{ab}$ (or the cross sections $\sigma_{ab}$) 
in lowest order perturbation theory.
These expressions are usually referred
to as 'box' results due to the diagram representing the tensor
\Wcontra, where the photons are attached
to a fermion box, see Fig.~\ref{fig:Box}.

\begin{figure}[h]
\begin{center}
\begin{picture}(320,100)(-20,0)

%upper left photon
\Photon(10,90)(40,70){-4}{5}
%lower left photon
\Photon(10,10)(40,30){4}{5}
% linewidth, default = 0.5
\SetWidth{1.0}
\ArrowLine(100,30)(40,30)
\ArrowLine(40,30)(40,70)
\ArrowLine(40,70)(100,70)
\SetWidth{0.5}
\Vertex(40,70){1.5}
\Vertex(40,30){1.5}

\Text(-15,90)[l]{\large $q,\mu$}
\Text(-15,10)[l]{\large $p,\nu$}
\Text(105,70)[l]{\large $k_1,m$}
\Text(105,30)[l]{\large $k_2,m$}
\Text(140,50)[l]{\Large $+$}
%
% u-channel
%
%upper left photon
\Photon(170,90)(200,70){-4}{5}
%lower left photon
\Photon(170,10)(200,30){4}{5}
% linewidth, default = 0.5
\SetWidth{1.0}
\ArrowLine(260,30)(230,50)
\Line(230,50)(200,70)
\ArrowLine(200,70)(200,30)
\Line(200,30)(230,50)
\ArrowLine(230,50)(260,70)
\SetWidth{0.5}
\Vertex(200,70){1.5}
\Vertex(200,30){1.5}

\Text(145,90)[l]{\large $q,\mu$}
\Text(145,10)[l]{\large $p,\nu$}
\Text(265,70)[l]{\large $k_1,m$}
\Text(265,30)[l]{\large $k_2,m$}
\end{picture}
\end{center}
\caption{\sf Amplitude $M^{\mu\nu}$ for the process
$\gamma^\star(q,\mu) \gamma^\star(p,\nu)\ \rightarrow
\mathrm{f}(k_1)\mathrm{\bar{f}}(k_2)$ 
where $\mathrm{f}$ is a fermion of mass $m$.}
\label{fig:Mmunu}
\end{figure}
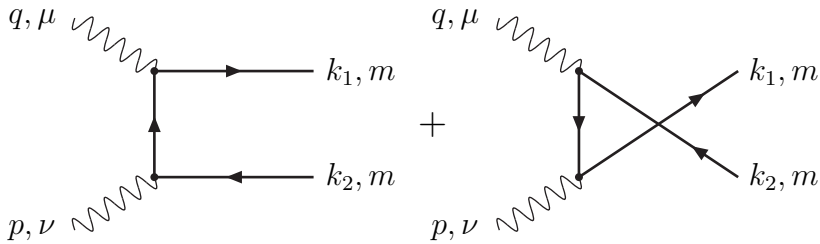

The calculation is a straightforward application of Feynman rules and 
will only be shortly outlined
here.
In order to obtain \Wcontra\ according to Eq.~(\ref{eq:defw})
one has to build up the amplitude $M^{\mu\nu}$ of the process
$\gamma^\star(q,\mu) \gamma^\star(p,\nu)\ \rightarrow \mathrm{f}(k_1)
\mathrm{\bar{f}}(k_2)$ 
shown in Fig.~\ref{fig:Mmunu} where $\mathrm{f}$ is either a lepton
or a quark of mass $m$.
The kinematics can be described 
with the help of the 4-momentum conservation relation $q + p = k_1 + k_2$
and the usual Mandelstam variables 
\begin{gather}
s = (q+p)^2 \ , \
t = (q-k_1)^2 = (p-k_2)^2\ , \ u = (p-k_1)^2 = (q-k_2)^2\ ,
\nonumber\\
s_1 \equiv 2 q \cdot p=s + Q^2 + P^2 \ , \
t_1 \equiv t - m^2\ , \ u_1 \equiv u - m^2 
\end{gather}
satisfying $s_1+t_1+u_1 = 0$.  

The external fermion lines are on-shell, i.e. $k_1^2 = k_2^2 = m^2$, whereas
the two photons are virtual (space-like): $q^2 = -Q^2 <0$, $p^2 = -P^2 <0$.
The Dirac traces occurring in \Wcontra\ (due to the closed fermion loop) have been
evaluated with the help of the \Mathematica\ \cite{cit:math} package
\Tracer\ \cite{Jamin:1993dp}.
Finally, the individual structure functions $W_{ab}$ have been projected
out using the projection operators given in Eq.~(\ref{eq:budprojectors}).
%Budnev;Box;naives QPM;QED
%
% recalculation
\subsubsection{Unintegrated Structure Functions}\label{sec:unintegrated}
The general structure of the boson-boson fusion cross 
section (in lowest order) 
is given by
\cite{Kretzer:1997pd,Schienbein:1997sb}
%\footnote{The $t-$channel amplitude has a propagator $1/t_1$ 
%and therefore has the form $(a+b t_1)/t_1$. The $u-$channel is obtained by
%$t_1 \leftrightarrow u_1 = -(s_1+t_1)$
%and thus reads $(a+b u_1)/u_1 = (c+d t_1)/(s_1+t_1)$.
%Squaring the amplitudes generates the structure 
%in Eq.~(\protect\ref{eq:bgf}).} 
\begin{equation}
\frac{dW_{ab}}{dz_1} = 16 \pi^2 \alpha^2 e_q^4 N_c
\left[\frac{A_{W_{ab}}}{(1-z_1)^2}+\frac{B_{W_{ab}}}{z_1^2}+
   \frac{C_{W_{ab}}}{1-z_1}+\frac{D_{W_{ab}}}{z_1}+E_{W_{ab}}\right] 
\label{eq:bgf}
\end{equation}
where $z_1 = 1+t_1/s_1$, $N_c=3$ is the number of colors
and $e_q$ is the quark charge.
The QED case can be obtained from Eq.~(\ref{eq:bgf}) by setting
$N_c e_q^4 \to 1$.
Note that $z_1$ is very similar to the fractional
momentum variables generally used to describe energy 
spectra of heavy quarks (mesons):
\begin{displaymath}
z_1 = \frac{k_1 \cdot p}{q \cdot p} - \frac{p^2}{2 q \cdot p}\ .
\end{displaymath}
On the other hand, $z_1=-u_1/s_1$ (or $1-z_1=-t_1/s_1$) can be viewed as a natural 
dimensionless combination of the Mandelstam variables of the problem.
%
% the following equations still have to be checked by
% comparing them with the original expressions in Mathematica !!! 
Furthermore, it is convenient to use the dimensionless variables
\begin{equation}
x = \frac{Q^2}{s_1} =\frac{Q^2}{2 \nu}\, ,\,
\xp = \frac{P^2}{s_1} =\frac{P^2}{2 \nu}\, ,\,
\lambda = \frac{4 m^2}{s}
\end{equation}
in order to write the coefficients in a form which is
manifestly symmetric under $x \leftrightarrow \xp$ 
(see Sec.~\ref{sec:construction})\footnote{Note however that 
neither $\WTS$ nor $\WST$ but only
the {\em sum} $\WTS+\WST$ is invariant under exchanging $x$ and $\xp$.}
and which
easily allows to read off the important 'massless limits' $P^2 \to 0$
or $m^2 \to 0$ to be discussed in 
Appendix \ref{app:limits}.
%Section \ref{sec:limits}.

With $\vtwo = \sqrt{1-4 x \xp}$ the coefficients read:
\begin{align}
\label{eq:ace}
\AWTT & =
\frac{-1}{32 \pi \vtwo^5}
\Big[2 x \vtwo^2 - (1-x-\xp)(4x\xp+\lambda \vtwo^2)\Big]
\Big[2 \xp \vtwo^2 
\nonumber\\*
&\phantom{=}
- (1-x-\xp)(4x\xp+\lambda \vtwo^2)\Big]
\nonumber
\\
\CWTT & =
\frac{-1}{16 \pi \vtwo^5}
\Big[\lambda^2 \vtwo^4 (1-x-\xp)^2 -2\lambda \vtwo^2 (1-x-\xp)^2  
\nonumber\\*
& \phantom{=\frac{-1}{16 \pi \vtwo^5}\Big[}
-2 \Big(1-2x(1-x)-2\xp(1-\xp)
-4 x \xp(x^2+\xp^2)
%\nonumber\\*  
%-4 x \xp(x^2+\xp^2)
\nonumber\\*
& \phantom{=\frac{-1}{16 \pi \vtwo^5}\Big[}
+8 x^2 \xp^2 \left[1+(1-x-\xp)^2\right]\Big)\Big]
\nonumber
\\
\EWTT & =
\frac{-1}{4 \pi \vtwo^5}\Big[1-2x(1-x)-2\xp(1-\xp)+4 x \xp(1-2x)(1-2\xp)\Big]
%\nonumber
\\
\nonumber\\
\AWTS & = 
\frac{\xp (1-x-\xp)}{4 \pi \vtwo^5}  
x \Big[2 x \vtwo^2 - (1-x-\xp)(4x\xp+\lambda \vtwo^2)\Big]
\nonumber
\\
\CWTS & = \frac{\xp(1-x-\xp)}{-4 \pi \vtwo^5}
\Big[\lambda \vtwo^2 \Big(1+2x(-1+x-\xp)\Big)+4x \Big(-1+2x
\nonumber
\\*
&\phantom{= \frac{\xp(1-x-\xp)}{-4 \pi \vtwo^5}\Big[}
+2\xp-2x\xp(1+x+\xp)\Big)\Big] 
\nonumber
\\
\EWTS & = \frac{\xp(1-x-\xp)}{\pi \vtwo^5}(1-2x)^2   
%\nonumber
\\
\nonumber\\
\AWSS & = -\frac{2 x^2 \xp^2 (1-x-\xp)^2}{\pi \vtwo^5} 
\nonumber
\\
\CWSS & = \frac{2 x \xp (1-x-\xp)^2}{\pi \vtwo^5}(1+ 2 x \xp) 
\nonumber
\\
\EWSS & = -\frac{8 x \xp (1-x-\xp)^2}{\pi \vtwo^5}
%\nonumber
\\
\nonumber\\
\AtTT & =
\frac{-1}{32 \pi \vtwo^5}(1-x-\xp)^2 [4 x \xp+\lambda \vtwo^2]^2
\nonumber
\\
\CtTT & =
\frac{-1}{16 \pi \vtwo^5}
\Big[\lambda^2 \vtwo^4 (1-x-\xp)^2 -4 \lambda \vtwo^2 (1-x-\xp)\Big(-x-\xp
\nonumber\\*
&\phantom{=\frac{-1}{16 \pi \vtwo^5}\Big[}
+2 x \xp(1+x+\xp)\Big)
-x \xp \Big(1-2x(1-x)-2\xp(1-\xp)
\nonumber\\*
&\phantom{=\frac{-1}{16 \pi \vtwo^5}\Big[}
+2x\xp(1-2x-2\xp-x^2-\xp^2+6 x \xp)\Big)\Big]
\nonumber
\\
\EtTT & =\frac{-1}{2 \pi \vtwo^5}(x+\xp-4 x \xp)^2
%\nonumber
\\
\nonumber\\
\AtTS & =
\frac{\sqrt{x\xp}(1-x-\xp)}{16 \pi \vtwo^5}
(1-2x)(1-2\xp)(4x\xp+\lambda \vtwo^2)
\nonumber
\\
\CtTS & =
\frac{\sqrt{x\xp}(1-x-\xp)}{4 \pi \vtwo^5}
\Big[4x\xp(-3+x+\xp)+2x+2\xp-\lambda \vtwo^2 (1-x-\xp)\Big]
\nonumber
\\
\EtTS & =
\frac{\sqrt{x\xp}(1-x-\xp)}{2 \pi \vtwo^5}(1-4x-4\xp+12 x\xp)
%\nonumber
\\
\nonumber\\
\AaTT & =
\frac{1}{16 \pi \vtwo^3}
\Big[-2x\xp\vtwo^2+(1-x-\xp)(4x\xp+\lambda \vtwo^2)\Big]
\nonumber
\\
\CaTT & = -\frac{(1-2x)(1-2\xp)}{8 \pi \vtwo^3}
\nonumber
\\
\EaTT & = \frac{(1-2x)(1-2\xp)}{4 \pi \vtwo^3}
%\nonumber
\\
\nonumber\\
\AaTS & =
\frac{\sqrt{x\xp}(1-x-\xp)}{16 \pi \vtwo^3}
(-4x\xp+\lambda \vtwo^2)
\nonumber
\\
\CaTS & =
\frac{\sqrt{x\xp}(1-x-\xp)}{\pi \vtwo^3} x\xp
\nonumber
\\
\EaTS & =
-\frac{\sqrt{x\xp}(1-x-\xp)}{2 \pi \vtwo^3}
%\nonumber
\\
\nonumber\\
\label{eq:ace2}
B_{W_{ab}} &=  A_{W_{ab}}\, , \qquad  D_{W_{ab}} =  C_{W_{ab}}\ .
\end{align}
The coefficients of $\WST$ can be obtained from the corresponding 
ones of $\WTS$ by exchanging $x \leftrightarrow \xp$:
$\AWST = \AWTS[x \leftrightarrow \xp]$,
$\CWST = \CWTS[x \leftrightarrow \xp]$, and
$\EWST = \EWTS[x \leftrightarrow \xp]$.

It is noteworthy that 
our results in Eq.~(\ref{eq:bgf}) generalize
the $z$-differential expressions for the (QED) structure functions
$F_2^\gamma$, $F_\mathrm{L}^\gamma$, and $F_\mathrm{T}^\gamma$
of real photons given in \cite{Nisius:1998ue} 
to the $P^2 \ne 0$ case.
\subsubsection{Inclusive Structure Functions}
The desired inclusive structure functions are obtained
by integrating over the kinematically allowed range in $z_1$.
The boundaries for the $z_1$-integration are given by
\begin{equation}\label{eq:zpm}
z_{1,\pm} = (1 \pm \vone\vtwo)/2
\end{equation}
with
$\vone^2 = 1- 4 m^2/{s}= 1-\lambda$
and $\vtwo^2 = 1-4 x^2 P^2/Q^2= 1-4 x \xp$.
%with
%$\vone = \sqrt{1- 4 m^2/{s}}= \sqrt{1-\lambda}=\sqrt{\lambda(s,m^2,m^2)}/s$
%and $\vtwo = \sqrt{1-4 x^2 P^2/Q^2}= \sqrt{1-4 x \xp}=\sqrt{\lambda(s,q^2,p^2)}/s_1$
%where $\lambda(x,y,z)$ denotes the triangle function.
%
%\begin{eqnarray*}
%\vone& = &  \sqrt{1- \frac{4 m^2}{s}}
%= \sqrt{1-\lambda}
%=\sqrt{\lambda(s,m^2,m^2)}/s
%\\
%\vtwo& = & \sqrt{1-\frac{4 Q^2 P^2}{s_1^2}}
%= \sqrt{1-4 x \xp}
%=\sqrt{\lambda(s,q^2,p^2)}/s_1
%\end{eqnarray*}

Noticing that 
\begin{gather}
z_{1,+} - z_{1,-}= \vone\vtwo\ , \qquad 
1-z_{1,+}=z_{1,-}\ , \qquad  1-z_{1,-}=z_{1,+}\ ,
\nonumber\\
z_{1,+}z_{1,-}=(1-z_{1,+})(1-z_{1,-})=\frac{4x\xp+\lambda \vtwo^2}{4} 
\end{gather}
the required integrals can be immediately obtained
\begin{equation}\label{eq:z1-integrals}
\begin{aligned}
\int_{z_{1,-}}^{z_{1,+}}\frac{dz_1}{(1-z_1)^2} &= 
\int_{z_{1,-}}^{z_{1,+}}\frac{dz_1}{z_1^2} 
= \frac{\vone\vtwo}{z_{1,+}z_{1,-}}
= \frac{4 \vone\vtwo}{4x\xp+\lambda \vtwo^2}
% = \frac{\vone\vtwo s_1^2}{T}
\\
\int_{z_{1,-}}^{z_{1,+}}\frac{dz_1}{(1-z_1)} &= 
\int_{z_{1,-}}^{z_{1,+}}\frac{dz_1}{z_1} = \ln \frac{z_{1,+}}{z_{1,-}} 
= \ln \frac{1+ \vone\vtwo}{1- \vone\vtwo}
\\
\int_{z_{1,-}}^{z_{1,+}} dz_1 &= \vone\vtwo \ .
\end{aligned}
\end{equation}
%\begin{eqnarray*}
%\int_{z_{1,-}}^{z_{1,+}}\frac{dz_1}{(1-z_1)^2} &=& 
%\int_{z_{1,-}}^{z_{1,+}}\frac{dz_1}{z_1^2} 
%= \frac{\vone\vtwo}{z_{1,+}z_{1,-}}
%= \frac{4 \vone\vtwo}{4x\xp+\lambda \vtwo^2}
%% = \frac{\vone\vtwo s_1^2}{T}
%\\
%\int_{z_{1,-}}^{z_{1,+}}\frac{dz_1}{(1-z_1)} &=& 
%\int_{z_{1,-}}^{z_{1,+}}\frac{dz_1}{z_1} = \ln \frac{z_{1,+}}{z_{1,-}} 
%= \ln \frac{1+ \vone\vtwo}{1- \vone\vtwo}
%\\
%\int_{z_{1,-}}^{z_{1,+}} dz_1 &=& \vone\vtwo \ .
%\end{eqnarray*}
Now Eq.~(\ref{eq:bgf}) can be integrated using Eq.~(\ref{eq:z1-integrals})
and we arrive at the following result for
the inclusive structure functions expressed
by the coefficients given in the previous Section~\ref{sec:unintegrated}:
\begin{equation}\label{eq:Wab}
\begin{aligned}
W_{ab} &= 
16 \pi^2 \alpha^2 e_q^4 N_c \Theta(\vone^2) \left\{ 2 C_{W_{ab}} L + \vone\vtwo 
\left(2 A_{W_{ab}} \frac{4}{4x\xp+\lambda \vtwo^2} + E_{W_{ab}}\right)\right\}
\end{aligned}
\end{equation} 
with
\begin{equation}\label{eq:log}
L =  \ln \frac{1+ \vone\vtwo}{1- \vone\vtwo}\ .
\end{equation} 
The $\Theta$-function guarantees that the physical threshold
condition $s \ge 4 m^2$ is satisfied. It is easy to see 
that $\vone^2 \ge 0$ also implies $\vtwo^2 \ge 0$.

Recalling the relation $\sigma_{ab} = \frac{1}{2 \nu \vtwo} W_{ab}$
we can rewrite Eq.~(\ref{eq:Wab}) for the photon-photon cross sections
\begin{equation}\label{eq:sigab}
\begin{aligned}
\sigma_{ab} &= N \left\{ 2 C_{W_{ab}} L + \vone\vtwo 
\left(2 A_{W_{ab}} \frac{4}{4x\xp+\lambda \vtwo^2} + E_{W_{ab}}\right)\right\}
\\
& \text{with} \quad
N\equiv \frac{16 \pi^2 \alpha^2 N_c e_q^4}{2 \nu \vtwo}\Theta(\vone^2)\ .
\end{aligned}
\end{equation} 
%with the normalization factor 
%$N=\frac{16 \pi^2 \alpha^2 N_c e_q^4}{2 \nu \vtwo}\Theta(\vone^2)$.
Inserting the coefficients given in Eqs.~\eqref{eq:ace}--\eqref{eq:ace2} 
into Eq.~(\ref{eq:sigab})
one obtains the \underline{final result} for the doubly virtual
box $\vgvg$ in leading order:
\begin{align}
%\label{eq:LO-box}
%
\sigtt &= \frac{N}{4 \pi}\frac{1}{\vtwo^5}
 \Biggl\{ \biggl[ 1-2x(1-x)-2\xp(1-\xp) - 4x\xp(x^2+\xp^2)  
\nonumber\\*
&  
+ 8 x^2\xp^2 \left[1+(1-x-\xp)^2\right]
+\lambda\vtwo^2(1-x-\xp)^2 -\frac{1}{2}\lambda^2\vtwo^4 (1-x-\xp)^2\biggr] \, L
\nonumber\\* 
& 
+ \beta\vtwo\, \Biggl[ 4x(1-x)-1
+4\xp(1-\xp)-8x\xp (1-x^2-\xp^2) 
\nonumber\\* 
& 
-(4x\xp+\lambda\vtwo^2)(1-x-\xp)^2  
-\frac{4x\xp\vtwo^4}{4x\xp + \lambda\,\vtwo^2}\Biggr] 
\Biggr\}
\label{eq:sigtt}
%\nonumber
\\
\nonumber\\
\siglt & = \frac{N}{4 \pi} \frac{4}{\vtwo^5} (1-x-\xp)\,
\Biggl\{ x\biggl[\, - \frac{1}{2}\lambda\,\vtwo^2 \Bigl( 1-2\xp(1+x-\xp)\Bigr) 
\nonumber\\* 
&  
-2\xp \Bigl(-1+2x
+2\xp 
-2x\xp(1+x+\xp) \Bigr) \biggr]\,  L
\nonumber\\* 
&  
+\beta\vtwo\bigg[ x(1-6\xp +6\xp^2 +2x\xp)  
+ \xp\vtwo^2\, \frac{4x\xp} {4x\xp +\lambda\, \vtwo^2} \bigg] \Biggr\} 
\label{eq:siglt}
%\nonumber
\\
\nonumber\\
\sigtl & = \siglt [x\leftrightarrow \xp] 
\label{eq:sigtl}
%\nonumber
\\
%& = N \frac{1-x-\xp}{-2 \pi \vtwo^5}
%\Bigg\{\xp[\lambda \vtwo^2(1+2x(-1+x-\xp))+4x(-1+2x+2\xp-2x\xp(1+x+\xp))]L
%\nonumber\\
%&-2 \vone\vtwo [x \vtwo^2 \frac{4x\xp}{4x\xp+\lambda \vtwo^2}+\xp(1-6x+6x^2+2x\xp)]
%\Bigg\}
%\nonumber\\
%
\nonumber\\
\sigll & =\frac{N}{4 \pi} \frac{16}{\vtwo^5} x \xp (1-x-\xp)^2
\left\{(1+ 2 x \xp)L -
2 \vone\vtwo \frac{6x\xp+\lambda \vtwo^2}{4x\xp+\lambda \vtwo^2}\right\}
\label{eq:sigll}
%\nonumber
\\
\nonumber\\
\tautt & = \frac{N}{4 \pi} \frac{1}{\vtwo^5}
\Bigg\{
\bigg[\frac{1}{2}x \xp \Big(1-2x(1-x)-2\xp(1-\xp)+2x\xp(1-2x-2\xp
\nonumber\\*
&
-x^2-\xp^2+6 x \xp)\Big)
-2 \lambda \vtwo^2 (1-x-\xp)\Big(x+\xp-2 x \xp(1+x+\xp)\Big)
\nonumber\\*
&
-\frac{1}{2}\lambda^2 \vtwo^4 (1-x-\xp)^2 \bigg]L
- \vone\vtwo \bigg[(1-x-\xp)^2 (4x\xp+\lambda\vtwo^2)
\nonumber\\*
&
+2 (x+\xp-4x\xp)^2\bigg]\Bigg\}
\label{eq:tautt}
%\nonumber
\\
\nonumber\\
\tautl & = \frac{N}{4 \pi} \frac{2}{\vtwo^5} \sqrt{x\xp}(1-x-\xp)
\Bigg\{\bigg[2x+2\xp-4x\xp(3-x-\xp)
\nonumber\\*
&
-\lambda \vtwo^2 (1-x-\xp)\bigg]L
+ \vone\vtwo (1-3x-3\xp+8x\xp)\Bigg\}
\label{eq:tautl}
%\nonumber
\\
\nonumber\\
\tauatt & = \frac{N}{4 \pi} \frac{1}{\vtwo^3}
\Bigg\{(2x-1)(1-2\xp)L
+ \vone\vtwo \bigg[\frac{-4x\xp \vtwo^2}{4x\xp+\lambda \vtwo^2}
+3-4x-4\xp+4x\xp\bigg]\Bigg\}
\label{eq:tauatt}
%\nonumber
\\
\nonumber\\
\tauatl & = \frac{N}{4 \pi} \frac{4}{\vtwo^3} \sqrt{x\xp}(1-x-\xp)
\left\{2x\xp L - \vone\vtwo \frac{4x\xp}{4x\xp+\lambda \vtwo^2}\right\}
\label{eq:tauatl}
\end{align}
This re-calculation is in agreement with the results of Ref.\
\cite{cit:Bud-7501} (with $N_c e_q^4 \to 1$)
with exception of a {\em relative} sign between the part containing
the logarithm $L$ and the part proportional
to $\vone\vtwo$ in $\tauatl$. This relative sign
has also been noted in \protect\cite{cit:Rossi-PhD} where in addition the
{\em overall} sign in $\tautl$ is different. Concerning the latter,
our calculation agrees with the results of \protect\cite{cit:Bud-7501}.

\subsubsection{Discussion of the results}
\begin{itemize}
\item The general structure of the final results in
\eqref{eq:sigtt}--\eqref{eq:tauatl} is given by
$\sigma_{ab}(x,\xp,\lambda) = \sigma_{1,ab}(x,\xp,\lambda) L + \vone \vtwo\
\sigma_{2,ab}(x,\xp,\lambda)$ as is evident from Eq.~\eqref{eq:sigab}.
The logarithm $L$ stems from the $1/z_1$ and $1/(1-z_1)$ terms in
Eq.~\eqref{eq:bgf}. It develops a mass singularity 
(collinear singularity) in the limit of small quark masses $m$ and
small photon virtualities $P^2$. 
This is related to the fact that the upper and lower integration bounds
in Eq.~\eqref{eq:zpm} approach one and zero in this limit:
$z_{1,+}\to 1$, $z_{1,-}\to 0$. 
As long as $m$ or $P^2$ is kept 
non-zero the singularity is regularized.
On the other hand, the $1/z_1^2$ (and $1/(1-z_1)^2$) parts give 
{\em finite} contributions 
because the coefficients $A_{W_{ab}}=B_{W_{ab}}$ in 
Eqs.~\eqref{eq:ace}--\eqref{eq:ace2} vanish
for $P^2, m \to 0$.
\item The photon photon cross sections 
$\sigtt$, $\sigll$, $\tautt$, $\tautl$, $\tauatt$ and $\tauatl$
in \eqref{eq:sigtt}--\eqref{eq:tauatl} 
are symmetric w.r.t.\ $x \leftrightarrow \delta$
whereas $\sigtl(x,\xp) = \siglt(\xp,x)$.
This is a direct consequence of the symmetries of the helicity
amplitudes as has been discussed in Sec.~\ref{sec:construction}.
\item As a rule each longitudinal polarisation vector 
results in a factor $\sqrt{P^2}$ or $\sqrt{Q^2}$ depending
on which of the photons, $\gamma(P^2)$ or  $\gamma(Q^2)$, is
longitudinal.
(It is helpful to recall that
in the helicity amplitudes $W_{m^\prime n^\prime,m n}$
the helicities $n^\prime$ and $n$ refer to $\gamma(P^2)$, whereas
$m^\prime$ and $m$ refer to $\gamma(Q^2)$.)
These factors guarantee that the contributions from longitudinal
photons vanish --with one exception which is discussed 
in App.~\ref{app:limits}-- 
in the limit when these photons become real.
This rule explains several prefactors $x$, $\xp$, $\sqrt{x}$ and
$\sqrt{\xp}$ in Eqs.~\eqref{eq:ace}--\eqref{eq:ace2}
and 
\eqref{eq:sigtt}--\eqref{eq:tauatl}.
\item Various limits can be easily derived from the general 
expressions in \eqref{eq:sigtt}--\eqref{eq:tauatl}, e.g.:
\begin{itemize}
\item $m^2 , P^2 \ll Q^2$
\item $m^2 = 0$ and $m^2 = 0, P^2 \ll Q^2$
\item $P^2= 0$ and $P^2 = 0, m^2 \ll Q^2$
\end{itemize}
A useful compilation of some of these limits
of the  doubly virtual box expressions 
%in \eqref{eq:sigtt}--\eqref{eq:tauatl}
can be found in App.~\ref{app:limits}.
\end{itemize}
\subsubsection{Applications}
The doubly virtual box expressions 
have several applications.
First of all, they can be applied in the QED case
(mainly $\vgvgmu$)
irrespective of $P^2$ and $Q^2$ as long as $Q^2 \ge P^2$ is not
too large, such that $Z$ or $W$-boson exchange can be neglected.
Indeed, the precise measurements
of the QED structure of the photon at LEP, see e.g.\ chapter 6 in 
\cite{Nisius:1999cv}, nicely agree with the perturbative 
QED box predictions.
%$F_2^{\gamma,{\rm QED}}(x,Q^2)$

In the QCD case ($\vgvg$) one can distinguish different 'phases' 
depending on the virtualities of the two photons, $Q^2$ and $P^2$, 
and on the invariant photon-photon mass $W^2 = (p+q)^2$.
These scales are
in comparison with a typical hadronic 
scale like the mass of the rho meson
$m_{\rho}= 770\ \mev$ or $\lamQCD = \Ord(200\ \mev)$.
Anyhow, the uncertainty in choosing the appropriate 
hadronic scale can be absorbed into the
ignorance of what '$\gg$' or '$\ll$' exactly means such that 
we can take $\lamQCD$ as our hadronic scale in the following examples:
\begin{itemize}
\item $Q^2,  P^2 \gg \lamQCD^2$:\\ 
In this region of phase space we have 
a purely perturbative ('golden') process within QCD where
the predictions of Eqs.~\eqref{eq:sigtt}--\eqref{eq:tauatl} are
generally expected to be applicable.
We can further subdivide:
\begin{itemize}
\item $Q^2 \simeq P^2 \gg \lamQCD^2$\\ 
%the photon tensor can be 
%calculated in fixed order perturbation theory (FOPT); in lowest order
%according to the virtual photon photon box 
%$\gam^\star(Q^2)\gam^\star(P^2)\to q \bar q$.
% Calculation of Cacciari et al.
\item $Q^2 \gg P^2 \gg \lamQCD^2$:\\
In this case we can think of a virtual target photon
$\gam(P^2)$ whose structure is resolved by a 
probe photon $\gam(Q^2)$. 
Since $P^2 \gg \lamQCD^2$ the photon structure
functions are {\em perturbatively calculable} either in FOPT or
by resumming potentially large logarithms $\ln Q^2/P^2$ 
occuring in the fixed order box calculation 
to all orders via renormalization group (RG) techniques
\cite{Gluck:2000sn}. 
This can be achieved by solving the
Altarelli-Parisi evolution equations for the photon supplemented with
the appropriate (perturbative) boundary 
conditions \cite{Uematsu:1981qy,*Uematsu:1982je,cit:GRS95}
which can be inferred from the
fixed order box calculation \cite{Schienbein:2001cd}.
%Measurements of the effective structure function
%$F_{eff} = $ in this kinematical region 
%So far the RG-resummed results not preferred over fixed order box
%\cite{Gluck:2000sn}.
\item $W^2 \gg Q^2, P^2 \gg \lamQCD^2$:\\
The high energy limit of virtual photon photon collisions
has attracted a lot of interest in the literature
\cite{Brodsky:1997sd,*Brodsky:1997sg,Bartels:1996ke,
*Bartels:1997er,*Bartels:2000sk,Kwiecinski:1999yx,*Kwiecinski:2000zs}
%\cite{Brodsky:1997sd,*Brodsky:1997sg,Bartels:1996ke,*Bartels:1997er}
since it might signal BFKL 
\cite{Fadin:1975cb,*Kuraev:1976ge,*Kuraev:1977fs,Balitsky:1978ic}
dynamics.
From this point of view the quark box results  
in Eqs.~\eqref{eq:sigtt}--\eqref{eq:tauatl} 
constitute a {\em background}.
On the other hand they can also be viewed as a competitive
theoretical approach to experimental measurements
\cite{Achard:2001kr,Abbiendi:2001tv}.
Recently, $\Ord(\alpha_s)$ corrections to 
the effective (measurable) combination 
%of the quark box results 
$\sigma_{\rm eff} = \sigtt + \vep_1 \sigtl + \vep_2 \siglt 
+ \vep_1 \vep_2 \sigll$ with $\vep_i = 2 (1-y_i)/(1+(1-y_i)^2)$ 
have been obtained in Ref.~\cite{Cacciari:2000cb}. 
For an overview of the current status see, e.g., Fig.\ 1 in
\cite{Brodsky:2001ye}.
%For the current status we refer to
%Fig.\ 1 in \cite{Brodsky:2001ye}.
%As is stands now
%$\text{LO-box} < \text{NLO-box} \lesssim \text{Data} 
%\gtrsim \text{NLO-BFKL} < \text{LO-BFKL}$.
\end{itemize}
\item $Q^2 \gg \lamQCD^2 \gtrsim P^2$:\\ 
If one of the photons is (quasi-)real or slightly virtual
perturbation theory is clearly not reliable due to
non-perturbative 'long distance' effects which are cut off
in the previous cases by the large virtuality $P^2$. 
For this reason the perturbative box calculation 
is in principle not directly applicable.
Nevertheless,
first analyses of real photon structure functions 
have been performed within the naive quark-parton model (QPM)
\cite{Walsh:1973mz,*Zerwas:1974tf,*Kingsley:1973wk,
*Chernyak:1974nv,*Worden:1974hc,*Ahmed:1975ff} 
where the so-called 'QPM' or 'Box' expressions 
for the photon structure functions have
been calculated according to the process 
$\gamma^\star(Q^2)\gamma \to q \bar q$ employing 
a {\em finite quark mass} $m = \Ord(300\ \mev)$.
Of course, this procedure to regularize the mass singularities
by some quark masses is ad hoc: the quark masses are unphysical
and, moreover, the long-distance dynamics is certainly not
well described by (massive) free quark propagators.
Therefore, in modern approaches to the structure of 
(quasi-)real and
(slightly) virtual target photons $\gam(P^2)$ 
\cite{cit:GRSc99,cit:GRS95,cit:SaS-9501,*Schuler:1996fc} 
the photon structure functions (or the photon tensor) 
are {\em factorized} 
\cite{Collins:1989gx,*Collins:1987pm}
into
non-perturbative parton distribution functions 
to be fixed by 
experimental information and calculable 
short distance (Wilson) coefficient functions.
\end{itemize}

\section{Photon Structure Functions}\label{sec:photonsfs}
It is the aim of this section to define structure functions
of a (virtual) {\em target photon} and to relate them 
to the  invariant functions
$W_{ab}$. The defining relations will be generally valid for arbitrary
$P^2$. However, they only have a meaningful interpretation as structure
functions of a target photon probed by a deeply virtual photon $\gamma(Q^2)$
in the limit $P^2 \ll Q^2$.

The structure tensor for 
time-like \cite{Hoodbhoy:1989am} 
and
space-like \cite{Mathews:1996yv}
spin-1 targets is described by 8 structure functions
(as long as e.m. interactions are considered).
Indeed, in Ref.~\cite{Mathews:1996yv} the case of a 
(space-like) virtual target photon has been considered.
However, for the discussion below 
(involving transversely and longitudinally polarized targets)
it is necessary to start
with the 4-th rank photon tensor given in Eq.~(\ref{eq:budtensor}).
It is possible to obtain the
2-nd rank structure tensor by 
contracting the (4-th rank) photon tensor 
with the polarization vectors of the target photons:
\begin{equation}
W_{\mu^\prime \mu}(q,p,\lambda,\lambda^\prime) \equiv
\epsp[p]{\star\nu^\prime\!}{\lambda^\prime} \epsp[p]{\nu}{\lambda}\Wco\ . 
\end{equation}
The spin-1 structure tensor can be decomposed into a
spin-averaged, a spin-dependent symmetric (singly-polarized) and
a spin-dependent antisymmetric part \cite{Hoodbhoy:1989am,Mathews:1996yv}.
This can be achieved by writing the polarization
vectors in the following manner:
\begin{equation}
E^{\nu^\prime}F^\nu = -\frac{1}{2} g^{\nu^\prime \nu}
+\frac{1}{2}(E^{\nu^\prime}F^\nu+F^{\nu^\prime}E^\nu+g^{\nu^\prime \nu})
+\frac{1}{2}(E^{\nu^\prime}F^\nu-F^{\nu^\prime}E^\nu)
\end{equation}
with $E^{\nu^\prime}=\epsp[p]{\star\nu^\prime\!}{\lambda^\prime}$ and 
$F^\nu =\epsp[p]{\nu}{\lambda}$.
The spin-averaged 
(structure functions $F_1$, $F_2$)
and spin-dependent antisymmetric parts 
(structure functions $g_1$, $g_2$)
are well-known
from spin-$1/2$ targets. On the other hand the spin-dependent 
symmetric tensor occurs in the spin-1 case for the first time.
It can be described in terms of 4 structure functions 
%usually denoted by 
$b_1,\ldots,b_4$ \cite{Hoodbhoy:1989am}.

In the rest of this section we discuss the spin-averaged 
part in more detail. 
In view of the recent literature on the parton content of 
virtual longitudinal photons
\cite{Chyla:2000hp,*Chyla:2000cu,*Chyla:2000ue,Friberg:2000nx}
and the factorization to be addressed in Sec.~\ref{sec:fact1}
we discuss also 
structure functions for transversely and
longitudinally polarized targets.
The following expressions are simplified if one introduces 
the transverse components 
of a four-vector $x_\mu$ and of 
the metric tensor
$g_{\mu\nu}$ 
\begin{equation}
x^T_\mu=x_\mu-\frac{q\cdot x}{q^2}q_\mu\, , \qquad
g^T_{\mu\nu}=g_{\mu\nu}-\frac{1}{q^2}q_\mu q_\nu \, ,
\end{equation}
where 'transverse' refers to $q$:
$q\cdot x^T =0$, $q^\mu g^T_{\mu\nu}=q^\nu g^T_{\mu\nu}=0$.
\subsection{Structure Functions for a Spin-Averaged Photon}\label{sec:spinav}
% Usually, in the literature ... are introduced.
Usually one introduces structure functions for a {\em spin-averaged}
target photon. The corresponding structure tensor can be obtained
by contracting $\Wco$ given in Eq.~(\ref{eq:budtensor}) with the metric 
tensor $g^{\nu\nu^\prime}$.
With the help of Eqs.~(\ref{eq:rmunu}) and (\ref{eq:orthogonality}) one obtains\footnote{Of 
course the virtual photon has three ($+,-,0$) degrees of freedom.
The factor $1/2$ guarantees the {\em conventional} normalization in the 
real photon limit with only two ($+,-$) transverse degrees of freedom.}:
\begin{equation}\label{eq:wsymspinav1}
\begin{aligned}
W_{\mu^\prime \mu}^{<\gamma>} & \equiv \frac{-g^{\nu{\nu^\prime}}}{2} \Wco
\\
&=R_{\mu^\prime\mu} \left[\WTT-\frac{1}{2} \WTS\right] + Q_{1\mu^\prime}Q_{1\mu} 
\left[\WST - \frac{1}{2} \WSS \right]
\\
&=-g_{\mu^\prime \mu}^T \left[\WTT-\frac{1}{2}\WTS\right] 
+p_{\mu^\prime}^T p_{\mu}^T \frac{Q^2}{\nu^2 \vtwo^2} 
\left[\WIIT-\frac{1}{2}\WIIS \right]
\end{aligned}
\end{equation}
where $\WIIT \equiv \WTT+\WST$ and $\WIIS \equiv \WTS +\WSS$.
Recall that the first index ($a = 2,{\rm T,L}$)
of the invariant functions $W_{ab}$ refers to the probe photon 
and the second one ($b = {\rm T,L}$) to the target photon.

Alternatively the spin-averaged tensor can be expressed 
in standard form in terms of the
structure functions $F_1\equiv W_1$ and $F_2 \equiv \nu W_2$:
\begin{equation}\label{eq:wsymspinav2}
\frac{1}{8 \pi^2 \alpha} W^{<\gamma>}_{\mu^\prime\mu}=-g_{\mu^\prime\mu}^T F_1^{<\gamma>} 
+ p_{\mu^\prime}^T p_{\mu}^T \frac{1}{\nu} F_2^{<\gamma>}\ .
\end{equation}

Comparing Eqs.~(\ref{eq:wsymspinav1}) and (\ref{eq:wsymspinav2})
we find
\begin{equation}
\begin{aligned}
2 x F_1^{<\gamma>} &=\frac{1}{8 \pi^2 \alpha}\frac{Q^2}{\nu}
\left[\WTT-\frac{1}{2}\WTS\right]
\\
F_2^{<\gamma>} &=
%\frac{1}{8 \pi^2 \alpha}
%\left[\WTT+\WST-\frac{1}{2}(\WTS+\WSS)\right]
%\frac{Q^2}{\nu \vtwo^2}
%\\
%&= &
\frac{1}{8 \pi^2 \alpha}\frac{Q^2}{\nu}\frac{1}{\vtwo^2} \left[\WIIT-\frac{1}{2} \WIIS\right]\ .
\end{aligned}
\end{equation}
These relations can be re-expressed in terms of the photon-photon cross sections 
$\sigma_{ab}= W_{ab}/(2 \nu \vtwo)\, (a,b = 2,\mathrm{L},\mathrm{T})$  
\cite{Nisius:1999cv,cit:Bud-7501,cit:Berger-Rep} 
\begin{equation}
\begin{aligned}\label{eq:sfs_av}
2 x F_1^{<\gamma>} &=\frac{Q^2}{4 \pi^2 \alpha}\vtwo
\left[\sigtt-\frac{1}{2}\sigtl\right]
\\
F_2^{<\gamma>} &=
%\frac{1}{8 \pi^2 \alpha}
%\left[\WTT+\WST-\frac{1}{2}(\WTS+\WSS)\right]
%\frac{Q^2}{\nu \vtwo^2}
%\\
%&= &
\frac{Q^2}{4 \pi^2 \alpha}\frac{1}{\vtwo} \left[\sigiit-\frac{1}{2} \sigiil\right]\ .
\end{aligned}
\end{equation}

Finally, $F_{\mathrm{L}}$ satisfies the usual relation
\begin{equation}
F_\mathrm{L}^{<\gamma>} = \vtwo^2 F_2^{<\gamma>} - 2 x F_1^{<\gamma>}\ .
\end{equation}
This can be seen by contracting
$W_{\mu^\prime \mu}^{<\gamma>}$ with the polarization vectors of longitudinal 
{\em probe} photons given in Eq.~(\ref{eq:photonpol}) thereby employing again
the orthogonality relations in Eq.~(\ref{eq:orthogonality})
\begin{displaymath}
\begin{aligned}
W_{\mathrm{L}}^{<\gamma>} &\equiv 
%{\varepsilon_{0}^{\star \mu^\prime}}\!\!(q) {\varepsilon_{0}^{\mu}}(q)
\epsp[q]{\star \mu^\prime\!\!}{0}\epsp[q]{\mu}{0}
W_{\mu^\prime \mu}^{<\gamma>}
= Q_1^{\mu^\prime}Q_1^{\mu}W_{\mu^\prime \mu}^{<\gamma>}=\WST-\frac{1}{2}\WSS
\\
&= 8 \pi^2 \alpha \left[-F_1^{<\gamma>}+ \frac{\vtwo^2}{2 x}F_2^{<\gamma>}\right]
\end{aligned}
\end{displaymath}
followed by the appropriate normalization:
\begin{displaymath}
F_\mathrm{L}^{<\gamma>} = \frac{1}{8 \pi^2 \alpha} 2 x W_{\mathrm{L}}^{<\gamma>}\ .
\end{displaymath}
%The factor $\vtwo^2 = 1 - 4 x^2 P^2/Q^2$ takes target mass corrections into account.
\subsection{Longitudinal and Transverse Target Photons}
Since the fluxes of transverse and longitudinal virtual photons 
will turn out to be {\em different} 
(see  Eq.~(\ref{eq:photonfluxes}) below)
it is most convenient to introduce structure
functions of transverse respectively longitudinal target photons
(instead of spin-averaged target photons).
The procedure is completely analogous to the one in the previous section 
using again Eqs.~(\ref{eq:photonpol})--(\ref{eq:orthogonality})
and for this reason the description will be brief.
% Transverse Photons

\noindent\underline{I. Transverse Photons}\\
With the help of Eq.~(\ref{eq:photonpol}) we can construct the structure tensor
for a transverse photon target which can
can be cast again into different forms
\begin{equation}
\begin{aligned}
W_{\mu^\prime \mu}^{\gam[T]} &\equiv 
%\frac{\epsilon_{+}^{*\nu^\prime}(p) \epsilon_{+}^\nu(p) +
%\epsilon_{-}^{*\nu^\prime}(p) \epsilon_{-}^{\nu}(p)}{2}\Wco
%\frac{\epsp[p]{\star\nu^\prime\!}{+} \epsp[p]{\nu}{+} 
%+\epsp[p]{\star\nu^\prime\!}{-} \epsp[p]{\nu}{-}}{2}
\frac{1}{2}\big[\epsp[p]{\star\nu^\prime\!}{+} \epsp[p]{\nu}{+} 
+\epsp[p]{\star\nu^\prime\!}{-} \epsp[p]{\nu}{-}\big]\Wco 
=\frac{1}{2} R^{\nu^\prime\nu}\Wco
\\
& = R_{\mu^\prime\mu} \WTT + Q_{1\mu^\prime}Q_{1\mu} \WST
=-g_{\mu^\prime \mu}^T \WTT +p_{\mu^\prime}^T p_{\mu}^T \frac{Q^2}{\nu^2 \vtwo^2} \WIIT
\\
&\overset{!}{=} 8 \pi^2 \alpha \left[ -g_{\mu^\prime\mu}^T F_1^{\gam[T]} 
+ p_{\mu^\prime}^T p_{\mu}^T \frac{1}{\nu} F_2^{\gam[T]} \right]
\end{aligned}
\end{equation}
%The projectors onto $F_1$ and $F_2$:
%\begin{eqnarray*}
%  P_1^{\mu^\prime\mu} &=&\frac{1}{2} 
%\left(-g^{\mu^\prime\mu}+\frac{-q^2}{X} p^{\mu^\prime} p^{\mu}\right)
%\\
%  P_2^{\mu^\prime\mu} &=& \frac{-q^2}{2 X} \nu 
%\left(-g^{\mu^\prime\mu} + 3 \frac{-q^2}{X} p^{\mu^\prime} p^{\mu}\right) 
%\end{eqnarray*}
and we can directly read off the structure functions: 
\begin{equation}\label{eq:sfsgamt}
\begin{aligned}
2 x F_1^{\gam[T]} &=\frac{1}{8 \pi^2 \alpha}\frac{Q^2}{\nu}\WTT
=\frac{Q^2}{4 \pi^2 \alpha} \vtwo \sigtt
\\
F_2^{\gam[T]} &=
\frac{1}{8 \pi^2 \alpha}\frac{Q^2}{\nu}\frac{1}{\vtwo^2} \WIIT
=\frac{Q^2}{4 \pi^2 \alpha}\frac{1}{\vtwo} \sigiit\ .
\end{aligned}
\end{equation}
Repeating the steps in Sec.~\ref{sec:spinav} to determine $F_\mathrm{L}^{<\gamma>}$, 
we obtain $W_{\mathrm{L}}^{\gam[T]}=\WST$ implying
\begin{equation}\label{eq:flgamt}
F_\mathrm{L}^{\gam[T]} = \frac{1}{8 \pi^2 \alpha} 2 x W_{\mathrm{L}}^{\gam[T]}
= \vtwo^2 F_2^{\gam[T]} - 2 x F_1^{\gam[T]}\ .
\end{equation}
 
% Longitudinal Photons
\noindent\underline{II. Longitudinal Photons}\\
The structure tensor for a longitudinal photon target 
is given by (using Eq.~(\ref{eq:photonpol}))
\begin{equation}
\begin{aligned}
W_{\mu^\prime\mu}^{\gam[L]} &\equiv
%\epsilon_{0}^{*\nu^\prime}(p) \epsilon_{0}^\nu(p) \Wco
\epsp[p]{\star \nu^\prime\!}{0}\epsp[p]{\nu}{0}\Wco
= Q_2^{\nu^\prime} Q_2^\nu \Wco
\\
& = R_{\mu^\prime\mu} \WTS + Q_{1\mu^\prime}Q_{1\mu} \WSS
=-g_{\mu^\prime \mu}^T \WTS +p_{\mu^\prime}^T p_{\mu}^T \frac{Q^2}{\nu^2 \vtwo^2} \WIIS
\\
&\overset{!}{=} 8 \pi^2 \alpha \left[ -g_{\mu^\prime\mu}^T F_1^{\gam[L]} 
+ p_{\mu^\prime}^T p_{\mu}^T \frac{1}{\nu} F_2^{\gam[L]} \right]
\end{aligned}
\end{equation}
and we find the following result for a longitudinal target photon:
\begin{equation}
\begin{aligned}\label{eq:sfsgaml}
2 x F_1^{\gam[L]} &=\frac{1}{8 \pi^2 \alpha}\frac{Q^2}{\nu}\WTS
=\frac{Q^2}{4 \pi^2 \alpha} \vtwo \sigtl
\\
F_2^{\gam[L]} &=
\frac{1}{8 \pi^2 \alpha}\frac{Q^2}{\nu}\frac{1}{\vtwo^2} \WIIS
=\frac{Q^2}{4 \pi^2 \alpha}\frac{1}{\vtwo} \sigiil\ .
\end{aligned}
\end{equation}
Finally, we have (as could be expected)
\begin{equation}\label{eq:flgaml}
W_{\mathrm{L}}^{\gam[L]}=\WSS \qquad \Rightarrow \qquad
F_\mathrm{L}^{\gam[L]} =\frac{1}{8 \pi^2 \alpha} 2 x W_{\mathrm{L}}^{\gam[L]}
= \vtwo^2 F_2^{\gam[L]} - 2 x F_1^{\gam[L]}\ .
\end{equation}

Further inspection of Eqs.~(\ref{eq:wsymspinav1})--(\ref{eq:flgaml})
reveals a relation 
[``$\displaystyle <\gamma> = \gam[T] - \tfrac{1}{2} \gam[L]$'']
between the spin-averaged, transverse
and longitudinal target photons
\begin{equation}
\begin{aligned}
W_{\mu^\prime \mu}^{<\gamma>}& = 
W_{\mu^\prime \mu}^{\gam[T]} - \frac{1}{2} W_{\mu^\prime \mu}^{\gam[L]} 
\\
F_{i}^{<\gamma>}& = F_{i}^{\gam[T]} - \frac{1}{2} F_{i}^{\gam[L]}\qquad (i = 1,2,\mathrm{L})
\end{aligned}
\end{equation}
which is a consequence of the completeness relation for {\em space-like} photons
%(with momentum $p$)
(cf.~\cite{cit:Bud-7501}, Eq.~(B.1)):
\begin{equation}
\begin{matrix}
%\underbrace{\epsilon_{+}^{*\mu}(p) \epsilon_{+}^{\nu}(p) + \epsilon_{-}^{*\mu}(p) \epsilon_{-}^{\nu}(p)}
\underbrace{\epsp[p]{\star\mu}{+} \epsp[p]{\nu}{+} +\epsp[p]{\star\mu}{-} \epsp[p]{\nu}{-}} 
&-& \underbrace{\epsp[p]{\star\mu}{0} \epsp[p]{\nu}{0}} & = & 
\underbrace{- g^{\mu \nu} + \frac{p^\mu p^\nu}{p^2}}&\ . \\
 2 \gam[T] &- & \gam[L] &=& 2 <\gamma>&
\end{matrix}
\end{equation}
\section{QED-Factorization}\label{sec:fact1}
\begin{figure}[htb]
\begin{center}
\setlength{\unitlength}{1pt}
\SetScale{1.8}
\vspace*{1.5cm}
\begin{picture}(80,70)
%\SetColor{CadetBlue}
\SetColor{Black}
%\SetColor{NavyBlue}
% upper electron
%\betont
   \Line(0,50)(25,50)
   \Line(25,50)(70,65)
% \Photon(x1,y1)(x2,y2){amplitude}{wiggles}
   \Photon(25,50)(50,40){-1.3}{6}
   \rText(45,75)[][]{\normalsize$q$}
% 50 * 1.3 (->SetScale) = 65
%   \rText(-25,85)[b][]{\normalsize e$(p_1)$}
% lower electron
%\SetColor{Black}
\SetColor{NavyBlue}
   \Line(0,15)(25,15)
   \Line(25,15)(70,10)
%   \rText(-25,35)[t][]{\normalsize e$(p_2)$}
   \Photon(25,15)(35,20){1.3}{2}
%\SetColor{CadetBlue}
%\SetColor{NavyBlue}
\SetColor{Black}
   \Photon(35,20)(50,27.5){1.3}{3}
% \rText(x,y)[b,t,l,r][rotatio]{text}
   \rText(45,40)[][]{\normalsize$p$}
   \Line(55,40)(70,45)
   \Line(55,35)(70,35)
   \Line(55,30)(70,25)
%
%\SetColor{NavyBlue}
   \GOval(55,35)(8.6,8.6)(0){0}

% cut
\SetColor{Black}
\SetWidth{1.0}
\Line(34,24)(38,17)
\Line(32,23)(36,16)
\SetWidth{1.0}
\end{picture}
\end{center}
\caption{\sf Factorization of the $\twogam$
cross section into a flux of ``target'' photons radiated off 
the lower lepton line times the cross section
for deep inelastic electron-photon scattering (black part). 
The cut in the photon line
indicates a time order between the two subprocesses 
(photon emission {\em followed} by deep inelastic $\e \gamma$ scattering)  
and implies also that these two factors are independent of each other.}
\label{fig:fact1}
\end{figure}
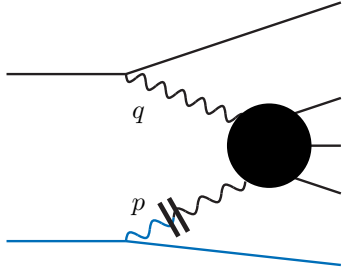
It is well known that for $P^2 \approx 0$ the general cross section
for the process $\twogam$ factorizes
into a product of a flux of target photons (radiated off the electron)
with the deep inelastic electron-photon scattering cross section
\cite{cit:Berger-Rep,Nisius:1999cv}, 
see Fig.~\ref{fig:fact1} for a graphical representation: 
\begin{equation}\label{eq:fact1}
\frac{d \sigma(ee \to ee X)}{dx dQ^2 dz dP^2} = 
\ft(z,P^2) \frac{d\sigma(e \gamma \to e X)}{dx dQ^2}\ \qquad (P^2 \approx 0)
\end{equation}
where $z= E_\gamma/E \approx y_2$ is the fraction of the lepton energy carried by the photon
(in the $\ee$-CMS).
The cross section for deep inelastic electron-photon scattering in Eq.~(\ref{eq:fact1})
reads
\begin{equation}\label{eq:egamWQ}
\begin{aligned}
\frac{d\sigma(e \gamma \to e X)}{dx dQ^2} & = 
\frac{4 \pi^2 \alpha y}{x Q^2}
\frac{\alpha}{2 \pi}
\left[\frac{1+(1-y)^2}{y} \frac{1}{Q^2}\right] 
\left[2 x \sfs[F]{1}{\gamma} + \frac{2 (1-y)}{1+(1-y)^2}\sfs{L}{\gamma}
\right]
\\
&=\frac{2 \pi \alpha^2}{x Q^4}
\left[(1+(1-y)^2) 2 x \sfs[F]{1}{\gamma} + 2 (1-y)\sfs{L}{\gamma}
\right]
\end{aligned}
\end{equation}
with the usual variable $\displaystyle y = \frac{p \cdot q}{p \cdot p_1} = y_1$.
Furthermore, $\ft$ denotes the flux factor
of {\em transversely} (or circularly) polarized photons with virtuality $P^2$.
For use below we also provide the  
flux factor $\fl$ of longitudinally polarized photons:
\begin{equation}\label{eq:photonfluxes}
\begin{aligned}
\ft(z,P^2) &= 
\frac{\alpha}{2 \pi}
\left[\frac{1+(1-z)^2}{z} \frac{1}{P^2}- \frac{2 \me^2 z}{P^4}\right] 
\\
\fl(z,P^2) &= \frac{\alpha}{2 \pi}
\left[\frac{2(1-z)}{z} \frac{1}{P^2}\right]\ .
\end{aligned}
\end{equation}
%
% short description ...
%See Fig.~\ref{fig:fact1} for a graphical representation of Eq.~(\ref{eq:fact1}):
%A ``target'' photon is radiated off the lower lepton and is subsequently 
%(indicated by the cut in the photon line)
%probed by a deep inelastic scattering process (black part).
%
%

The factorization in (\ref{eq:fact1}) is {\em essential}\ for 
relating the concept of the structure of a (real) photon 
to experimental measurements of two-photon processes.
For this reason we want to generalize Eq.~(\ref{eq:fact1}) for photons 
with virtuality $P^2 \ne 0$ and show that, in the Bjorken limit, factorization 
holds for virtual ``target'' photons as well.
For definiteness, as the Bjorken limit we consider
\begin{equation}
Q^2\equiv Q_1^2 \to \infty\ , \ \nu \to \infty\ , \ 
x = Q^2/2 \nu = \mathrm{fixed}\ .
\end{equation}
Practically, this means $P^2\equiv Q_2^2 \ll Q^2,\nu$ such that 
$\xp \equiv P^2/2 \nu = x P^2/Q^2$ is a small quantity which can be neglected.

% ``Proof'':
The starting point is the general cross section in Eq.~(\ref{eq:eeXsec}).
Employing Eqs.~(\ref{eq:sfsgamt}), (\ref{eq:flgamt}), 
(\ref{eq:sfsgaml}), and (\ref{eq:flgaml})
%and rearranging the square bracketed terms 
and rearranging the terms inside the square brackets 
it can be written as
\begin{equation}\label{eq:eeXsec2}
\begin{aligned}
d^6\sigma 
&=  \frac{d^3 p^{\prime}_1 d^3 p^{\prime}_2}{E^{\prime}_1 E^{\prime}_2}
\frac{\alpha^2}{16 \pi^4 Q^2 P^2}\frac{F_{\gamma\gamma}}{F_{ee}} 
%\\
\Bigg[\frac{4 \pi^2 \alpha}{Q^2 \vtwo}
\times
\\
&\phantom{=} 
\Big(2 \rtwopp 2 \ronepp 
[2 x \sfs{1}{\gam[T]} + \vep \sfs{L}{\gam[T]}]
 +  \rtwozz 2 \ronepp [2 x \sfs{1}{\gam[L]} + \vep \sfs{L}{\gam[L]}]\Big)
\\
&\phantom{=}  
+2 |\ronepm \rtwopm| \tautt \cos 2 \bar{\phi}
  -8 |\ronepz \rtwopz| \tautl \cos \bar{\phi} \Bigg]
\\
& \text{with} \qquad \vep = \frac{\ronezz}{2 \ronepp}\ .
\end{aligned}
\end{equation}
% with $\vep = \tfrac{\ronezz}{2 \ronepp}$.

The general strategy will be to demonstrate that
$\rtwopp$ and $\rtwozz$ are proportional to the flux factors of 
transverse and longitudinal photons, respectively, radiated off an electron
and that the interference terms disappear after having performed
an appropriate angular integration.

In the Bjorken limit it is useful to perform a ``light cone decomposition''
of the 4-momenta of the two photons \cite{cit:Bud-7501}:\\
\begin{equation}\label{eq:lcd}
\begin{gathered}
q  = \qp p_1 + \qm p_2 + \qt\,
\qquad \,
p = \ppl p_1 + \pmi p_2 + \pt
\\
\text{with} 
\\
p_i^2 = 0,\ p_i\cdot \qt = p_i\cdot \pt = 0,\ p_1 \cdot p_2 = S/2
\end{gathered}
\end{equation}
where $S$ is the square of the $\e\e$-CMS energy.
The momentum fractions $\qp$, $\qm$, $\ppl$ and $\pmi$ and
the transverse momenta can be easily calculated:
\begin{equation}\label{eq:lcd2}
\begin{gathered}
\begin{array}{lcl}
2 p_1 \cdot q = S \qm = -Q^2 &\qquad \Rightarrow &\qquad \qm = -Q^2/S
\\
2 p_2 \cdot q = S \qp & \qquad \Rightarrow & \qquad
\qp = 2 p_2\cdot q/S\, (\approx y_1)
\\
2 p_1 \cdot p  = S \pmi & \qquad \Rightarrow & \qquad
\pmi = 2 p_1\cdot p/S\, (\approx y_2)
\\
2 p_2 \cdot p  = S \ppl = -P^2 & \qquad \Rightarrow &\qquad \ppl = -P^2/S
\\
Q^2 =- S \qp \qm - \qt^2 & \qquad \Rightarrow &\qquad \qt^2 = -Q^2 (1-\qp)
\\
P^2 =- S \ppl \pmi - \pt^2 & \qquad \Rightarrow &\qquad \pt^2 = -P^2 (1-\pmi)
\end{array}
\\
\qt \cdot \pt  \equiv - \sqrt{\qt^2 \pt^2} \cos \phi 
                = - \sqrt{Q^2 P^2 (1-\qp)(1-\pmi)} \cos \phi \ .
\end{gathered}
\end{equation}
Here $\phi$ is
the angle between the scattering planes of
the $\e^-$ and $\e^+$ in the $\ee$-CMS.
Obviously $\ppl$ is negligibly small such that we can use\footnote{Of course, in the calculation 
of quantities which are themselves small of the order $P^2$ 
(e.g. $\pt^2$ in the $\ee$-CMS or $\piit^2$ in the $\gamma \gamma$-CMS) $\ppl$ must be 
taken into account.} 
\begin{equation}\label{eq:lcone}
p = \pmi p_2 + \pt \ .
\end{equation}
(On the other hand $\qm$ {\em cannot} be neglected in the Bjorken limit.)
In the $\ee$-CMS the 4-momenta of the incoming leptons can be written as
$p_1 = (E,0,0,E)$ and $p_2 = (E,0,0,-E)$ where $E = \tfrac{\sqrt{S}}{2}$
(neglecting terms of the order $\Ord(\tfrac{\me^2}{S})$).
Since the transverse 4-vector is given by $\pt = (0,{\pt}_x,{\pt}_y,0)$
we can infer from Eq.~(\ref{eq:lcone}) that $E_\gamma = \pmi E$, i.e., in the $\ee$-CMS 
$\pmi$ is the energy fraction of the lepton energy transferred to the photon. 
% $p = \pmi p_2 + \pt$.
For a {\em real} ($P^2 = 0$) photon we recover the familiar relation
$p = \pmi p_2$ between the {\em 4-momenta} $p$ and $p_2$ 
of the collinearly radiated photon and its (massless) ``parent'' lepton, respectively.

Before turning to the photon density matrix elements in the Bjorken limit it
is helpful to relate the variable 
$\nu \equiv p \cdot q$ to $\qp$, $\pmi$ and the 
transverse momenta:
\begin{equation}\label{eq:nu}
2\nu = S \pmi \qp (1+\rho),\,  \rho \equiv 
\frac{2 \pt \cdot \qt}{S \pmi \qp}
\approx - 2 \sqrt{x \xp (1-\qp)(1-\pmi)} \cos \phi 
\end{equation}
where $\rho \propto \sqrt{\xp}$ is small in the Bjorken limit.
Employing Eq.~(\ref{eq:nu}) we find in addition
\begin{equation}\label{eq:y1y2}
y_1 \equiv \frac{\nu}{p \cdot p_1} = \qp (1+\rho)\ , \qquad 
y_2 \equiv \frac{\nu}{q \cdot p_2} = \pmi (1+\rho)\ . 
\end{equation} 

% Phase Space:
Introducing the variables 
$\omega_1 \equiv q\cdot (p_1+p_2)/\sqrt{S} = \tfrac{\sqrt{S}}{2} (\qm+\qp)$
and 
$\omega_2 \equiv p\cdot (p_1+p_2)/\sqrt{S} = \tfrac{\sqrt{S}}{2} (\pmi+\ppl)$
the phase space can be written as \cite{cit:Bud-7501} [Eq.(5.15b)]
(up to terms of the order $\Ord(\tfrac{\me^2}{S})$) 
\begin{equation}\label{eq:PS}
\begin{aligned}
\frac{d^3 p_1^\prime}{E_1^\prime} \frac{d^3 p_2^\prime}{E_2^\prime} 
&=\frac{2\pi}{S} dQ^2 dP^2 d\omega_1 d\omega_2 d\phi
%& = E_1^\prime dE_1^\prime d\phi_1 d\cos \theta_1
%    E_2^\prime dE_2^\prime d\phi_2 d\cos \theta_2
=\frac{\pi}{2} dQ^2 d\qp dP^2 d\pmi d\phi
\\
&=\frac{\pi}{2} dQ^2 dy_1 dP^2 d\pmi d\phi 
(1+\Ord(\rho))
\end{aligned}
\end{equation}
where the third equality can be understood with the help of Eq.~(\ref{eq:y1y2}).

In the Bjorken limit the photon density matrix elements 
in Eq.~(\ref{eqn:rhos}) can be cast in a very compact form
using the symmetric notation $Q_1^2 = Q^2$ and $Q_2^2 = P^2$:
\begin{equation}\label{eq:rhos2}
\begin{aligned}
2 \ripp & = 
\frac{2}{y_i} Q_i^2 \left[\frac{1+(1-y_i)^2}{y_i} \frac{1}{Q_i^2}
- \frac{2 \me^2 y_i}{Q_i^4}\right] + \Ord(\xp)
\\
\rizz & =
\frac{2}{y_i} Q_i^2 \left[\frac{2(1-y_i)}{y_i} \frac{1}{Q_i^2}\right]
+ \Ord(\xp)
\\
|\ripm| & = 
\frac{1}{y_i} Q_i^2 \left[\frac{2(1-y_i)}{y_i} \frac{1}{Q_i^2}
- \frac{2 \me^2 y_i}{Q_i^4}\right] + \Ord(\xp)\ .
\end{aligned}
\end{equation}
These results have to be expressed by the independent integration
variables $Q^2$, $y_1$, $P^2$, and $\pmi$. Furthermore, from here
on we identify $y_1 \equiv y$.

Obviously $\rtwopp$ and $\rtwozz$ are proportional to the flux factors
of transverse respectively longitudinal photons 
in Eq.~(\ref{eq:photonfluxes})
\begin{equation}\label{eq:photonfluxes2}
\begin{aligned}
2 \rtwopp &= \frac{2}{\pmi} P^2 \frac{2\pi}{\alpha} \ft(\pmi,P^2) 
+ \Ord(\rho)
\\
\rtwozz & = \frac{2}{\pmi} P^2 \frac{2\pi}{\alpha} \fl(\pmi,P^2) 
+ \Ord(\rho)
\end{aligned}
\end{equation}
%\begin{equation}\label{eq:photonfluxes2}
%\begin{aligned}
%2 \rtwopp &= \frac{2}{y_2} P^2 \frac{2\pi}{\alpha} \ft(y_2,P^2) 
%+ \Ord(\xp)
%\\
%\rtwozz & = \frac{2}{y_2} P^2 \frac{2\pi}{\alpha} \fl(y_2,P^2) 
%+ \Ord(\xp)\ .
%\end{aligned}
%\end{equation}
and similarly we can write
\begin{equation}\label{eq:rho1}
\begin{aligned}
2 \ronepp & = 
\frac{2}{y} Q^2 \left[\frac{1+(1-y)^2}{y} \frac{1}{Q^2}\right] 
\, , \qquad
\ronezz =
\frac{2}{y} Q^2 \left[\frac{2(1-y)}{y} \frac{1}{Q^2}\right]
\\
\vep &= \frac{\ronezz}{2 \ronepp}= \frac{2 (1-y)}{1+(1-y)^2}
\end{aligned}
\end{equation}
where we have discarded the mass terms due to $Q^2 \gg \me^2$ .

Inserting Eqs.~(\ref{eq:photonfluxes2}), 
(\ref{eq:rhos2}), and (\ref{eq:PS}) into Eq.~(\ref{eq:eeXsec2})
one straightforwardly obtains 
(using $\frac{F_{\gamma\gamma}}{F_{ee}}=2 \nu \vtwo/S = \pmi y \vtwo$) 
\begin{equation}
\begin{aligned}
d\sigma 
&= dQ^2 dy dP^2 d\pmi \frac{d\phi}{2 \pi} 
\bigg\{\ft(\pmi,P^2) 
\frac{d\sigma(e \gam[T] \to e X)}{dy dQ^2}|_{\hat{S}=\pmi S}
+ [\gam[T] \rightarrow \gam[L]] 
\\
&\phantom{= dQ^2 dy dP^2 d\pmi \frac{d\phi}{2 \pi} \bigg\{} 
+ \Ord(\rho)
+ \text{interference-terms} 
\bigg\}
%\\
%&\qquad +2 |\ronepm \rtwopm| \tautt \cos 2 \bar{\phi}
%  -8 |\ronepz \rtwopz| \tautl \cos \bar{\phi} \Bigg]\ .
%
\end{aligned}
\end{equation}
with the cross sections for deep inelastic electron-photon scattering given
by [cf. Eq.~(\ref{eq:egamWQ})]
\begin{equation}\label{eq:egamWQ2}
\begin{aligned}
\frac{d\sigma(e \gam[T,L] \to e X)}{dy dQ^2}|_{\hat{S}=\pmi S}&=
\frac{4 \pi^2 \alpha}{Q^2} \frac{\alpha}{2 \pi} 
\left[ \frac{1+(1-y)^2}{y} \frac{1}{Q^2}\right]\times
\\
&\phantom{=}
\left[2 x \sfs{1}{\gam[T,L]}(x,Q^2) + \vep \sfs{L}{\gam[T,L]}(x,Q^2)
\right]
\end{aligned}
\end{equation}
where $\vep$ can be found in Eq.~(\ref{eq:rho1}).
Note that only two of the variables $x$, $y$, and $Q^2$ are independent since
they are related via $Q^2 = \hat{S} x y$.
The terms proportional to $\rho$ vanish after $\phi$-integration like terms
of the order $\Ord(\xp)$.

Unfortunately, the interference terms are proportional to
$\cos \bar{\phi}$ and $\cos 2\bar{\phi}$ 
where
$\bar{\phi}$ is the angle of the electron scattering 
planes in the $\gamma\gamma$-CMS while $\phi$ is  
the angle of the electron scattering planes in the $e^+e^-$-CMS
($\equiv$ laboratory system).
However, we show below that
\begin{equation}\label{eq:cosphi}
\cos \bar{\phi} = \cos \phi\ \times\ (1 + \Ord(\rho))\ .
\end{equation}
Therefore, we can also get rid of the interference terms 
(proportional to $\cos \bar{\phi} \approx \cos \phi$)
by integrating over $\phi$, leaving a remainder 
of the order $\Ord(\sqrt{\xp})$.
The latter becomes obvious if we remember that the 
variable $\rho$ defined in (\ref{eq:nu})
is proportional to $\sqrt{\xp} \cos \phi$ such that terms
$\rho \cos \phi$ occurring for example in Eq.~(\ref{eq:cosphi}) 
are proportional to
$\sqrt{\xp} \cos^2 \phi$ which do {\em not} vanish
by integrating over $\phi$. 

Before calculating $\cos \bar{\phi}$ let us state the final
factorization formula:
\begin{equation}\label{eq:fact2}
\begin{aligned}
\frac{d \sigma(ee \to ee X)}{dy dQ^2 d\pmi dP^2} &= \phantom{+{}}
\ft(\pmi,P^2) \frac{d\sigma(e \gam[T] \to e X)}{dy dQ^2}|_{\hat{S}=\pmi S} + \Ord(\xp)
\\
&\phantom{={}}+\fl(\pmi,P^2) \frac{d\sigma(e \gam[L] \to e X)}{dy dQ^2}|_{\hat{S}=\pmi S}
+ \Ord(\xp)
\\
&\phantom{={}} +\Ord(\sqrt{\xp})\ .
\end{aligned}
\end{equation}

To complete our derivation we still have to show that 
$\cos \bar{\phi} = \cos \phi + \Ord(\rho)$
where $\cos \bar{\phi}$ is given by Eq.~(A.4) in
\cite{cit:Bud-7501}
\begin{equation}\label{eq:phibar}
\cos \bar{\phi} \equiv \frac{- \pit \cdot \piit}{\sqrt{\pit^2 \piit^2}}
\qquad \text{with} \qquad {p_i}_\perp^\mu = - {p_i}_\nu R^{\mu \nu}(q,p)
\end{equation}
and where $R^{\mu \nu}(q,p)$ has been defined in Eq.~(\ref{eq:rmunu}).
Using the decomposition in Eq.~(\ref{eq:lcd}) it is straightforward to obtain
\begin{equation}
\begin{aligned}
\pit \cdot  \piit &= \frac{\pt \cdot \qt}{\pmi \qp}
\\
\pit^2 &=-\frac{Q^2}{{\qp}^2}(1-\qp) + \Ord(\rho) 
\\
\piit^2 &=-\frac{P^2}{{\pmi}^2}(1-\pmi) + \Ord(\rho) \ .
\end{aligned}
\end{equation}
Inserting these relations into Eq.~(\ref{eq:phibar}) and comparing with
the definition of $\cos \phi$ in Eq.~(\ref{eq:lcd2}) we find the above stated result 
\begin{equation}
\cos \bar{\phi} = \frac{- \pt \cdot \qt}{\sqrt{\pt^2 \qt^2}} + \Ord(\rho)
= \cos \phi + \Ord(\rho)\ .
\end{equation} 

A few comments are in order:
\begin{itemize}
% Real Photon Limit
\item In the limit $P^2 \to P^2_{\text{min}} \approx 0$ the 
contributions from longitudinal target photons
have to vanish since a real photon has only two transverse physical 
degrees of freedom and one recovers Eq.~(\ref{eq:fact1}).
Indeed, the 
doubly virtual box results
for $\sigtl$ and $F_2^{\gam[L]}(x,Q^2)$
vanish like $\propto P^2/m_q^2$ where $m_q$ is
the quark mass.
Clearly, the box expressions are not reliable in this limit.
However, assuming that 
the structure functions of a longitudinal target photon vanish like 
$P^2/\Lambda^2$ where $\Lambda$
is a typical hadronic scale, say $\Lambda = m_\rho$, we 
obtain a {\em non-zero} result 
because $\fl \propto 1/P^2$ which is, however, negligible 
compared to the contribution from
transverse target photons.
% Numerical check
\item The factorized result in Eq.~(\ref{eq:fact2}) is valid up to terms
$\Ord(\sqrt{\xp})$ which formally go to zero in the Bjorken limit.
However, for practical purposes it is not clear when 
$\xp= x P^2/Q^2$ is small enough for Eq.~(\ref{eq:fact2}) to be a 
good approximation of the exact
cross section in Eq.~(\ref{eq:eeXsec}). Here, a numerical comparison 
of the factorization formula with the exact cross section 
would be interesting.
% comment:ep-scattering 
\item As is well known, a similar 
factorization into transverse and longitudinal photon
fluxes also holds in $ep$ scattering
such that the parton distributions for transverse and 
longitudinal virtual photons are useful also in the description
of these reactions. 
Here 'transverse' and 'longitudinal' refer to the $\gamma^\star p$-CMS.
% t-channel bremsstrahlung
\item One should keep in mind that the factorization 
discussed here only applies to two-photon
processes as specified in the introduction and 
in Sec.\ \ref{sec:kinematics}. 
%It basically depends on the ratio $P^2/Q^2$.
%At larger virtualities $P^2$, the 
%$t$-channel bremsstrahlung contributions 
%to the observable process $\eetoeeX$ 
%might no longer be negligible, see the discussion in 
%Sec.~\ref{sec:kinematics}.
\end{itemize}
%%%%%%%%%%%%%%%%%%%%%%%%%%%%%%%%%%%%%%%%%%%%%%%%%%%%%%%%%%%%%%%%%%%%%%%%%%%%%%%%%%%%%
% discussion
%\item space-time picture? Physical reason, why factorization in Bj-lim
%%%%%%%%%%%%%%%%%%%%%%%%%%%%%%%%%%%%%%%%%%%%%%%%%%%%%%%%%%%%%%%%%%%%%%%%%%%%%%%%%%%%%
%
\section{Conclusions}\label{sec:conclusions}
In this work two-photon processes and their relation to the structure
of real and virtual photons have been reviewed.
Sec.~\ref{sec:twogam} contains a comprehensible 
introduction to two-photon processes including a re-calculation
of the virtual photon-photon box ($\vgvg$).
Thereby two discrepancies in the literature could be clarified.
A hopefully useful collection of several limits of the most
general results has been provided in the Appendix \ref{app:limits}.
Many of these results are wide-spread over the literature 
%utilizing
with
different notations and conventions.

In the 'generalized' ($P^2 \ne 0$) Bjorken limit, 
which practically means $P^2 \ll Q^2$, two-photon processes 
can be described in terms
of virtual photon structure functions.
In contrast to the case of real target photons with only 
transverse ($=$ spin-averaged) degrees of freedom 
%which also coincide with the spin-averaged case 
one can distinguish between spin-averaged,
transverse and longitudinal target photons. The latter two being
more directly related to observable quantities.
In Sec.~\ref{sec:photonsfs} virtual photon structure functions
have been defined for spin-averaged, transverse and longitudinal
target photons and their relation has been discussed.

Finally, we have demonstrated the factorization 
of two-photon processes in the generalized Bjorken limit
in Sec.~\ref{sec:fact1}.
The factorization is into fluxes of {\em transverse} and 
{\em longitudinal} target photons times the corresponding cross
sections for deep inelastic scattering off these targets.

These results being model independent may serve as a starting 
point for parton model calculations.
In the parton model the partonic content of 
both spin-averaged and transverse photons
is governed by {\em inhomogeneous} evolution equations.
The difference is in the boundary conditions.
On the other hand, parton distributions of longitudinal 
target photons obey {\em homogeneous} evolution equations.
Since the transverse and longitudinal photon fluxes are different
it appears to be advantageous to analyze the parton content of
transverse and longitudinal target photons.
This is also supported by the fact that
a similar factorization holds in $ep$ scattering.
%where the notions 'transverse' and 'longitudinal'
%again refer to the probe-target-CMS.
%
Measurements of triple differential dijet cross sections
at the $ep$ collider HERA 
indicate that the contributions from resolved longitudinal
photons 
in addition to transverse resolved photons
improve the description of the data \cite{Sedlak:2001pr}.
Moreover, the $y$-dependence of these data 
favors a resolved $\gam[L]$ contribution instead of
simply enhancing the $\gam[T]$ resolved processes.
On the other hand, due to experimental limitations
measurements of $\DISeg$ are performed at small $y$ where
the transverse and longitudinal fluxes are equal.
However, it is not the spin-averaged 
combination '$\gam[T]-\tfrac{1}{2}\gam[L]$' which is 
probed but the effective combination '$\gam[T]+\gam[L]$'.

So far models for the parton content of virtual photons
\cite{cit:GRSc99,cit:GRS95,cit:SaS-9501,*Schuler:1996fc}
do not clearly distinguish between 'transverse' and 'spin-averaged'.
Furthermore, the only analysis
of the parton content of longitudinal target photons
\cite{Chyla:2000hp,*Chyla:2000cu,*Chyla:2000ue}
is valid in the perturbative realm
$P^2 \gg \Lambda^2$ where $\Lambda$ is a typical hadronic scale 
and does not take into account 
any hadronic contributions.
This leaves room for future investigations of the parton
content of virtual photons.
\appendix
%\section{Limits of the Doubly Virtual Box}\label{sec:limits}
\section{Limits of the Doubly Virtual Box}\label{app:limits}
Due to the particular choice of variables, the results 
in Eqs.~\eqref{eq:sigtt}--\eqref{eq:tauatl}
%(\ref{eq:LO-box}) 
are especially well suited for deriving 
various important limits %(which partly overlap) 
wide-spread over the literature.
\begin{itemize}
\item Most important is the Bjorken limit 
($Q^2 \to \infty, \nu \to \infty, x={\rm const}$) 
in which we can study structure functions of the real and virtual photon.
%as has already been discussed in Chapter \ref{chap:twogam}. 
\item Another important case is the real photon $P^2=0$ ($\xp = 0$) case 
in which the general virtual box results in 
\eqref{eq:sigtt}--\eqref{eq:tauatl}
%(\ref{eq:LO-box}) 
reduce to the
standard box-diagram $\vgg$ expressions
for a {\em real} photon $\gamma \equiv \gamma(P^2 = 0)$. 
Keeping the full mass dependence in 
\eqref{eq:sigtt}--\eqref{eq:tauatl}
%(\ref{eq:LO-box}) 
we obtain expressions relevant
for the heavy quark contribution to the photon structure functions. 
\item Finally, the general light quark mass 
limit (for arbitrary $P^2$, $Q^2$) can be
easily obtained from 
\eqref{eq:sigtt}--\eqref{eq:tauatl}
%(\ref{eq:LO-box}) 
by setting $m=0$ ($\lambda = 0$, $\vone = 1$) and
needs no separate discussion.
\end{itemize}

% Outline
In the following, our main concern will lie on the Bjorken limit.
%$Q^2 \to \infty, \nu \to \infty, x={\rm const}$.
Practically this limit means that $Q^2$ is much larger than the other scales $m^2$ and $P^2$.
Beside the general case $m^2,P^2 \ll Q^2$ which will be studied in Section \ref{sec:bjorken1}
additional orderings 
$m^2=0, P^2 \ll Q^2$ (Section \ref{sec:bjorken2}) and $P^2=0, m^2 \ll Q^2$ (Section \ref{sec:bjorken3})
are of interest and the expressions further simplify 
under these circumstances.
One must be careful in handling these limits because for some of the
box cross sections the result depends on which of the two limits $m \to 0$ and
$P^2 \to 0$ is taken first.
For example, below we will find the following 
asymptotic expressions for $\sigtl$ derived
from 
\eqref{eq:sigtt}--\eqref{eq:tauatl}:
%(\ref{eq:LO-box}): 
\begin{equation*}
0 = \lim_{m^2 \to 0} \lim_{P^2 \to 0} \sigtl \neq \lim_{P^2 \to 0} \lim_{m^2 \to 0} \sigtl 
= N_c e_q^4 \frac{4 \pi \alpha^2}{Q^2} 4 x^2 (1-x)  
\end{equation*}
Mathematically, the origin of such a behavior is easily identified.
Terms like $\displaystyle \frac{4 x \xp}{4 x \xp + \lambda \vtwo^2}$ 
occurring in 
\eqref{eq:sigtt}--\eqref{eq:tauatl}
%(\ref{eq:LO-box})
(not only as the argument of the logarithm) require a careful treatment.  
In general they are {\em not} negligible even for small $m^2$ and $P^2$ 
and, viewed as a function of $m^2$ and $P^2$, they are discontinuous at
$(m^2,P^2)=(0,0)$.
%\footnote{anomaly, physically no problem}
Finally, in Section \ref{sec:bjorken3} we also derive expressions
for the heavy quark contributions to the photon structure
functions in the real photon limit by setting $P^2 = 0$ ($\xp=0$) 
but keeping the full mass 
dependence in 
\eqref{eq:sigtt}--\eqref{eq:tauatl}.
%(\ref{eq:LO-box}). 

All results for the photon-photon cross sections will be given 
for a single quark with charge $e_q$ and mass $m$.
The photon structure functions 
$\sfs{i}{},\, (\mathrm{i}=1,2,\mathrm{L})$ can be obtained from
these expressions with help of the relations in 
Section \ref{sec:photonsfs} which
simplify in the Bjorken limit ($P^2 \ll Q^2 \Rightarrow \vtwo \simeq 1$):
\begin{alignat}{2}\label{eq:sfs1}
\sfs{2}{\gam[T]} &= \Normc \sigiit\, , & \qquad \sfs{L}{\gam[T]} 
&= \Normc \siglt 
\nonumber\\
\sfs{2}{\gam[L]} &= \Normc \sigiil\, , & \qquad \sfs{L}{\gam[L]} 
&= \Normc \sigll 
\end{alignat} 
with $\sigiit = \sigtt + \siglt$, $\sigiil = \sigtl + \sigll$.
% used, employed, utilized
The commonly utilized expressions for a spin-averaged target photon 
are given by
\begin{equation}\label{eq:sfs2}
\sfs{i}{<\gam>}=\sfs{i}{\gam[T]} - \tfrac{1}{2}\sfs{i}{\gam[L]}, \, 
(\mathrm{i}=1,2,\mathrm{L})\ .
\end{equation}
Finally, the structure function $\sfs{1}{}$ can be deduced from 
$\sfs{L}{} = \sfs{2}{}- 2 x \sfs{1}{}$.
Since the structure functions are 
(apart from the normalization factor $Q^2/4 \pi^2 \alpha$)
simple linear combinations of the photon-photon cross sections they will only 
be written out in some special cases.
% $\tTT$ and $\tTS$ related to $W_3$; $\aTT$, $\aTS$ related to 
%the polarized sf $g_1$
%
\subsection{General Bjorken Limit: $m^2,P^2 \ll Q^2$}\label{sec:bjorken1}
In the general Bjorken limit the normalization factor $N$ given 
in Eq.~(\ref{eq:sigab}) and
the logarithm $L$ from Eq.~(\ref{eq:log}) are given by
\begin{equation}
\begin{gathered}
N = 4\pi N_c e_q^4 \frac{4 \pi \alpha^2}{Q^2} x\ (1+\Ord(\xp))
\\
L = \ln \frac{4}{4x\xp+\lambda}+\Ord(\xp,\lambda)\ .
\end{gathered}
\end{equation}
Keeping this in mind and using
$\vone = 1 + \Ord(\lambda)$, $\vtwo = 1 + \Ord(\xp)$
the following results can be easily deduced from
Eqs.~\eqref{eq:sigtt}--\eqref{eq:tauatl}:
%(\ref{eq:LO-box}): 
%\begin{equation}
\begin{align}\label{eq:Bj1}
\sigtt & = 
\Norm
\bigg\{\Big[x^2+(1-x)^2\Big]L+4x(1-x)-1
-\frac{4 x \xp}{4x\xp+\lambda}
%\nonumber\\
%&\phantom{= \Norm \bigg\{}
+\Ord(\xp,\lambda)\bigg\}
\nonumber\\
\siglt & =\Norm \left\{4x(1-x)+\Ord(\xp,\lambda)\right\}
\nonumber\\
\sigtl & = \Norm
\left\{4 x(1-x) \frac{4x\xp}{4x\xp+\lambda}+\Ord(\xp,\lambda)\right\}
\nonumber\\
%\sigll & = 0+\Ord(\xp,\lambda)
%\nonumber\\
\tautt & = \Norm
\left\{-2 x^2+\Ord(\xp,\lambda)\right\}
\nonumber\\
%\tautl & = \frac{4 \pi \alpha^2}{Q^2} N_c e_q^4 x\sqrt{x\xp}(1-x)
%\left\{4x L + 2(1-3x)+\Ord(\xp,\lambda)\right\}
%\nonumber\\
\tauatt & = \Norm
\left\{(2x-1)L+3-4x-\frac{4x\xp}{4x\xp+\lambda}
+\Ord(\xp,\lambda)\right\}
%\nonumber\\
%\tauatl & = -\frac{4 \pi \alpha^2}{Q^2} N_c e_q^4 x \sqrt{x\xp}4(1-x)
%\left\{\frac{4x\xp}{4x\xp+\lambda}+\Ord(\xp,\lambda)\right\}
%\ .
\end{align}
%\end{equation}
with $\Norm = N_c e_q^4 \frac{4 \pi \alpha^2}{Q^2} x$ where $N_c = 3$
is the number of colors and $e_q$ is the quark charge.
The remaining expressions are suppressed by powers of $\xp = x P^2/Q^2$ 
(see Eqs.~\eqref{eq:sigtt}--\eqref{eq:tauatl}). 
%(\ref{eq:LO-box})). 

At the price of a slightly worse approximation (at larger $x$) 
one can further use
\begin{equation}
\begin{gathered}
\frac{4 x \xp}{4 x \xp + \lambda} = \frac{P^2 x (1-x)}{P^2 x(1-x) + m^2} 
+\Ord(\tfrac{P^2}{W^2},\lambda)\, ,
\\
L = \ln \frac{Q^2 (1-x)}{x [P^2 x (1-x) + m^2]} 
+\Ord(\tfrac{P^2}{W^2},\lambda)\, .
\end{gathered}
\end{equation}
For example we can write: 
\begin{align}\label{eq:Bj1b}
\sigtt & = \Norm
\bigg\{\Big[1-2x(1-x)\Big]\ln \frac{Q^2 (1-x)}{x [P^2 x (1-x) + m^2]} 
+4 x (1-x)
\nonumber\\
& \phantom{=\Norm\bigg\{}
-1 
-\frac{P^2 x (1-x)}{P^2 x(1-x) + m^2} 
+\Ord(\tfrac{P^2}{W^2},\lambda)\bigg\}\ .
%\nonumber\\
%\sigtl & = \frac{4 \pi \alpha^2}{Q^2} N_c e_q^4 x(1-x)
%\Bigg\{4 x \frac{4x\xp}{4x\xp+\lambda}+\Ord(\xp,\lambda)\Bigg\}
%\nonumber\\
%\tauatt & =  \frac{4 \pi \alpha^2}{Q^2} N_c e_q^4 x
%\Bigg\{(2 x-1)(L-1)+(1-x) \frac{2 m^2- P^2 x(2x-1)}{m^2+ P^2 x(1-x)}
%+\Ord(\tfrac{P^2}{W^2},\lambda)\Bigg\}
%\nonumber\\
%\tauatl & = -\frac{4 \pi \alpha^2}{Q^2} N_c e_q^4 x \sqrt{x\xp}4(1-x)
%\left\{\frac{4x\xp}{4x\xp+\lambda}+\Ord(\xp,\lambda)\right\}
\end{align}
%$\tauatt$ in Eq.~(\ref{eq:Bj1b}) agrees up to normalization 
%with $g_1^{\gamma,Box}$ of \cite{cit:Bass-9201}.
% 
\subsection{$m^2=0, P^2 \ll Q^2$}\label{sec:bjorken2}
In this section we consider the
asymptotic virtual ($P^2 \ne 0$) box expressions for light quarks 
in the Bjorken limit
for which the expressions in Eq.~(\ref{eq:Bj1}) further reduce.
Noticing that the logarithm $L$ is given by $L = \ln \frac{Q^2}{P^2 x^2}$ we can
write in this case (neglecting terms of the order $\Ord(\xp)$): 
\begin{align}
\sigtt & \simeq  \Norm \left\{[x^2+(1-x)^2]\ln \frac{Q^2}{P^2 x^2} +4x(1-x)-2\right\}
\nonumber\\
\sigtl & \simeq \siglt \simeq \Norm [4 x (1-x)]
\nonumber\\
%\sigll & \simeq  0
%\nonumber\\
\tautt & \simeq \Norm [-2 x^2]
\nonumber\\
%\tautl & \simeq   N_c e_q^4 \frac{4 \pi \alpha^2}{Q^2}x\sqrt{x\xp}(1-x)
%\left\{4x \ln \frac{Q^2}{P^2 x^2} + 2(1-3x)\right\}
%\nonumber\\
\tauatt & \simeq  \Norm \left\{(2 x-1)\ln \frac{Q^2}{P^2 x^2}+2-4x\right\}\ .
%\nonumber\\
%\tauatl & \simeq - N_c e_q^4 \frac{4 \pi \alpha^2}{Q^2}x\sqrt{x\xp}4(1-x)
\end{align}
(Recall the normalization factor 
$\Norm = N_c e_q^4 \frac{4 \pi \alpha^2}{Q^2} x$.) 

Summing over $q=u,d,s$ and utilizing Eqs.~(\ref{eq:sfs1}) and (\ref{eq:sfs2})
we recover the well known asymptotic results for the virtual ($P^2 \ne 0$) box
structure functions for the light $q=u,d,s$ quarks in the Bjorken limit $P^2/Q^2 \ll 1$:
%\begin{align}
%\sfs{2,box}{\gam[T],\ell}(x,Q^2,P^2) &\simeq 
%N_c \sum e_q^4\, \frac{\alpha}{\pi} x 
%  \bigg\{ \left[ x^2+(1-x)^2\right] \, \ln \frac{Q^2}{P^2x^2} 
%\nonumber\\
%& \phantom{\simeq N_c\sum e_q^4\, \frac{\alpha}{\pi} x\bigg\{} 
%+ 8x(1-x)
%-2 \bigg\}
%\label{eq:f2boxt}\\
%\sfs{2,box}{\gam[L],\ell}(x,Q^2,P^2) &\simeq 
%N_c \sum e_q^4\, \frac{\alpha}{\pi} x \left\{ 4 x (1-x) \right\} 
%\label{eq:f2boxl}\\
%\sfs{2,box}{<\gam>,\ell}(x,Q^2,P^2) &\simeq 
%N_c \sum e_q^4\, \frac{\alpha}{\pi} x 
%  \bigg\{ \left[ x^2+(1-x)^2\right] \, \ln \frac{Q^2}{P^2x^2} 
%\nonumber\\
%& \phantom{\simeq N_c\sum e_q^4\, \frac{\alpha}{\pi} x\bigg\{} 
%+ 6x(1-x)
%-2 \bigg\}\, . 
%\label{eq:f2boxav}
%\end{align}
\begin{equation}
\begin{aligned}
\sfs{2,box}{\gam[T],\ell}(x,Q^2,P^2) &\simeq 
N_c \sum e_q^4\, \frac{\alpha}{\pi} x 
  \bigg\{ \left[ x^2+(1-x)^2\right] \, L
+ 8x(1-x)
-2 \bigg\}
\\
\sfs{2,box}{\gam[L],\ell}(x,Q^2,P^2) &\simeq 
N_c \sum e_q^4\, \frac{\alpha}{\pi} x \left\{ 4 x (1-x) \right\} 
\\
\sfs{2,box}{<\gam>,\ell}(x,Q^2,P^2) &\simeq 
N_c \sum e_q^4\, \frac{\alpha}{\pi} x 
  \bigg\{ \left[ x^2+(1-x)^2\right] L
+ 6x(1-x)
-2 \bigg\}\, . 
\end{aligned}
\end{equation}

%\begin{equation}
%\begin{aligned}
%\sigiit &\equiv \sigtt + \siglt \simeq
%N_c e_q^4 \frac{4 \pi \alpha^2}{Q^2} x
%\{(x^2+(1-x)^2)\ln \frac{Q^2}{P^2 x^2}+8 x (1-x)-2 \}%+\Ord(P^2/Q^2)
%\\
%\sigiil &\equiv \sigtl + \sigll \simeq
%N_c e_q^4 \frac{4 \pi \alpha^2}{Q^2} 4 x^2 (1-x)%+\Ord(P^2/Q^2)
%\\
%\sigma_{2<\gamma>} &\equiv \sigiit - \frac{1}{2} \sigiil \simeq
%N_c e_q^4 \frac{4 \pi \alpha^2}{Q^2} x
%\{(x^2+(1-x)^2)\ln \frac{Q^2}{P^2 x^2}+6 x (1-x)-2 \}%+\Ord(P^2/Q^2)
%\end{aligned}
%\end{equation}
%
%\section{$P^2=0, m^2 \ll Q^2$}\label{sec:bjorken3}
\subsection{Real Photon Limit: $P^2=0$}\label{sec:bjorken3}
%$+\Ord(\xp,\lambda,\xp/\lambda)$
For $P^2=0$ the virtual box results in 
\eqref{eq:sigtt}--\eqref{eq:tauatl}
%(\ref{eq:LO-box}) 
reduce to
the standard box-diagram $\vgg$ expressions
for a {\em real} photon $\gam \equiv \gam(P^2=0)$.
\subsubsection{Heavy Quark Contribution}
The heavy quark contribution becomes,
utilizing $\xp = 0$, $\vtwo = 1$ and $\displaystyle \lambda = \frac{4 m_h^2 x}{Q^2 (1-x)}$ and the normalization factor
$\Normh = N_c e_h^4 \frac{4 \pi \alpha^2}{Q^2} x$: 
\begin{align}\label{eq:really-heavy}
%\sigtt 
%& = N_c e_h^4 \frac{4 \pi \alpha^2}{Q^2}x
%\Theta(\vone^2)\ \bigg\{[x^2+(1-x)^2+\lambda (1-x)^2 -\frac{1}{2}\lambda^2 (1-x)^2 ]L
%\nonumber\\
%&+ \vone[4x(1-x)-1-(1-x)^2 \lambda]+\Ord(\xp,\tfrac{P^2}{m_h^2})\bigg\}
%\nonumber\\
\sigtt & = \Normh
\Theta(\vone^2)\ 
\bigg\{\left[x^2+(1-x)^2+x(1-x) \frac{4 m_h^2}{Q^2} 
-x^2\frac{8 m_h^4}{Q^4}\right]
\times
\nonumber\\
&\phantom{=}
\ln \frac{1+\vone}{1-\vone}
+ \vone\left[4x(1-x)-1-x(1-x)\frac{4 m_h^2}{Q^2}\right]
%+\Ord(\tfrac{P^2}{W^2},\tfrac{P^2}{m_h^2})
\bigg\}
\nonumber\\
%
%\siglt 
%& =N_c e_h^4 \frac{4 \pi \alpha^2}{Q^2} 4 x
%\Theta(\vone^2)\ \bigg\{\vone x(1-x)-x(1-x)\frac{\lambda}{2}L+\Ord(\xp,\tfrac{P^2}{m_h^2})\bigg\}
%\nonumber\\
\siglt & =\Normh
\Theta(\vone^2)\ 
\bigg\{-x^2 \frac{8 m_h^2}{Q^2} \ln \frac{1+\vone}{1-\vone}
+\vone\ 4x(1-x)
%+\Ord(\tfrac{P^2}{W^2},\tfrac{P^2}{m_h^2})
\bigg\}
\nonumber\\
%
%\sigtl & = 0 + \Ord(\xp,\tfrac{P^2}{m_h^2})
%\nonumber\\
%
%\sigll & = 0 + \Ord(\xp,\tfrac{P^2}{m_h^2})
%\nonumber\\
%
%\tautt & =N_c e_h^4 \frac{4 \pi \alpha^2}{Q^2}x\ 
%\Theta(\vone^2)\ \bigg\{
%\left[-2 \lambda (1-x)x -\frac{1}{2}\lambda^2 (1-x)^2\right]L-
%\vone \Big[2 x^2+(1-x)^2 \lambda\Big]
%+ \Ord(\xp,\tfrac{P^2}{m_h^2})\bigg\}
%\nonumber\\
%
\tautt & =\Normh
\Theta(\vone^2)\ \bigg\{
\left[-x^2 \frac{8 m_h^2}{Q^2}-x^2 \frac{8 m_h^4}{Q^4}\right]
\ln \frac{1+\vone}{1-\vone}
\nonumber\\
&\phantom{=}
-\vone \left[2 x^2+x(1-x) \frac{4 m_h^2}{Q^2}\right]
%+ \Ord(\tfrac{P^2}{W^2},\tfrac{P^2}{m_h^2})
\bigg\}
\nonumber\\
%
%\tautl & = N_c e_h^4 \frac{4 \pi \alpha^2}{Q^2}x\sqrt{x\xp}2(1-x)
%\Theta(\vone^2)\ \bigg\{[2x-\lambda (1-x)]L+ \vone(1-3x)+ \Ord(\xp,\tfrac{P^2}{m_h^2})\bigg\}
%\nonumber\\
%
\tauatt & = \Normh
\Theta(\vone^2)\ \bigg\{(2x-1)\ln \frac{1+\vone}{1-\vone}
+ \vone(3-4x)
%+ \Ord(\tfrac{P^2}{W^2},\tfrac{P^2}{m_h^2})
\bigg\}\ .
%\nonumber\\
%
%\tauatl & = 0 + \Ord(\xp,\tfrac{P^2}{m_h^2})
\end{align}
i.e., according to (\ref{eq:sfs2}) (or (\ref{eq:sfs1}))
\begin{align}\label{eq:vgamA.8}  
\sfs{2,box}{\gam,h}(x,Q^2)  & =  
%3\, e_h^4\, \frac{\alpha}{\pi}x\, \Theta(\vone^2) 
3 e_h^4 \frac{\alpha}{\pi}x\ \Theta(\vone^2) 
\bigg\{ \bigg[ x^2+(1-x)^2+x(1-3x)\, \frac{4m_h^2}{Q^2}
\nonumber\\ 
& 
-x^2\, \frac{8m_h^4}{Q^4}\bigg] \ln \frac{1+\vone}{1-\vone} 
+ \vone\bigg[ 8x(1-x)-1-x(1-x) \frac{4m_h^2}{Q^2}\bigg] \bigg\}
\nonumber\\ 
\sfs{L,box}{\gam,h}(x,Q^2)  & =  
%3\, e_h^4\, \frac{\alpha}{\pi}x\, \Theta(\vone^2) 
3 e_h^4 \frac{\alpha}{\pi}x\ \Theta(\vone^2) 
\left\{-x^2 \frac{8 m_h^2}{Q^2} \ln \frac{1+\vone}{1-\vone}
+\vone\ 4x(1-x)
%+\Ord(\tfrac{P^2}{W^2},\tfrac{P^2}{m_h^2})
\right\}
\end{align} 
which are the 
familiar massive Bethe-Heitler expressions 
\cite{cit:Wit-7601,*cit:GR-7901} relevant for the  
heavy quark contributions to the structure functions of real photons 
(cf.\ \cite{cit:GRSc99}, for example).  

%\begin{equation}
%\begin{aligned}
%\sigiit & =  \sigtt +\siglt %= \sigma_{2<\gamma>} + \Ord(\xp,\tfrac{P^2}{m_h^2}) 
%\\
%&= N_c e_h^4 \frac{4 \pi \alpha^2}{Q^2}x\ 
%\Theta(\vone^2)\ \bigg\{\left[x^2+(1-x)^2+x(1-3x) \frac{4 m_h^2}{Q^2} 
%-x^2\frac{8 m_h^4}{Q^4}\right]\ln \frac{1+\vone}{1-\vone}
%\\
%&+ \vone\left[8x(1-x)-1-x(1-x)\frac{4 m_h^2}{Q^2}\right]
%%+\Ord(\tfrac{P^2}{W^2},\tfrac{P^2}{m_h^2})
%\bigg\}
%\end{aligned}
%\end{equation}
\subsubsection{Light Quark Contribution}
In the light quark sector where 
$\lambda \ll 1$, i.e. $m^2\equiv m_q^2 \ll Q^2$, 
the logarithm can be written as
\begin{displaymath}
L = \ln \frac{1+\vone}{1-\vone} = 
\ln \frac{Q^2 (1-x)}{m_q^2 x} +\Ord(\lambda)\ .
\end{displaymath}
and  we obtain from (\ref{eq:Bj1}) or (\ref{eq:really-heavy}) 
the following 
results (neglecting terms of the order $\Ord(\lambda)$):
\begin{align}
\sigtt & \simeq \Norm
 \bigg\{[x^2+(1-x)^2]\ln \frac{Q^2 (1-x)}{m_q^2 x} + 4 x (1-x) - 1\bigg\}
\nonumber\\
\siglt & \simeq \Norm [4 x (1-x)]
\nonumber\\
%\sigtl & \simeq N_c e_q^4 \frac{4 \pi \alpha^2}{Q^2}
%\frac{P^2}{m_q^2}4 x^3 (1-x)^2 %+ \Ord(P^2/Q^2) 
%= 0 +\Ord(\xp,\lambda,\xp/\lambda)
%\\
%
%\sigll & \simeq 0
%\\
%
\tautt & \simeq \Norm [-2 x^2]
\nonumber\\
%
%\tautl & \simeq  N_c e_q^4 \frac{4 \pi \alpha^2}{Q^2}x\sqrt{x\xp}(1-x)
%\Big\{4x L + 2(1-3x)+\Ord(\xp,\lambda)\Big\}
%\\
%
\tauatt & \simeq \Norm \bigg\{(2x-1)\ln \frac{Q^2 (1-x)}{m_q^2 x}+3-4x\bigg\}\ ,
%\\
%
%\tauatl & \simeq 0 
%N_c e_q^4 \frac{4 \pi \alpha^2}{Q^2}\sqrt{x\xp} 8x(1-x) L
\end{align}
i.e., according to (\ref{eq:sfs2}) (or (\ref{eq:sfs1}))
\begin{equation}\label{eq:f2boxreal}
\sfs{2,box}{\gam,\ell}(x,Q^2)  \simeq N_c \sum e_q^4\, \frac{\alpha}{\pi} x  
 \bigg\{[x^2+(1-x)^2]\ln \frac{Q^2 (1-x)}{m_q^2 x} + 8 x (1-x) - 1\bigg\}\ .
\end{equation}
Note also that $\sigtl$ vanishes like $\sigtl \propto P^2/m_q^2$. 
%\begin{eqnarray}
%\sigiit &\equiv& \sigtt + \siglt \simeq 
%N_c e_q^4 \frac{4 \pi \alpha^2}{Q^2} x 
%\Big\{(x^2+(1-x)^2)L + 8 x (1-x) - 1\Big\}
%\end{eqnarray}
%\subsection{Heavy Quark Contribution: $0 \simeq P^2 \ll m^2, Q^2$}\label{sec:hqlim}
%Noticing that
%$\displaystyle L = \ln \frac{1+\vone}{1-\vone}+\Ord(\xp)$ and
%$\displaystyle \lambda = \frac{4 m^2 x}{Q^2 (1-x)}+\Ord(\tfrac{P^2}{W^2})$
%(with $W^2 \ge  4 m^2 \gg P^2$)
%the heavy quark contribution becomes (accurate 
%up to terms of the order $\Ord(\tfrac{P^2}{W^2},\tfrac{P^2}{m^2})$):

\begin{acknowledge}
I am grateful to E.\ Reya for suggesting this work and for
continuous encouragement.
I also wish to thank C.\ Sieg for a pleasant collaboration
on parts of Secs.~\ref{sec:photonsfs} and \ref{sec:fact1}.
This work has been supported in part by the 'Bundesministerium
f{\"u}r Bildung und Forschung', Berlin/Bonn.
\end{acknowledge}

%\begin{noteinproof}
%A note added in proof, if there is one, should be the final text before 
%the references.
%\end{noteinproof}

%\newpage
\bibliographystyle{/home/zylon/schien/Bibliography/test}
\bibliography{/home/zylon/schien/Bibliography/heavyquarks,/home/zylon/schien/Bibliography/photon}}

\begin{thebibliography}{10}
\expandafter\ifx\csname bibnamefont\endcsname\relax
  \def\bibnamefont#1{#1}\fi
\expandafter\ifx\csname bibfnamefont\endcsname\relax
  \def\bibfnamefont#1{#1}\fi
\expandafter\ifx\csname url\endcsname\relax
  \def\url#1{\texttt{#1}}\fi
\expandafter\ifx\csname urlprefix\endcsname\relax\def\urlprefix{URL }\fi
\expandafter\ifx\csname bibinfo\endcsname\relax \def\bibinfo#1#2{#2}\fi
\expandafter\ifx\csname eprint\endcsname\relax \def\eprint#1{#1}\fi

\bibitem{Brodsky:1971vm}
\bibinfo{author}{\bibfnamefont{S.~J.} \bibnamefont{Brodsky}},
  \bibinfo{author}{\bibfnamefont{T.}~\bibnamefont{Kinoshita}},
  \bibnamefont{and} \bibinfo{author}{\bibfnamefont{H.}~\bibnamefont{Terazawa}},
  \bibinfo{journal}{Phys. Rev. Lett.} \textbf{\bibinfo{volume}{27}},
  \bibinfo{pages}{280} (\bibinfo{year}{1971}).

\bibitem{Walsh:1971xy}
\bibinfo{author}{\bibfnamefont{T.~F.} \bibnamefont{Walsh}},
  \bibinfo{journal}{Phys. Lett.} \textbf{\bibinfo{volume}{B36}},
  \bibinfo{pages}{121} (\bibinfo{year}{1971}).

\bibitem{Nisius:1999cv}
\bibinfo{author}{\bibfnamefont{R.}~\bibnamefont{Nisius}},
  \bibinfo{journal}{Phys. Rept.} \textbf{\bibinfo{volume}{332}},
  \bibinfo{pages}{165} (\bibinfo{year}{2000}).

\bibitem{Krawczyk:2000mf}
\bibinfo{author}{\bibfnamefont{M.}~\bibnamefont{Krawczyk}},
  \bibinfo{author}{\bibfnamefont{A.}~\bibnamefont{Zembrzuski}},
  \bibnamefont{and} \bibinfo{author}{\bibfnamefont{M.}~\bibnamefont{Staszel}},
  \bibinfo{journal}{Phys. Rept.} \textbf{\bibinfo{volume}{345}},
  \bibinfo{pages}{265} (\bibinfo{year}{2001}).

\bibitem{cit:GRV-9201}
\bibinfo{author}{\bibfnamefont{M.}~\bibnamefont{Gl{\"u}ck}},
  \bibinfo{author}{\bibfnamefont{E.}~\bibnamefont{Reya}}, \bibnamefont{and}
  \bibinfo{author}{\bibfnamefont{A.}~\bibnamefont{Vogt}},
  \bibinfo{journal}{Phys. Rev.} \textbf{\bibinfo{volume}{D45}},
  \bibinfo{pages}{3986} (\bibinfo{year}{1992}).

\bibitem{cit:GRV-9202}
\bibinfo{author}{\bibfnamefont{M.}~\bibnamefont{Gl{\"u}ck}},
  \bibinfo{author}{\bibfnamefont{E.}~\bibnamefont{Reya}}, \bibnamefont{and}
  \bibinfo{author}{\bibfnamefont{A.}~\bibnamefont{Vogt}},
  \bibinfo{journal}{Phys. Rev.} \textbf{\bibinfo{volume}{D46}},
  \bibinfo{pages}{1973} (\bibinfo{year}{1992}).

\bibitem{cit:AFG-9401}
\bibinfo{author}{\bibfnamefont{P.}~\bibnamefont{Aurenche}},
  \bibinfo{author}{\bibfnamefont{M.}~\bibnamefont{Fontannaz}},
  \bibnamefont{and} \bibinfo{author}{\bibfnamefont{J.-P.}
  \bibnamefont{Guillet}}, \bibinfo{journal}{Z. Phys.}
  \textbf{\bibinfo{volume}{C64}}, \bibinfo{pages}{621} (\bibinfo{year}{1994}).

\bibitem{cit:SaS-9501}
\bibinfo{author}{\bibfnamefont{G.~A.} \bibnamefont{Schuler}} \bibnamefont{and}
  \bibinfo{author}{\bibfnamefont{T.}~\bibnamefont{Sj{\"o}strand}},
  \bibinfo{journal}{Z. Phys.} \textbf{\bibinfo{volume}{C68}},
  \bibinfo{pages}{607} (\bibinfo{year}{1995}).

\bibitem{Schuler:1996fc}
\bibinfo{author}{\bibfnamefont{G.~A.} \bibnamefont{Schuler}} \bibnamefont{and}
  \bibinfo{author}{\bibfnamefont{T.}~\bibnamefont{Sj{\"o}strand}},
  \bibinfo{journal}{Phys. Lett.} \textbf{\bibinfo{volume}{B376}},
  \bibinfo{pages}{193} (\bibinfo{year}{1996}).

\bibitem{Gordon:1997pm}
\bibinfo{author}{\bibfnamefont{L.~E.} \bibnamefont{Gordon}} \bibnamefont{and}
  \bibinfo{author}{\bibfnamefont{J.~K.} \bibnamefont{Storrow}},
  \bibinfo{journal}{Nucl. Phys.} \textbf{\bibinfo{volume}{B489}},
  \bibinfo{pages}{405} (\bibinfo{year}{1997}).

\bibitem{cit:GRSc99}
\bibinfo{author}{\bibfnamefont{M.}~\bibnamefont{Gl{\"u}ck}},
  \bibinfo{author}{\bibfnamefont{E.}~\bibnamefont{Reya}}, \bibnamefont{and}
  \bibinfo{author}{\bibfnamefont{I.}~\bibnamefont{Schienbein}},
  \bibinfo{journal}{Phys. Rev.} \textbf{\bibinfo{volume}{D60}},
  \bibinfo{pages}{054019} (\bibinfo{year}{1999}), \bibinfo{note}{{Erratum: {\bf
  D62} (2000) 019902}}.

\bibitem{Gluck:1992fy}
\bibinfo{author}{\bibfnamefont{M.}~\bibnamefont{Gl{\"u}ck}} \bibnamefont{and}
  \bibinfo{author}{\bibfnamefont{W.}~\bibnamefont{Vogelsang}},
  \bibinfo{journal}{Z. Phys.} \textbf{\bibinfo{volume}{C55}},
  \bibinfo{pages}{353} (\bibinfo{year}{1992}).

\bibitem{Gluck:1993zq}
\bibinfo{author}{\bibfnamefont{M.}~\bibnamefont{Gl{\"u}ck}} \bibnamefont{and}
  \bibinfo{author}{\bibfnamefont{W.}~\bibnamefont{Vogelsang}},
  \bibinfo{journal}{Z. Phys.} \textbf{\bibinfo{volume}{C57}},
  \bibinfo{pages}{309} (\bibinfo{year}{1993}).

\bibitem{Gluck:1994ee}
\bibinfo{author}{\bibfnamefont{M.}~\bibnamefont{Gl{\"u}ck}},
  \bibinfo{author}{\bibfnamefont{M.}~\bibnamefont{Stratmann}},
  \bibnamefont{and}
  \bibinfo{author}{\bibfnamefont{W.}~\bibnamefont{Vogelsang}},
  \bibinfo{journal}{Phys. Lett.} \textbf{\bibinfo{volume}{B337}},
  \bibinfo{pages}{373} (\bibinfo{year}{1994}).

\bibitem{Stratmann:1996an}
\bibinfo{author}{\bibfnamefont{M.}~\bibnamefont{Stratmann}} \bibnamefont{and}
  \bibinfo{author}{\bibfnamefont{W.}~\bibnamefont{Vogelsang}},
  \bibinfo{journal}{Phys. Lett.} \textbf{\bibinfo{volume}{B386}},
  \bibinfo{pages}{370} (\bibinfo{year}{1996}).

\bibitem{Gluck:2001az}
\bibinfo{author}{\bibfnamefont{M.}~\bibnamefont{Gl{\"u}ck}},
  \bibinfo{author}{\bibfnamefont{E.}~\bibnamefont{Reya}}, \bibnamefont{and}
  \bibinfo{author}{\bibfnamefont{C.}~\bibnamefont{Sieg}},
  \bibinfo{journal}{Phys. Lett.} \textbf{\bibinfo{volume}{B503}},
  \bibinfo{pages}{285} (\bibinfo{year}{2001}).

\bibitem{Gluck:2001rn}
\bibinfo{author}{\bibfnamefont{M.}~\bibnamefont{Gl{\"u}ck}},
  \bibinfo{author}{\bibfnamefont{E.}~\bibnamefont{Reya}}, \bibnamefont{and}
  \bibinfo{author}{\bibfnamefont{C.}~\bibnamefont{Sieg}},
  \bibinfo{journal}{Eur. Phys. J.} \textbf{\bibinfo{volume}{C20}},
  \bibinfo{pages}{271} (\bibinfo{year}{2001}).

\bibitem{Stratmann:1999bv}
\bibinfo{author}{\bibfnamefont{M.}~\bibnamefont{Stratmann}} \bibnamefont{and}
  \bibinfo{author}{\bibfnamefont{W.}~\bibnamefont{Vogelsang}},
  \bibinfo{journal}{Nucl. Phys. Proc. Suppl.} \textbf{\bibinfo{volume}{82}},
  \bibinfo{pages}{400} (\bibinfo{year}{2000}).

\bibitem{Stratmann:1997xy}
\bibinfo{author}{\bibfnamefont{M.}~\bibnamefont{Stratmann}} \bibnamefont{and}
  \bibinfo{author}{\bibfnamefont{W.}~\bibnamefont{Vogelsang}},
  \bibinfo{journal}{Z. Phys.} \textbf{\bibinfo{volume}{C74}},
  \bibinfo{pages}{641} (\bibinfo{year}{1997}).

\bibitem{Stratmann:2000yd}
\bibinfo{author}{\bibfnamefont{M.}~\bibnamefont{Stratmann}},
  \emph{\bibinfo{title}{Determining the spin structure of the photon at future
  colliders}}, \bibinfo{note}{{talk presented at the 8th International Workshop
  on Deep Inelastic Scattering and QCD (DIS 2000), Liverpool, England, 25-30
  Apr 2000}}, \eprint{hep-ph/0006285}.

\bibitem{Borzumati:1993za}
\bibinfo{author}{\bibfnamefont{F.~M.} \bibnamefont{Borzumati}}
  \bibnamefont{and} \bibinfo{author}{\bibfnamefont{G.~A.}
  \bibnamefont{Schuler}}, \bibinfo{journal}{Z. Phys.}
  \textbf{\bibinfo{volume}{C58}}, \bibinfo{pages}{139} (\bibinfo{year}{1993}).

\bibitem{Drees:1994eu}
\bibinfo{author}{\bibfnamefont{M.}~\bibnamefont{Drees}} \bibnamefont{and}
  \bibinfo{author}{\bibfnamefont{R.~M.} \bibnamefont{Godbole}},
  \bibinfo{journal}{Phys. Rev.} \textbf{\bibinfo{volume}{D50}},
  \bibinfo{pages}{3124} (\bibinfo{year}{1994}).

\bibitem{cit:GRS95}
\bibinfo{author}{\bibfnamefont{M.}~\bibnamefont{Gl{\"u}ck}},
  \bibinfo{author}{\bibfnamefont{E.}~\bibnamefont{Reya}}, \bibnamefont{and}
  \bibinfo{author}{\bibfnamefont{M.}~\bibnamefont{Stratmann}},
  \bibinfo{journal}{Phys. Rev.} \textbf{\bibinfo{volume}{D51}},
  \bibinfo{pages}{3220} (\bibinfo{year}{1995}).

\bibitem{Uematsu:1981qy}
\bibinfo{author}{\bibfnamefont{T.}~\bibnamefont{Uematsu}} \bibnamefont{and}
  \bibinfo{author}{\bibfnamefont{T.~F.} \bibnamefont{Walsh}},
  \bibinfo{journal}{Phys. Lett.} \textbf{\bibinfo{volume}{B101}},
  \bibinfo{pages}{263} (\bibinfo{year}{1981}).

\bibitem{Uematsu:1982je}
\bibinfo{author}{\bibfnamefont{T.}~\bibnamefont{Uematsu}} \bibnamefont{and}
  \bibinfo{author}{\bibfnamefont{T.~F.} \bibnamefont{Walsh}},
  \bibinfo{journal}{Nucl. Phys.} \textbf{\bibinfo{volume}{B199}},
  \bibinfo{pages}{93} (\bibinfo{year}{1982}).

\bibitem{Rossi:1984xz}
\bibinfo{author}{\bibfnamefont{G.}~\bibnamefont{Rossi}},
  \bibinfo{journal}{Phys. Rev.} \textbf{\bibinfo{volume}{D29}},
  \bibinfo{pages}{852} (\bibinfo{year}{1984}).

\bibitem{cit:Rossi-PhD}
\bibinfo{author}{\bibfnamefont{G.}~\bibnamefont{Rossi}},
  \emph{\bibinfo{title}{Virtual Photon Structure Functions in Quantum
  Chromodynamics}}, Ph.D. thesis, \bibinfo{school}{UC San Diego}
  (\bibinfo{year}{1983}), \bibinfo{note}{{UC San Diego report UCSD--10P10-227
  (unpublished)}}.

\bibitem{Ibes:1990pj}
\bibinfo{author}{\bibfnamefont{W.}~\bibnamefont{Ibes}} \bibnamefont{and}
  \bibinfo{author}{\bibfnamefont{T.~F.} \bibnamefont{Walsh}},
  \bibinfo{journal}{Phys. Lett.} \textbf{\bibinfo{volume}{B251}},
  \bibinfo{pages}{450} (\bibinfo{year}{1990}).

\bibitem{Sasaki:1998vb}
\bibinfo{author}{\bibfnamefont{K.}~\bibnamefont{Sasaki}} \bibnamefont{and}
  \bibinfo{author}{\bibfnamefont{T.}~\bibnamefont{Uematsu}},
  \bibinfo{journal}{Phys. Rev.} \textbf{\bibinfo{volume}{D59}},
  \bibinfo{pages}{114011} (\bibinfo{year}{1999}).

\bibitem{Sasaki:1999py}
\bibinfo{author}{\bibfnamefont{K.}~\bibnamefont{Sasaki}} \bibnamefont{and}
  \bibinfo{author}{\bibfnamefont{T.}~\bibnamefont{Uematsu}},
  \bibinfo{journal}{Phys. Lett.} \textbf{\bibinfo{volume}{B473}},
  \bibinfo{pages}{309} (\bibinfo{year}{2000}).

\bibitem{Sasaki:2000zc}
\bibinfo{author}{\bibfnamefont{K.}~\bibnamefont{Sasaki}} \bibnamefont{and}
  \bibinfo{author}{\bibfnamefont{T.}~\bibnamefont{Uematsu}},
  \bibinfo{journal}{Eur. Phys. J.} \textbf{\bibinfo{volume}{C20}},
  \bibinfo{pages}{283} (\bibinfo{year}{2001}).

\bibitem{cit:Bud-7501}
\bibinfo{author}{\bibfnamefont{V.~M.} \bibnamefont{Budnev}},
  \bibinfo{author}{\bibfnamefont{I.~F.} \bibnamefont{Ginzburg}},
  \bibinfo{author}{\bibfnamefont{G.~V.} \bibnamefont{Meledin}},
  \bibnamefont{and} \bibinfo{author}{\bibfnamefont{V.~G.} \bibnamefont{Serbo}},
  \bibinfo{journal}{Phys. Rep.} \textbf{\bibinfo{volume}{15}},
  \bibinfo{pages}{181} (\bibinfo{year}{1975}).

\bibitem{Chyla:2000hp}
\bibinfo{author}{\bibfnamefont{J.}~\bibnamefont{Ch{\'y}la}},
  \bibinfo{journal}{Phys. Lett.} \textbf{\bibinfo{volume}{B488}},
  \bibinfo{pages}{289} (\bibinfo{year}{2000}).

\bibitem{Chyla:2000cu}
\bibinfo{author}{\bibfnamefont{J.}~\bibnamefont{Ch{\'y}la}} \bibnamefont{and}
  \bibinfo{author}{\bibfnamefont{M.}~\bibnamefont{Ta{\v s}evsk{\'y}}},
  \bibinfo{journal}{Eur. Phys. J.} \textbf{\bibinfo{volume}{C16}},
  \bibinfo{pages}{471} (\bibinfo{year}{2000}).

\bibitem{Chyla:2000ue}
\bibinfo{author}{\bibfnamefont{J.}~\bibnamefont{Ch{\'y}la}} \bibnamefont{and}
  \bibinfo{author}{\bibfnamefont{M.}~\bibnamefont{Ta{\v s}evsk{\'y}}},
  \bibinfo{journal}{Eur. Phys. J.} \textbf{\bibinfo{volume}{C18}},
  \bibinfo{pages}{723} (\bibinfo{year}{2001}).

\bibitem{Friberg:2000nx}
\bibinfo{author}{\bibfnamefont{C.}~\bibnamefont{Friberg}} \bibnamefont{and}
  \bibinfo{author}{\bibfnamefont{T.}~\bibnamefont{Sj{\"o}strand}},
  \bibinfo{journal}{Phys. Lett.} \textbf{\bibinfo{volume}{B492}},
  \bibinfo{pages}{123} (\bibinfo{year}{2000}).

\bibitem{Bhattacharya:1977re}
\bibinfo{author}{\bibfnamefont{R.}~\bibnamefont{Bhattacharya}},
  \bibinfo{author}{\bibfnamefont{J.}~\bibnamefont{Smith}}, \bibnamefont{and}
  \bibinfo{author}{\bibfnamefont{G.}~\bibnamefont{Grammer}},
  \bibinfo{journal}{Phys. Rev.} \textbf{\bibinfo{volume}{D15}},
  \bibinfo{pages}{3267} (\bibinfo{year}{1977}).

\bibitem{Brown:1971pk}
\bibinfo{author}{\bibfnamefont{R.~W.} \bibnamefont{Brown}} \bibnamefont{and}
  \bibinfo{author}{\bibfnamefont{I.~J.} \bibnamefont{Muzinich}},
  \bibinfo{journal}{Phys. Rev.} \textbf{\bibinfo{volume}{D4}},
  \bibinfo{pages}{1496} (\bibinfo{year}{1971}).

\bibitem{Carlson:1971pk}
\bibinfo{author}{\bibfnamefont{C.~E.} \bibnamefont{Carlson}} \bibnamefont{and}
  \bibinfo{author}{\bibfnamefont{W.-K.} \bibnamefont{Tung}},
  \bibinfo{journal}{Phys. Rev.} \textbf{\bibinfo{volume}{D4}},
  \bibinfo{pages}{2873} (\bibinfo{year}{1971}).

\bibitem{Manohar:1992tz}
\bibinfo{author}{\bibfnamefont{A.~V.} \bibnamefont{Manohar}},
  \emph{\bibinfo{title}{An introduction to spin dependent deep inelastic
  scattering}}, \bibinfo{note}{{lectures given at Lake Louise Winter Inst.,
  Lake Louise, Canada, Feb 23-29, 1992}}, \eprint{hep-ph/9204208}.

\bibitem{Sasaki:2001pc}
\bibinfo{author}{\bibfnamefont{K.}~\bibnamefont{Sasaki}},
  \bibinfo{author}{\bibfnamefont{J.}~\bibnamefont{Soffer}}, \bibnamefont{and}
  \bibinfo{author}{\bibfnamefont{T.}~\bibnamefont{Uematsu}},
  \bibinfo{journal}{Phys. Lett.} \textbf{\bibinfo{volume}{B522}},
  \bibinfo{pages}{22} (\bibinfo{year}{2001}).

\bibitem{cit:Berger-Rep}
\bibinfo{author}{\bibfnamefont{C.}~\bibnamefont{Berger}} \bibnamefont{and}
  \bibinfo{author}{\bibfnamefont{W.}~\bibnamefont{Wagner}},
  \bibinfo{journal}{Phys. Rep.} \textbf{\bibinfo{volume}{146}},
  \bibinfo{pages}{1} (\bibinfo{year}{1987}).

\bibitem{cit:math}
\bibinfo{author}{\bibfnamefont{S.}~\bibnamefont{Wolfram}},
  \emph{\bibinfo{title}{Mathematica --- Ver. 3 or higher}},
  \bibinfo{organization}{Wolfram Research} (\bibinfo{year}{1997}).

\bibitem{Jamin:1993dp}
\bibinfo{author}{\bibfnamefont{M.}~\bibnamefont{Jamin}} \bibnamefont{and}
  \bibinfo{author}{\bibfnamefont{M.~E.} \bibnamefont{Lautenbacher}},
  \bibinfo{journal}{Comput. Phys. Commun.} \textbf{\bibinfo{volume}{74}},
  \bibinfo{pages}{265} (\bibinfo{year}{1993}).

\bibitem{Kretzer:1997pd}
\bibinfo{author}{\bibfnamefont{S.}~\bibnamefont{Kretzer}} \bibnamefont{and}
  \bibinfo{author}{\bibfnamefont{I.}~\bibnamefont{Schienbein}},
  \bibinfo{journal}{Phys. Rev.} \textbf{\bibinfo{volume}{D56}},
  \bibinfo{pages}{1804} (\bibinfo{year}{1997}).

\bibitem{Schienbein:1997sb}
\bibinfo{author}{\bibfnamefont{I.}~\bibnamefont{Schienbein}},
  \bibinfo{journal}{Phys. Rev.} \textbf{\bibinfo{volume}{D59}},
  \bibinfo{pages}{013001} (\bibinfo{year}{1999}).

\bibitem{Nisius:1998ue}
\bibinfo{author}{\bibfnamefont{R.}~\bibnamefont{Nisius}} \bibnamefont{and}
  \bibinfo{author}{\bibfnamefont{M.~H.} \bibnamefont{Seymour}},
  \bibinfo{journal}{Phys. Lett.} \textbf{\bibinfo{volume}{B452}},
  \bibinfo{pages}{409} (\bibinfo{year}{1999}).

\bibitem{Gluck:2000sn}
\bibinfo{author}{\bibfnamefont{M.}~\bibnamefont{Gl{\"u}ck}},
  \bibinfo{author}{\bibfnamefont{E.}~\bibnamefont{Reya}}, \bibnamefont{and}
  \bibinfo{author}{\bibfnamefont{I.}~\bibnamefont{Schienbein}},
  \bibinfo{journal}{Phys. Rev.} \textbf{\bibinfo{volume}{D63}},
  \bibinfo{pages}{074008} (\bibinfo{year}{2001}).

\bibitem{Schienbein:2001cd}
\bibinfo{author}{\bibfnamefont{I.}~\bibnamefont{Schienbein}},
  \emph{\bibinfo{title}{Heavy quark production in CC and NC DIS and the
  structure of real and virtual photons in NLO QCD}}, Ph.D. thesis,
  \bibinfo{school}{University of Dortmund} (\bibinfo{year}{2001}),
  \eprint{hep-ph/0110292}.

\bibitem{Brodsky:1997sd}
\bibinfo{author}{\bibfnamefont{S.~J.} \bibnamefont{Brodsky}},
  \bibinfo{author}{\bibfnamefont{F.}~\bibnamefont{Hautmann}}, \bibnamefont{and}
  \bibinfo{author}{\bibfnamefont{D.~E.} \bibnamefont{Soper}},
  \bibinfo{journal}{Phys. Rev.} \textbf{\bibinfo{volume}{D56}},
  \bibinfo{pages}{6957} (\bibinfo{year}{1997}).

\bibitem{Brodsky:1997sg}
\bibinfo{author}{\bibfnamefont{S.~J.} \bibnamefont{Brodsky}},
  \bibinfo{author}{\bibfnamefont{F.}~\bibnamefont{Hautmann}}, \bibnamefont{and}
  \bibinfo{author}{\bibfnamefont{D.~E.} \bibnamefont{Soper}},
  \bibinfo{journal}{Phys. Rev. Lett.} \textbf{\bibinfo{volume}{78}},
  \bibinfo{pages}{803} (\bibinfo{year}{1997}), \bibinfo{note}{{Erratum: {\bf
  79} (1997) 3544}}.

\bibitem{Bartels:1996ke}
\bibinfo{author}{\bibfnamefont{J.}~\bibnamefont{Bartels}},
  \bibinfo{author}{\bibfnamefont{A.}~\bibnamefont{De~Roeck}}, \bibnamefont{and}
  \bibinfo{author}{\bibfnamefont{H.}~\bibnamefont{Lotter}},
  \bibinfo{journal}{Phys. Lett.} \textbf{\bibinfo{volume}{B389}},
  \bibinfo{pages}{742} (\bibinfo{year}{1996}).

\bibitem{Bartels:1997er}
\bibinfo{author}{\bibfnamefont{J.}~\bibnamefont{Bartels}},
  \bibinfo{author}{\bibfnamefont{A.}~\bibnamefont{De~Roeck}},
  \bibinfo{author}{\bibfnamefont{C.}~\bibnamefont{Ewerz}}, \bibnamefont{and}
  \bibinfo{author}{\bibfnamefont{H.}~\bibnamefont{Lotter}},
  \eprint{hep-ph/9710500}.

\bibitem{Bartels:2000sk}
\bibinfo{author}{\bibfnamefont{J.}~\bibnamefont{Bartels}},
  \bibinfo{author}{\bibfnamefont{C.}~\bibnamefont{Ewerz}}, \bibnamefont{and}
  \bibinfo{author}{\bibfnamefont{R.}~\bibnamefont{Staritzbichler}},
  \bibinfo{journal}{Phys. Lett.} \textbf{\bibinfo{volume}{B492}},
  \bibinfo{pages}{56} (\bibinfo{year}{2000}).

\bibitem{Kwiecinski:1999yx}
\bibinfo{author}{\bibfnamefont{J.}~\bibnamefont{Kwieci{\'n}ski}}
  \bibnamefont{and} \bibinfo{author}{\bibfnamefont{L.}~\bibnamefont{Motyka}},
  \bibinfo{journal}{Phys. Lett.} \textbf{\bibinfo{volume}{B462}},
  \bibinfo{pages}{203} (\bibinfo{year}{1999}).

\bibitem{Kwiecinski:2000zs}
\bibinfo{author}{\bibfnamefont{J.}~\bibnamefont{Kwieci{\'n}ski}}
  \bibnamefont{and} \bibinfo{author}{\bibfnamefont{L.}~\bibnamefont{Motyka}},
  \bibinfo{journal}{Eur. Phys. J.} \textbf{\bibinfo{volume}{C18}},
  \bibinfo{pages}{343} (\bibinfo{year}{2000}).

\bibitem{Fadin:1975cb}
\bibinfo{author}{\bibfnamefont{V.~S.} \bibnamefont{Fadin}},
  \bibinfo{author}{\bibfnamefont{E.~A.} \bibnamefont{Kuraev}},
  \bibnamefont{and} \bibinfo{author}{\bibfnamefont{L.~N.}
  \bibnamefont{Lipatov}}, \bibinfo{journal}{Phys. Lett.}
  \textbf{\bibinfo{volume}{B60}}, \bibinfo{pages}{50} (\bibinfo{year}{1975}).

\bibitem{Kuraev:1976ge}
\bibinfo{author}{\bibfnamefont{E.~A.} \bibnamefont{Kuraev}},
  \bibinfo{author}{\bibfnamefont{L.~N.} \bibnamefont{Lipatov}},
  \bibnamefont{and} \bibinfo{author}{\bibfnamefont{V.~S.} \bibnamefont{Fadin}},
  \bibinfo{journal}{Sov. Phys. JETP} \textbf{\bibinfo{volume}{44}},
  \bibinfo{pages}{443} (\bibinfo{year}{1976}).

\bibitem{Kuraev:1977fs}
\bibinfo{author}{\bibfnamefont{E.~A.} \bibnamefont{Kuraev}},
  \bibinfo{author}{\bibfnamefont{L.~N.} \bibnamefont{Lipatov}},
  \bibnamefont{and} \bibinfo{author}{\bibfnamefont{V.~S.} \bibnamefont{Fadin}},
  \bibinfo{journal}{Sov. Phys. JETP} \textbf{\bibinfo{volume}{45}},
  \bibinfo{pages}{199} (\bibinfo{year}{1977}).

\bibitem{Balitsky:1978ic}
\bibinfo{author}{\bibfnamefont{I.~I.} \bibnamefont{Balitsky}} \bibnamefont{and}
  \bibinfo{author}{\bibfnamefont{L.~N.} \bibnamefont{Lipatov}},
  \bibinfo{journal}{Sov. J. Nucl. Phys.} \textbf{\bibinfo{volume}{28}},
  \bibinfo{pages}{822} (\bibinfo{year}{1978}).

\bibitem{Achard:2001kr}
\bibinfo{author}{\bibfnamefont{P.}~\bibnamefont{Achard}} \emph{et~al.},
  \bibinfo{collaboration}{L3} Collaboration, \bibinfo{journal}{Phys. Lett.}
  \textbf{\bibinfo{volume}{B531}}, \bibinfo{pages}{39} (\bibinfo{year}{2002}).

\bibitem{Abbiendi:2001tv}
\bibinfo{author}{\bibfnamefont{G.}~\bibnamefont{Abbiendi}} \emph{et~al.},
  \bibinfo{collaboration}{OPAL} Collaboration,
  \emph{\bibinfo{title}{Measurement of the hadronic cross-section for the
  scattering of two virtual photons at LEP}}, \eprint{hep-ex/0110006}.

\bibitem{Cacciari:2000cb}
\bibinfo{author}{\bibfnamefont{M.}~\bibnamefont{Cacciari}},
  \bibinfo{author}{\bibfnamefont{V.}~\bibnamefont{{Del Duca}}},
  \bibinfo{author}{\bibfnamefont{S.}~\bibnamefont{Frixione}}, \bibnamefont{and}
  \bibinfo{author}{\bibfnamefont{Z.}~\bibnamefont{Tr{\'o}cs{\'a}nyi}},
  \bibinfo{journal}{JHEP} \textbf{\bibinfo{volume}{02}}, \bibinfo{pages}{029}
  (\bibinfo{year}{2001}).

\bibitem{Brodsky:2001ye}
\bibinfo{author}{\bibfnamefont{S.~J.} \bibnamefont{Brodsky}},
  \bibinfo{author}{\bibfnamefont{V.~S.} \bibnamefont{Fadin}},
  \bibinfo{author}{\bibfnamefont{V.~T.} \bibnamefont{Kim}},
  \bibinfo{author}{\bibfnamefont{L.~N.} \bibnamefont{Lipatov}},
  \bibnamefont{and} \bibinfo{author}{\bibfnamefont{G.~B.}
  \bibnamefont{Pivovarov}}, \emph{\bibinfo{title}{High-energy asymptotics of
  photon photon collisions in QCD}} (\bibinfo{year}{2001}),
  \bibinfo{note}{{talk given at the 14th International Workshop on
  Photon-Photon Collisions (Photon 2001), Ascona, Switzerland, 2-7 Sep 2001}},
  \eprint{hep-ph/0111390}.

\bibitem{Walsh:1973mz}
\bibinfo{author}{\bibfnamefont{T.~F.} \bibnamefont{Walsh}} \bibnamefont{and}
  \bibinfo{author}{\bibfnamefont{P.}~\bibnamefont{Zerwas}},
  \bibinfo{journal}{Phys. Lett.} \textbf{\bibinfo{volume}{B44}},
  \bibinfo{pages}{195} (\bibinfo{year}{1973}).

\bibitem{Zerwas:1974tf}
\bibinfo{author}{\bibfnamefont{P.~M.} \bibnamefont{Zerwas}},
  \bibinfo{journal}{Phys. Rev.} \textbf{\bibinfo{volume}{D10}},
  \bibinfo{pages}{1485} (\bibinfo{year}{1974}).

\bibitem{Kingsley:1973wk}
\bibinfo{author}{\bibfnamefont{R.~L.} \bibnamefont{Kingsley}},
  \bibinfo{journal}{Nucl. Phys.} \textbf{\bibinfo{volume}{B60}},
  \bibinfo{pages}{45} (\bibinfo{year}{1973}).

\bibitem{Chernyak:1974nv}
\bibinfo{author}{\bibfnamefont{V.~L.} \bibnamefont{Chernyak}} \bibnamefont{and}
  \bibinfo{author}{\bibfnamefont{V.~G.} \bibnamefont{Serbo}},
  \bibinfo{journal}{Nucl. Phys.} \textbf{\bibinfo{volume}{B71}},
  \bibinfo{pages}{395} (\bibinfo{year}{1974}).

\bibitem{Worden:1974hc}
\bibinfo{author}{\bibfnamefont{R.~P.} \bibnamefont{Worden}},
  \bibinfo{journal}{Phys. Lett.} \textbf{\bibinfo{volume}{B51}},
  \bibinfo{pages}{57} (\bibinfo{year}{1974}).

\bibitem{Ahmed:1975ff}
\bibinfo{author}{\bibfnamefont{M.~A.} \bibnamefont{Ahmed}} \bibnamefont{and}
  \bibinfo{author}{\bibfnamefont{G.~G.} \bibnamefont{Ross}},
  \bibinfo{journal}{Phys. Lett.} \textbf{\bibinfo{volume}{B59}},
  \bibinfo{pages}{369} (\bibinfo{year}{1975}).

\bibitem{Collins:1989gx}
\bibinfo{author}{\bibfnamefont{J.~C.} \bibnamefont{Collins}},
  \bibinfo{author}{\bibfnamefont{D.~E.} \bibnamefont{Soper}}, \bibnamefont{and}
  \bibinfo{author}{\bibfnamefont{G.}~\bibnamefont{Sterman}}, in
  \emph{\bibinfo{booktitle}{Perturbative Quantum Chromodynamics}}, edited by
  \bibinfo{editor}{\bibnamefont{{A. H. Mueller}}} (\bibinfo{publisher}{World
  Scientific}, \bibinfo{year}{1989}).

\bibitem{Collins:1987pm}
\bibinfo{author}{\bibfnamefont{J.~C.} \bibnamefont{Collins}} \bibnamefont{and}
  \bibinfo{author}{\bibfnamefont{D.~E.} \bibnamefont{Soper}},
  \bibinfo{journal}{Ann. Rev. Nucl. Part. Sci.} \textbf{\bibinfo{volume}{37}},
  \bibinfo{pages}{383} (\bibinfo{year}{1987}).

\bibitem{Hoodbhoy:1989am}
\bibinfo{author}{\bibfnamefont{P.}~\bibnamefont{Hoodbhoy}},
  \bibinfo{author}{\bibfnamefont{R.~L.} \bibnamefont{Jaffe}}, \bibnamefont{and}
  \bibinfo{author}{\bibfnamefont{A.}~\bibnamefont{Manohar}},
  \bibinfo{journal}{Nucl. Phys.} \textbf{\bibinfo{volume}{B312}},
  \bibinfo{pages}{571} (\bibinfo{year}{1989}).

\bibitem{Mathews:1996yv}
\bibinfo{author}{\bibfnamefont{P.}~\bibnamefont{Mathews}} \bibnamefont{and}
  \bibinfo{author}{\bibfnamefont{V.}~\bibnamefont{Ravindran}},
  \bibinfo{journal}{Int. J. Mod. Phys.} \textbf{\bibinfo{volume}{A11}},
  \bibinfo{pages}{2783} (\bibinfo{year}{1996}).

\bibitem{Sedlak:2001pr}
\bibinfo{author}{\bibfnamefont{K.}~\bibnamefont{Sedl{\'a}k}},
  \bibinfo{collaboration}{H1} Collaboration, \emph{\bibinfo{title}{Structure of
  virtual photons at HERA}} (\bibinfo{year}{2001}), \bibinfo{note}{{talk given
  at the 14th International Workshop on Photon-Photon Collisions (Photon 2001),
  Ascona, Switzerland, 2-7 Sep 2001}}, \eprint{hep-ex/0111019}.

\bibitem{cit:Wit-7601}
\bibinfo{author}{\bibfnamefont{E.}~\bibnamefont{Witten}},
  \bibinfo{journal}{Nucl. Phys.} \textbf{\bibinfo{volume}{B104}},
  \bibinfo{pages}{445} (\bibinfo{year}{1976}).

\bibitem{cit:GR-7901}
\bibinfo{author}{\bibfnamefont{M.}~\bibnamefont{Gl{\"u}ck}} \bibnamefont{and}
  \bibinfo{author}{\bibfnamefont{E.}~\bibnamefont{Reya}},
  \bibinfo{journal}{Phys. Lett.} \textbf{\bibinfo{volume}{83B}},
  \bibinfo{pages}{98} (\bibinfo{year}{1979}).

\end{thebibliography}

\end{document}